# Controllable Thermal Conductivity in Twisted Homogeneous Interfaces of Graphene and Hexagonal Boron Nitride

Wengen Ouyang,[1] Huasong Qin,[2] Michael Urbakh,[1*] and Oded Hod[1]

[1]*Department of Physical Chemistry, School of Chemistry, The Raymond and Beverly Sackler Faculty of Exact Sciences and The Sackler Center for Computational Molecular and Materials Science, Tel Aviv University, Tel Aviv 6997801, Israel.*

[2]*State Key Laboratory for Strength and Vibration of Mechanical Structures, School of Aerospace, Xi'an Jiaotong University, Xi'an 710049, China*

ABSTRACT

Thermal conductivity of homogeneous twisted stacks of graphite is found to strongly depend on the misfit angle. The underlying mechanism relies on the angle dependence of phonon-phonon couplings across the twisted interface. Excellent agreement between the calculated thermal conductivity of narrow graphitic stacks and corresponding experimental results indicates the validity of the predictions. This is attributed to the accuracy of interlayer interactions descriptions obtained by the dedicated registry-dependent interlayer potential used. Similar results for *h*-BN stacks indicate overall higher conductivity and reduced misfit angle variation. This opens the way for the design of tunable heterogeneous junctions with controllable heat-transport properties ranging from substrate-isolation to efficient heat evacuation.

Graphene is considered to be one of the most promising heat dissipating materials in nanoelectronics [1] due to its ultrahigh in-plane room-temperature thermal conductivity of ~3000-5000 Wm$^{-1}$K$^{-1}$ [2,3]. This, however, can be hindered by graphene-substrate interactions that may lead to a substantial reduction of the heat-transport due to phonon leakage across the graphene-substrate interface and strong interfacial scattering of flexural phonon modes [4]. Such undesirable substrate effects can be reduced by considering multilayer graphene stacks. These are expected to effectively isolate the top graphene layers from the substrate, due to the considerably lower cross-plane thermal conductivity (~6.8 Wm$^{-1}$K$^{-1}$) [5], while exhibiting high in-plane conductivity that can be tuned via the stack thickness [6-13]. Anisotropic thermal conductivity is also observed for bulk hexagonal boron nitride (*h*-BN), with the in-plane and cross-plane thermal conductivity in the range of 390-420 Wm$^{-1}$K$^{-1}$ and 2.5-4.8 Wm$^{-1}$K$^{-1}$, respectively [14,15].

Efficient is-situ tuning of the thermal conductivity of such graphitic structures can be achieved by controlling the twist angle between adjacent layers within the stack. This has been recently computationally demonstrated for finite-sized nanoscale few-layer graphene junctions [16,17]. Two factors, however, limit the applicability of these results: (i) the simulations were performed using simplistic isotropic interlayer potentials that are known to be inaccurate for simulating the interlayer interactions in layered materials [18-21]; and (ii) the relevance of the results for large-scale interfaces is questionable due to significant edge scattering effects inherent to the small finite-sized model systems studied.

To address these issues, we investigate the interlayer thermal conductivity of graphene and *h*-BN stacks of varying thicknesses and twist angles. This allows us to gain fundamental understanding of the heat transport mechanisms in layered materials stacks and identify feasible means to control it. Our model system consists of two contacting identical AB (AA')-stacked graphite (*h*-BN) slabs, whose interfacing graphene (*h*-BN) layers are twisted with respect to each other to create a stacking fault of misfit angle $\theta$ (see Figure 1). Recent experiments demonstrated fine control over the misfit angle in such setups [22,23]. The thickness of the entire construction is varied between 2.7 nm-35 nm (8-104 layers) and periodic boundary conditions are applied in all directions. Heat transport simulations are performed using state-of-the-art anisotropic interlayer potentials [18-21] applied to the twisted stacks. These potentials were shown to capture well the structural, dynamic, heat dissipation, and phonon spectrum of graphitic and *h*-BN layered systems [24-27]. A thermal bias is induced by applying Langevin thermostats with different temperatures to two layers residing on opposite sides away from the twisted interface (see Sec. 1 of the Supporting Information (SI) for further details).

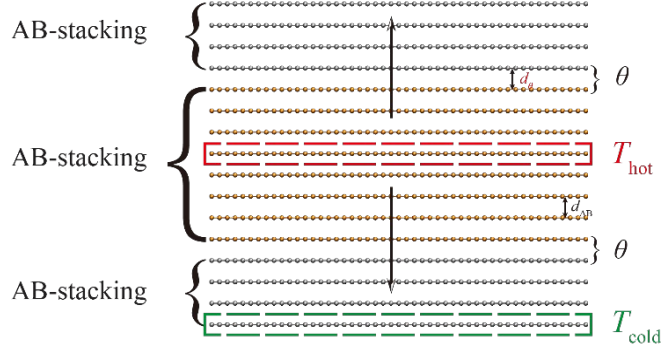

Figure 1. Schematic representation of the simulation setup. Two identical AB-stacked graphite slabs (gray and orange respectively) are twisted with respect to each other to create a stacking fault of misfit angle $\theta$. A thermal bias is induced by applying Langevin thermostats to the two layers marked by dashed red ($T_{hot}$) and green ($T_{cold}$) rectangles. The arrows indicate the direction of the vertical heat flux. Since periodic boundary conditions are applied also in the vertical direction, two twisted interfaces are shown across which heat flows in opposite directions.

We start by studying the effect of the misfit angle on the cross-plane thermal conductivity of the twisted graphite and $h$-BN stacks. Figure 2 presents the dependence of the cross-plane thermal conductivity of the entire stack on the misfit angle for model systems consisting of 8 (red circles) and 16 (black triangles) layers for (a) graphite and (b) $h$-BN. A pronounced dependence of the cross-plane thermal conductivity ($\kappa_{CP}$) of the entire graphitic stack is clearly evident, which above a misfit angle of $\sim 5°$ $\kappa_{CP}$ drops by a factor of 3-4 with respect to the value obtained for the aligned contact. Similar misfit-angle dependence of $\kappa_{CP}$ is obtained for twisted bilayer graphene (tBLG) using the transient MD simulation approach (see Sec. 2 of the SI). We note that this sharp drop for graphite is steeper and that the overall reduction is higher than those previously obtained using Lennard-Jones interlayer potentials in finite model systems [16,17]. The corresponding cross-plane thermal conductivity of the commensurate $h$-BN stack is found to be approximately double that of graphite for the same number of layers. Notably, it reduces more gradually with the twist angle and saturates at $\sim 15°$, with an overall two-three fold reduction.

The thermal conductivity of both graphite and $h$-BN stacks is found to increase when doubling their thickness. To identify the source of this thickness dependence we plot in Figure 2(c-d) the interfacial thermal resistance (ITR) (see Sec. 1.2 of the SI for the definition) associated with the twisted junction formed between the contacting graphene or $h$-BN layers of the two optimally-stacked slabs. Note that unlike $\kappa_{CP}$, which measures the conductivity of the entire stack, the ITR corresponds to the heat transport resistance of the two adjacent layers forming the twisted interface. Two important observations can be made: (i) the ITR is weakly dependent on the stack thickness, indicating that the thickness dependence arises from the conductivity of the optimally-stacked interfacing slabs. Specifically, in the thickness range considered the heat conductivity grows with slab thickness due to reduction of phonon-phonon interactions and increased contribution of long wave-length phonons below the mean-free path [28-30] (ii) the ITR strongly depends on the twist angle demonstrating a ~10-fold (4-fold) increase when the twist angle at the graphene (*h*-BN) interface is varied from 0° to 15°. This clearly indicates that the twist angle can be utilized to control the cross-plane thermal conductivity of hexagonal two-dimensional (2D) materials and to effectively thermally isolate the top layers from the underlying substrate.

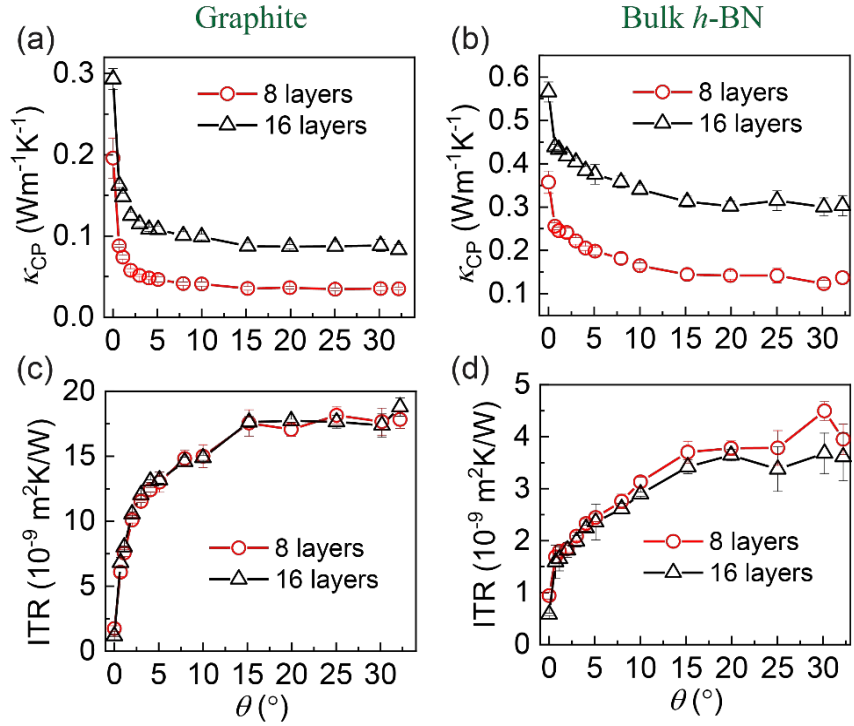

Figure 2. Twist-angle dependence of the cross-plane thermal conductivity of the entire stack (a, b), and the interfacial thermal resistance (c, d) of the twisted contact formed between the optimally-stacked slabs of graphite (a, c) and bulk *h*-BN (b, d), respectively. Red circles and black triangles correspond to the results obtained using 8 and 16 layer models, respectively.

The strong dependence of the cross-plane thermal conductivity of graphene and *h*-BN on the stacking fault twist angle is related to the degree of coupling between the phonon modes of the two contacting layers at the twisted interface. Note that the term "coupling" used herein is not related to the standard notion of phonon-phonon couplings due to anharmonic effects. Instead, we regard to the off-diagonal terms of the Hessian when represented in the basis of the harmonic phonon modes of the isolated layers. To demonstrate this, we write the dynamical matrix (the mass-reduced Fourier transform of the force constant matrix) in block form as follows:

$$\boldsymbol{\Phi}(\boldsymbol{q}) = \begin{pmatrix} \boldsymbol{\Phi}_{11}(\boldsymbol{q}) & \boldsymbol{\Phi}_{12}(\boldsymbol{q}) \\ \boldsymbol{\Phi}_{21}(\boldsymbol{q}) & \boldsymbol{\Phi}_{22}(\boldsymbol{q}) \end{pmatrix}, \quad (1)$$

where $\boldsymbol{\Phi}_{11}(\boldsymbol{q})$ and $\boldsymbol{\Phi}_{22}(\boldsymbol{q})$ are the block matrices relating to the first and second layer and $\boldsymbol{\Phi}_{12}(\boldsymbol{q})$ and $\boldsymbol{\Phi}_{21}(\boldsymbol{q}) = \boldsymbol{\Phi}_{12}^\dagger(\boldsymbol{q})$, all evaluated at wave-vector $\boldsymbol{q}$. The interlayer phonon-phonon couplings are obtained by diagonalizing separately $\boldsymbol{\Phi}_{11}(\boldsymbol{q})$ and $\boldsymbol{\Phi}_{22}(\boldsymbol{q})$ such that $\widetilde{\boldsymbol{\Phi}}_{11}(\boldsymbol{q}) = \boldsymbol{U}_1^\dagger(\boldsymbol{q})\boldsymbol{\Phi}_{11}(\boldsymbol{q})\boldsymbol{U}_1(\boldsymbol{q})$ and $\widetilde{\boldsymbol{\Phi}}_{22}(\boldsymbol{q}) = \boldsymbol{U}_2^\dagger(\boldsymbol{q})\boldsymbol{\Phi}_{22}(\boldsymbol{q})\boldsymbol{U}_2(\boldsymbol{q})$ are diagonal matrices containing the frequencies $(\omega_i)$ of the phonon modes of the two layers and $\boldsymbol{U}_1(\boldsymbol{q})$ and $\boldsymbol{U}_2(\boldsymbol{q})$ are unitary matrices of the corresponding eigenvectors. We now construct a global block diagonal transformation matrix of the form:

$$\boldsymbol{U}(\boldsymbol{q}) = \begin{pmatrix} \boldsymbol{U}_1(\boldsymbol{q}) & 0 \\ 0 & \boldsymbol{U}_2(\boldsymbol{q}) \end{pmatrix}, \quad (2)$$

and transform the full dynamical matrix as follows:

$$\boldsymbol{U}^\dagger(\boldsymbol{q})\boldsymbol{\Phi}(\boldsymbol{q})\boldsymbol{U}(\boldsymbol{q}) = \begin{pmatrix} \widetilde{\boldsymbol{\Phi}}_{11}(\boldsymbol{q}) & \widetilde{\boldsymbol{\Phi}}_{12}(\boldsymbol{q}) \\ \widetilde{\boldsymbol{\Phi}}_{21}(\boldsymbol{q}) & \widetilde{\boldsymbol{\Phi}}_{22}(\boldsymbol{q}) \end{pmatrix}, \quad (3)$$

where $\widetilde{\boldsymbol{\Phi}}_{12}(\boldsymbol{q}) = \boldsymbol{U}_1^\dagger(\boldsymbol{q})\boldsymbol{\Phi}_{12}(\boldsymbol{q})\boldsymbol{U}_2(\boldsymbol{q})$ and $\widetilde{\boldsymbol{\Phi}}_{21}(\boldsymbol{q}) = \widetilde{\boldsymbol{\Phi}}_{12}^\dagger(\boldsymbol{q})$ are the interlayer phonon-phonon coupling blocks. Naturally, when the two layers are infinitely separated, these coupling blocks vanish and the diagonal blocks converge to those of the isolated layers.

The overall coupling between the two layers can be obtained from the individual phonon-phonon coupling matrix elements via Fermi's golden rule [31], which reads as (see Sec. 4 of SI for a detailed derivation):

$$\Gamma_{tot} = \frac{\pi \hbar^3}{2} \sum_{q\lambda} \frac{e^{-\beta E_{q\lambda}}}{Z} \frac{\rho(E_{q\lambda})\left|V_{\lambda,\lambda+\frac{3r}{2}}(\boldsymbol{q})\right|^2}{E_{q\lambda}^2}, \quad (4)$$

where $Z = \sum_{q\lambda} e^{-\beta E_{q\lambda}}$ is the partition function, $E_{q\lambda}$ is the energy of phonons at branch $\lambda$ with wave number $\boldsymbol{q}$, $\rho(E_{q\lambda})$ is the density-of-states (DOS) at $E_{q\lambda}$, and $\left|V_{\lambda,\lambda+\frac{3r}{2}}(\boldsymbol{q})\right|^2$ is the coupling matrix element between branches of phonons of similar energy in the two layers, whose number of atoms in one unit cell is $r$.

Using Eq. (4), we can rationalize the misfit angle dependence of the heat flux across the twisted interface from the calculated inter-phonon coupling. To that end, we performed room temperature (300K) simulations (technical details can be found in Sec. 3 of SI) for tBLG with different misfit angles using the Green's function molecular dynamics (GFMD) developed by Kong et al [32] as implemented in LAMMPS [33]. The simulations allow us to evaluate the dynamical matrix from which the phonon-phonon couplings can be extracted (see details in Sec. 3 of SI) and the overall heat transfer rate calculated. Figure 3 shows the resulting heat transfer rate (normalized to is value for the aligned contact $(\theta = 0)$) as a function of the misfit angle compared to the interfacial thermal conductivity defined as the inverse of the ITR presented in Figure 2(c), $\text{ITC} \equiv 1/\text{ITR}$. The remarkable agreement between the calculated interfacial thermal conductivity and Fermi's golden rule results indicate that the dependence of the interlayer phonon-phonon couplings on the misfit angle is responsible for the strong angle dependence of the interfacial conductivity. Notably, the sharp heat conductivity drop at misfit angles in the range of 0°-5° as well as the small conductivity for larger misfit angles are well captured by Fermi's golden rule.

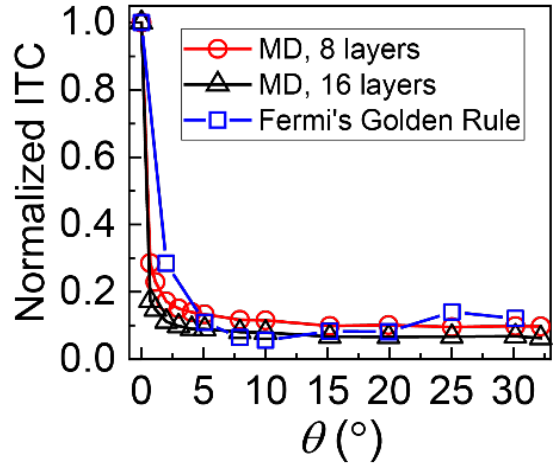

Figure 3. Comparison between Fermi's golden rule results (open blue squares) for the interfacial heat-transfer rate of a tBLG and the calculated interfacial thermal conductivity at various misfit angles. ITC simulation results are presented for both 8 layers (open red circles) and 16 layers (open black triangles) showing similar behavior. For comparison purposes, all data sets are normalized to their value obtained for the aligned contact.

To correlate our results with experimentally measured thermal conductivities that are often obtained for thick samples, we repeated our calculations for increasing stack thicknesses at fixed misfit angles. Figure 4 presents results for the calculated heat conductivity of (a) graphite and (b) $h$-BN stacks either aligned (open red circles) or twisted by $\theta = 30.16°$ (open black diamond symbols) as a function of number of layers in the stack. As discussed above, for both systems the misoriented stack exhibits lower heat conductivity compared to the aligned system, however, its thickness dependence is considerably stronger. This can be attributed to the significantly higher interface resistance of the twisted interface that, when plugged in Eq. (S2) of the SI for the overall conductivity, induces stronger thickness dependence.

Comparing our calculated heat conductivities for the aligned contact (open red circles) to available experimental data for ~35 nm thick graphite slabs [34] (dashed green line) we find that at the thickest model system considered of 104 graphene layers (~34 nm thick) the calculated value of 0.85±0.05 W/m·K is in remarkable agreement with the measured value of ~0.7 W/m·K. Furthermore, experimental values for bulk graphite [5] indicate that the thermal conductivity continues to grow up to ~6.8 W/m·K (black dash-dotted line), which is consistent with the general trend of the calculated heat conductivity that does not saturate for the thickest model system considered. These results strongly enforce the validity of our force-field and model systems to model the heat conductivity of twisted layered materials interfaces. Available experimental results for the heat conductivity of bulk $h$-BN are marked by the dashed-dotted black and dashed-green lines in Figure 4(b). In line with our findings for the graphitic interface, our calculated finite slab heat conductivities for the aligned interface (open red circles) continue to grow with the number of layers and are consistently below the bulk value.

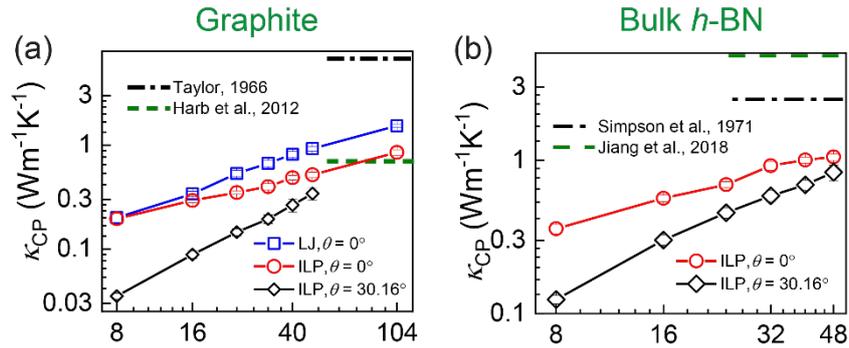

Figure 4. Thickness dependence of the thermal conductivity, $\kappa_{CP}$, of aligned (open red circles) and twisted by 30.16° (open black diamond symbols) graphite (a) and $h$-BN (b) stacks. Blue squares represent results obtained using the isotropic Lennard-Jones potential for the aligned contacts. The green dashed and black dash-dotted lines represent experimental results measured for graphite (a) (Refs. [5,34]) and bulk $h$-BN (b) (Refs. [14,15]). Note that both axis scales are logarithmic. Error bar estimation procedure is discussed in Sec. 1 of the SI.

Another important factor that may affect the interlayer thermal transport properties of 2D material stacks, is the average temperature of the system, which was taken to be ~300K in all abovementioned simulations. To evaluate the sensitivity of our results towards this parameter we repeated the heat conductivity and interfacial resistance calculations of optimally stacked graphite and $h$-BN stacks for an average temperature of 400 K. The results, presented in see Sec. 4 of the SI, indicate that $\kappa_{CP}$ and ITR weakly depend on the average temperature over the entire thickness range considered. This conclusion is consistent with recent experimental findings [11]. We may attribute this to the fact that the thickness of the graphite and *h*-BN stack model considered is much smaller than their phonon mean-free path (~200 nm for graphite [10,11,13,35,36] and ~100 nm for bulk *h*-BN [15]) such that the phonon transport is dominated by phonon-boundary scattering, which weakly depends on temperature in the range considered. Therefore, temperature dependent Umklapp processes have only marginal contribution to our results [11].

Finally, we note that previous calculations of the heat conductivity of twisted graphitic interfaces relied on Lennard-Jones (LJ) potentials describing the interlayer interactions [8-10,13]. To demonstrate the importance of using registry-dependent interlayer potentials, we have repeated our calculations of the heat conductivity of graphitic slabs with the REBO intralayer potential augmented by LJ interlayer interactions [37] ($\varepsilon = 2.84$ meV, $\sigma = 3.4$ Å). We find that the calculated heat conductivities obtained using the LJ interlayer potential are consistently higher than those obtained by our ILP and that the difference between them grows with the model system thickness. Notably, the heat conductivity obtained using the LJ potential for a graphitic slab of thickness ~34 nm is 1.54 W/m · K, overestimating the experimental value by more than a factor of 2.

The excellent agreement of our ILP calculations with experimental data of nanoscale graphitic stacks, therefore, demonstrates the reliability of our predictions for the strong interfacial misfit angle dependence of cross-layer thermal conductivity in graphite and *h*-BN. The observed sharp conductivity decrease of twisted graphitic interfaces at misfit angles < 5° opens the way to control the thermal evacuation rate and thermal isolation of active layers in graphene-based electronic and mechanical devices. The revealed underlying mechanism, suggests that design rules can be obtained by carefully tailoring the phonon-phonon couplings across the twisted interface. While the misfit angle dependence of *h*-BN is found to be weaker than that of graphite, the overall thermal conductivity of the former is found to be higher. This may be utilized to achieve higher conductivity and controllability in twisted heterogeneous junctions of layered materials.

ASSOCIATED CONTENT

**Supporting Information**.

The supporting Information section includes a description of: the methodology; thermal conductivity of twisted bilayer graphene; theory for calculating the phonon coupling of twisted bilayer graphene;

derivation of Fermi's golden rule, and temperature dependence of the cross-plane thermal conductivity.

AUTHOR INFORMATION

**Corresponding Author**

* E-mail: urbakh@tauex.tau.ac.il.

**Notes**

The authors declare no competing financial interest.

ACKNOWLEDGMENTS

The authors would like to thank Prof. Abraham Nitzan, Dr. Guy Cohen and Dr. Yiming Pan for helpful discussions. W.O. acknowledges the financial support from a fellowship program for outstanding postdoctoral researchers from China and India in Israeli Universities and the support from the National Natural Science Foundation of China (Nos. 11890673 and 11890674). H.Q. acknowledges the financial support from the National Natural Science Foundation of China (No. 11890674). M.U. acknowledges the financial support of the Israel Science Foundation, Grant No. 1141/18 and the ISF-NSFC joint grant 3191/19. O.H. is grateful for the generous financial support of the Israel Science Foundation under grant no. 1586/17 and the Naomi Foundation for generous financial support via the 2017 Kadar Award. This work is supported in part by COST Action MP1303.

# Supporting information for "Controllable Thermal Conductivity of Twisted Graphene and Hexagonal Boron Nitride Interfaces"

Wengen Ouyang,[1] Huasong Qin,[2] Michael Urbakh,[1*] and Oded Hod[1]

[1]*Department of Physical Chemistry, School of Chemistry, The Raymond and Beverly Sackler Faculty of Exact Sciences and The Sackler Center for Computational Molecular and Materials Science, Tel Aviv University, Tel Aviv 6997801, Israel.*

[2]*State Key Laboratory for Strength and Vibration of Mechanical Structures, School of Aerospace, Xi'an Jiaotong University, Xi'an 710049, China*

The Supporting information includes following sections:

1. Methodology
2. Thermal conductivity of twisted bilayer graphene
3. Theory for calculating the phonon coupling of twisted bilayer graphene
4. Derivation of Fermi's golden rule
5. Temperature dependence of cross-plane thermal conductivity

# 1 Methodology

## 1.1 Model system

The initial interlayer distance across the layered stack was set equal to 3.4 Å and 3.3 Å for graphite and bulk *h*-BN, respectively. Periodic boundary conditions were applied in all directions. It should be noted that the lattice structure is rigorously periodic only at some specific twist angles, the values of which are listed in Table S1 in section 3 below. While the cross-sectional area for each misfit angle, $\theta$, is different, all systems considered have a contact area exceeding 12 nm$^2$, which was shown to provide converged results with respect to unit-cell dimensions [1]. The intralayer interactions within each graphene and *h*-BN layer were modeled via the second generation REBO potential [2] and Tersoff potential [3], respectively. The interlayer interactions between the layers of graphite and bulk *h*-BN were described via our dedicated interlayer potential (ILP) [4], which is implemented in the LAMMPS [5] suite of codes [6].

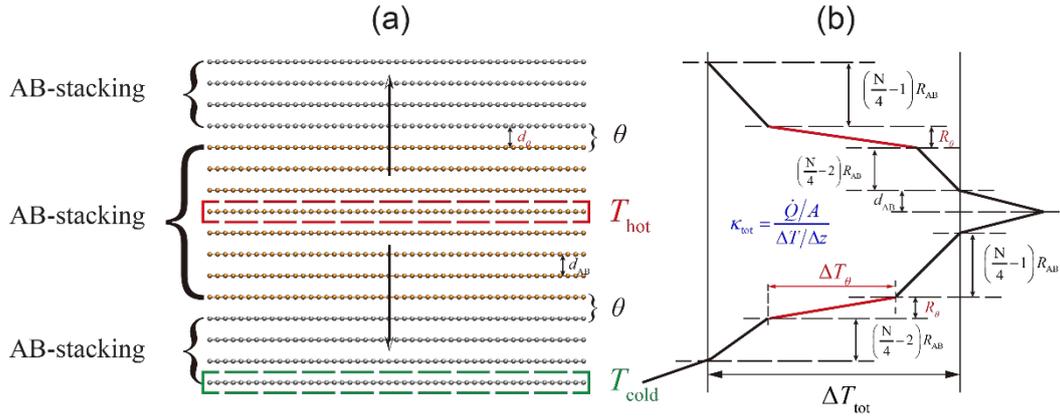

Fig. S1. Schematic representation of the simulation setup (a) and steady-state temperature profile (b), respectively. In panel (a), two identical AB-stacked graphite slabs (gray and orange respectively) are twisted with respect to each other to create a stacking fault of misfit angle $\theta$. A thermal bias is induced by applying Langevin thermostats to the two layers marked by dashed red ($T_{\text{hot}}$) and green ($T_{\text{cold}}$) rectangles. The arrows indicate the direction of the vertical heat flux. Since periodic boundary conditions are applied also in the vertical direction, two twisted interfaces are, shown across which heat flows in opposite directions. The steady-state temperature profiles are illustrated in panel (b), where $N$ is the total number of layers in the model system and $R_{\text{AB}}$, $d_{\text{AB}}$ and $R_\theta$, and $d_\theta$ mark the interfacial Kapitza resistance [8,9] and interlayer distance for contacting graphene layers with AB-stacking and misfit angle $\theta$, respectively. The red lines in panel (b) mark the temperature variation across the twisted interface, where the vertical axis corresponds to the position of the various layers along the stack and the horizontal axis marks the temperature of the various layers.

## 1.2 Simulation Protocol

All MD simulations were performed with the LAMMPS simulation package [5]. The velocity-Verlet algorithm with a time-step of 0.5 fs was used to propagate the equations of motion. A Nosé-Hoover thermostat with a time constant of 0.25 ps was used for constant temperature simulations. To maintain a specified hydrostatic pressure, the three translational vectors of the simulation cell were adjusted independently by a Nosé-Hoover barostat with a time constant of 1.0 ps [7]. To relax the box, we first equilibrated the systems in the NPT ensemble at a temperature of $T = 300$ K and zero pressure for 250 ps (see Fig. S2). After equilibration, Langevin thermostats with damping coefficients 1.0 ps$^{-1}$ were applied to the bottom and middle layer of the graphene stack (see Fig. S1) with target temperature $T_{\text{hot}} = 375$ K (hot reservoir) and $T_{\text{cold}} = 225$ K (cold reservoir), respectively. Then the system was allowed to reach steady-state over a subsequent simulation period of 750 ps (see Fig. S2), during which the dynamics of all non-thermostated layers followed the NVE ensemble. For the larger model systems, the length of the NPT and Langevin stages was doubled (for the 32 and 48 layers systems) or tripled (for the 104 layers graphitic system) to ensure convergence of the obtained steady-state. Once steady-state was obtained, the last 500 ps were used to calculate the thermal conductivity of the twisted graphite and bulk $h$-BN. The statistical errors were estimated using ten different data sets, each calculated over a time interval of 50 ps.

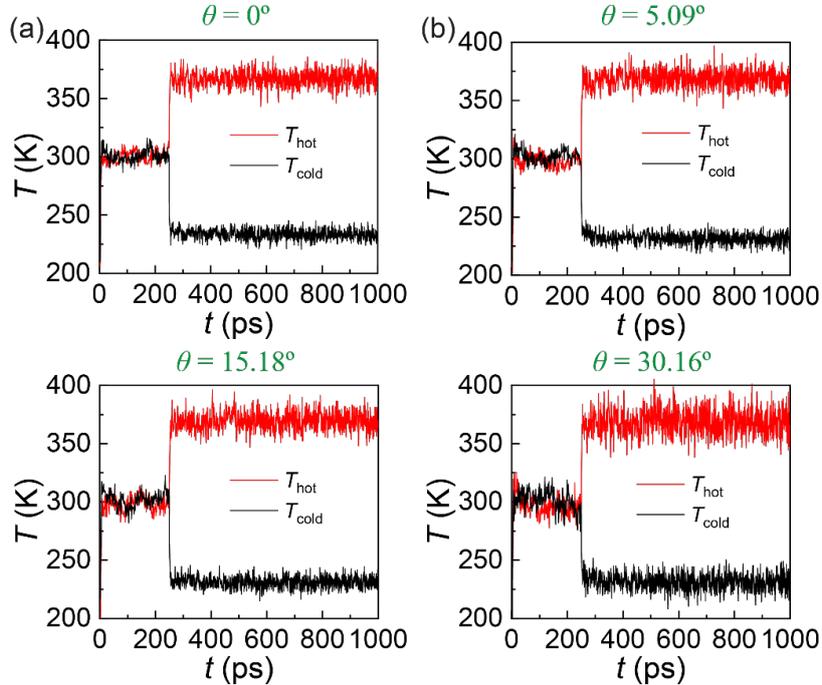

Fig. S2. Time evolution of the temperature of thermostated layers for 16 layers twisted graphite with misfit angle (a) $\theta = 0°$, (b) $\theta = 5.09°$, (c) $\theta = 15.18°$, and (d) $\theta = 30.16°$. Note that the thermal fluctuations increase with increasing the misfit angle due to the growing interfacial thermal resistance that enhances phonon back scattering at the twisted junction.

## 1.3 Calculation of the interfacial thermal resistance

According to Fourier's law, the cross-plane thermal conductivity ($\kappa_{CP}$) of a twisted graphitic interface of misfit angle $\theta$ can be calculated as:

$$\kappa_{CP} = \frac{\dot{Q}/A}{\Delta T/\Delta z}, \tag{S1}$$

where $\dot{Q}$ is the heat flux, $A$ is the cross-section area and $\Delta T/\Delta z$ is the temperature gradient along the direction of heat flux (perpendicular to the basal plane in our case). Fig. S1(b) shows a schematic temperature profile along the z-direction, where the vertical axis corresponds to the position of the various layers along the stack and the horizontal axis marks the temperature of the various layers. The actual temperature profiles extracted from the NEMD simulations for twisted graphite with different number of layers can be found in Fig. S3. For Bernal-stacked graphite (i.e., $\theta = 0°$, red circles in Fig. S3), only the linear region of the temperature profile was used to calculate $\kappa_{CP}$ and the points corresponding to the layers where the thermostats were applied were omitted (marked with green triangle in Fig. S3). The $\kappa_{CP}$ of the system was calculated using Eq. (S1) by averaging over the two linear regions of the temperature profiles. For the twisted case ($\theta \neq 0°$), we found a sudden temperature decrease $\Delta T_\theta$ at the position of the twisted interface (see black squares in Fig. S3). $\kappa_{CP}$, in this case, was calculated using the temperature gradient calculated for the same layer range as that for $\theta = 0°$. To characterize the thermal properties of the twisted interface, the concept of interfacial thermal resistance (ITR), i.e., Kapitza resistance [8,9], was introduced. According the definition of the Kapitza resistance [8], $R = A\Delta T/\dot{Q}$ and noticing that $\Delta T_{tot} = (N/2 - 3)\Delta T_{AB} + \Delta T_\theta$ and $\Delta z = (N/2 - 3)d_{AB} + d_\theta$, Eq. (1) can be rewritten considering two-resistors in series as [see Fig. S1(b)]:

$$(N/2 - 3)R_{AB} + R_\theta = [(N/2 - 3)d_{AB} + d_\theta]/\kappa_{CP}. \tag{S2}$$

Here $R_{AB}$, $d_{AB}$, $\Delta T_{AB}$ and $R_\theta$, $d_\theta$, $\Delta T_\theta$ are the ITR, interlayer distance and temperature difference for adjacent AB-stacked and twisted graphene layers, respectively. For the aligned contact ($\theta = 0°$), the ITR can be simply calculated as $R_{AB} = d_{AB}/\kappa_{AB}$. Once $R_{AB}$ is known, $R_\theta$ can be calculated from Eq. (S2). We note that the sharp temperature drop at the twisted interface indicates that $R_\theta$ should be much larger than $R_{AB}$.

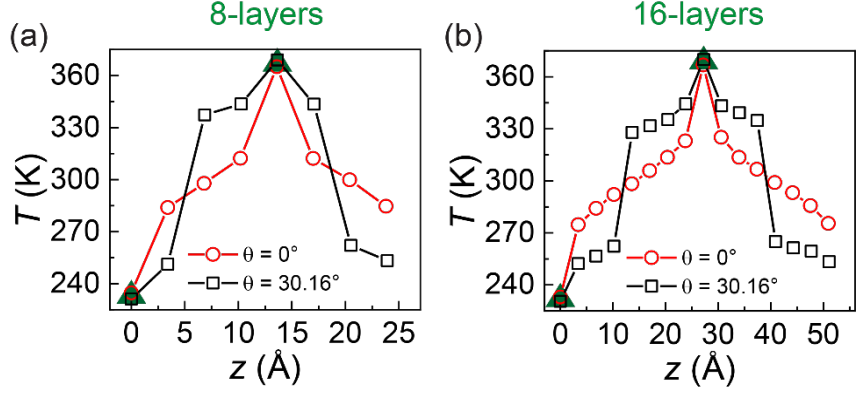

Fig. S3. Temperature profiles for graphitic stacks consisting of (a) 8 and (b) 16 layers. The red circles and black squares represent the temperature profiles for the aligned ($\theta = 0°$) and twisted ($\theta = 30.16°$) junctions, respectively. Green triangles represent data points that were omitted in the $\kappa_{CP}$ calculation.

## 2  Thermal conductivity of twisted bilayer graphene

As comparison, we also calculated the interfacial thermal conductivity (ITC) and ITR of twisted bilayer graphene (tBLG) with the transient MD simulation approach [15-17] since the NEMD simulation protocol used in the main text becomes invalid in this case. In this protocol, the system was first equilibrated within the NPT ensemble at $T = 200$ K and zero pressure for 100 ps, which was followed by a 100 ps NVT ensemble equilibration stage and a 100 ps of NVE ensemble equilibration stage. After the system reached steady-state, an ultrafast heat impulse was imposed on the top layer of the t-BLG for 50 fs to increase the temperature of the top layer from 200 K to 400 K, while that of bottom layer of tBLG remained unchanged. After the external heat source was removed, thermal energy flowed from the top layer to the bottom layer due to the temperature difference and the temperature of both layers approached 300 K when quasi-steady-state was reached. During the thermal relaxation time interval (500 ps), the temperature and energy of the system sections were recorded. The ITR could then be extracted using the following equation [15-17]:

$$\frac{\partial E_t}{\partial t} = \frac{A}{R}[T_{bot}(t) - T_{top}(t)], \tag{S3}$$

where $E_t$ is the total energy of the top graphene layer, $R$ is the ITR of the tBLG, $A$ is the interfacial cross-section area, and $T_{bot}$ and $T_{top}$ are the instantaneous temperatures measured for the bottom and top layers of the tBLG, respectively. Note that in Eq. S3 we assume a linear dependence of the heat flux on the temperature difference between the layers. The ITC of the tBLG is simply defined as ITC $\equiv d/$ITR, where d is the average interlayer distance.

The ITC and ITR of tBLG as functions of misfit angle calculated with the transient MD simulation protocol are illustrated in Fig. S4, demonstrating similar misfit-angle dependence as that for the NEMD protocol with Langevin thermostat exercised to obtain the results presented in the main text. This further validates the reliability of the simulation protocol adopted in the main text, which is more suitable to treat thick slabs and allows to obtain a true stead-state.

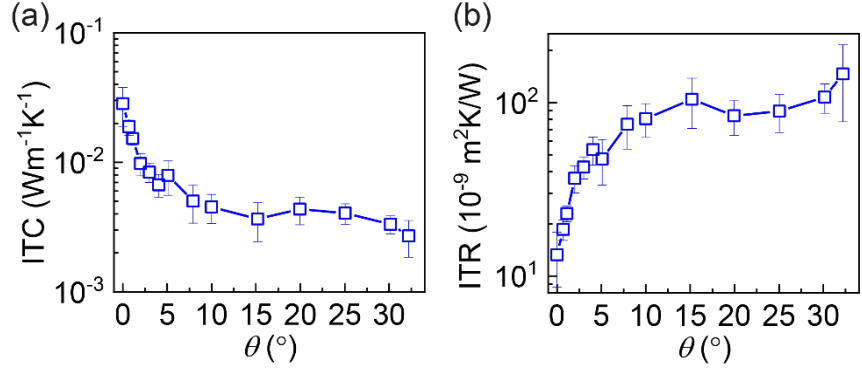

Fig. S4. Misfit-angle dependence of (a) ITC and (b) ITR for a twisted bilayer graphene obtained using the transient MD simulation approach.

## 3 Theory for calculating the phonon-phonon coupling of twisted bilayer graphene

### 3.1 Brillouin Zone of supercell in tBLG

For tBLG, the lattice structure is rigorously periodic only at some specific misfit angles, $\theta$, where the lattice vector $\boldsymbol{L}_1 = n_1\boldsymbol{a}_1 + n_2\boldsymbol{a}_2$ in the bottom layer equals the vector $\boldsymbol{L}_2 = m_1\boldsymbol{a}_1 + m_2\boldsymbol{a}_2$ in the top layer with certain integers $m_1$, $m_2$ and $n_1$, $n_2$. Here, $\boldsymbol{a}_1 = a(1,0)$ and $\boldsymbol{a}_2 = a(1/2, \sqrt{3}/2)$ are the primitive lattice vectors of the bottom layer and $a$ is the lattice constant of monolayer graphene. Thus, the exact superlattice period is then given by [18]:

$$L = |n_1\boldsymbol{a}_1 + n_2\boldsymbol{a}_2| = a\sqrt{n_1^2 + n_2^2 + n_1 n_2} = \frac{|n_1-n_2|a}{2\sin(\theta/2)}, \quad (S4)$$

where $\theta$ is the angle between two lattice vectors $\boldsymbol{L}_1$ and $\boldsymbol{L}_2$. In the simulations below, we always rotated the supercell such that its lattice vector is $\boldsymbol{L}_1 = L(1,0)$ and $\boldsymbol{L}_2 = L(1/2, \sqrt{3}/2)$. In this case, the corresponding reciprocal lattice vector of the moiré superlattice satisfies the relation $\boldsymbol{G}_i \cdot \boldsymbol{L}_j = 2\pi\delta_{ij}$, such that:

$$\boldsymbol{G}_1 = \frac{4\pi}{\sqrt{3}L}\left(\frac{\sqrt{3}}{2}, -\frac{1}{2}\right), \quad \boldsymbol{G}_2 = \frac{4\pi}{\sqrt{3}L}(0,1). \quad (S5)$$

Both the lattice vectors and the corresponding reciprocal lattice vectors of the superlattice of the tBLG are presented in Fig. S5. Table S1 reports the parameters used to construct rhombus periodic supercells of different misfit angles that can be duplicated to construct a rectangular periodic supercell.

Table S1. The parameters used to construct periodic supercells of various misfit angles.

| $\theta$ (°) | $A$ (nm$^2$) | $n_1$ | $n_2$ | $m_1$ | $m_2$ |
|---|---|---|---|---|---|
| 0 | 60.383688 | 24 | 0 | 24 | 0 |
| 0.696407 | 709.613169 | 48 | 47 | 47 | 48 |
| 1.121311 | 273.718420 | 30 | 29 | 29 | 30 |
| 2.000628 | 85.648391 | 17 | 16 | 16 | 17 |
| 3.006558 | 152.322047 | 23 | 21 | 21 | 23 |
| 4.048894 | 189.013524 | 26 | 23 | 23 | 26 |
| 5.085849 | 53.255058 | 14 | 12 | 12 | 14 |
| 7.926470 | 49.376245 | 14 | 11 | 11 | 14 |
| 9.998709 | 86.277388 | 19 | 14 | 14 | 19 |
| 15.178179 | 31.554670 | 15 | 4 | 4 | 15 |
| 19.932013 | 42.876612 | 15 | 8 | 8 | 15 |
| 25.039660 | 13.942761 | 9 | 4 | 4 | 9 |
| 30.158276 | 18.974735 | 11 | 4 | 4 | 11 |
| 32.204228 | 21.805221 | 12 | 4 | 4 | 12 |

## 3.2 *Special points for Brillouin Zone integration*

The calculation of the sum over wave vector $q$ in Eq. (4) in the main text can be transformed to an integral using the relation $\sum_q(\cdots) = \frac{1}{V_b}\int_{BZ}(\cdots)\,d\boldsymbol{q}$, where $V_b = (2\pi)^3/V$ is the volume of the Brillouin Zone (BZ) and $V$ is volume of the real-space unit-cell. The calculation of integral is usually inefficient since it requires calculating the value of the function over a large set of $k$ points in the first BZ. To calculate such integrations more efficiently, simple k-point meshes can be replaced by a carefully selected set of special points in the BZ, $\boldsymbol{q}_i$, [19-22] over which the function is evaluated. The integral can then be estimated via:

$$I = \frac{1}{V_b}\int_{BZ} f(\boldsymbol{q})\,d\boldsymbol{q} \approx \frac{1}{N}\sum_i w_i f(\boldsymbol{q}_i), \tag{S6}$$

where $V_b$ is the volume of the BZ, $w_i$ is the weight of the $i^{th}$ data point, and $N$ normalizes the weighting factors to unity: $N = \sum_i w_i$. The set of selected $\{q_i\}$ forms a grid in the irreducible Brillouin zone (IBZ), as is illustrated by the red points in Fig. S5(b). The coordinates of these points for a hexagonal lattice are presented in Eq. (S7).

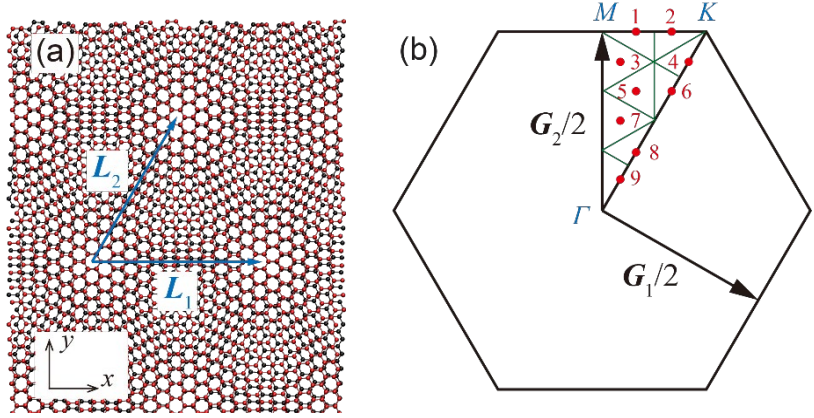

Fig. S5. (a) Twisted bilayer graphene of misfit angle $\theta = 5.09°$. $L_1$ and $L_2$ are the superlattice vectors. (b) The corresponding first Brillouin Zone of (a). $G_1$ and $G_2$ are the reciprocal lattice vectors of the superlattice. The triangle $\Delta \Gamma MK$ represents the irreducible Brillouin zone. Red circles mark the position of the special points used to evaluate the integral over the first Brillouin Zone.

$$\begin{cases} q_1 = \left(\frac{1}{9}, \frac{t}{3}\right), q_2 = \left(\frac{2}{9}, \frac{t}{3}\right), q_3 = \left(\frac{1}{18}, \frac{5t}{18}\right) \\ q_4 = \left(\frac{5}{18}, \frac{5t}{18}\right), q_5 = \left(\frac{1}{9}, \frac{2t}{9}\right), q_6 = \left(\frac{2}{9}, \frac{2t}{9}\right). \\ q_7 = \left(\frac{1}{18}, \frac{3t}{18}\right), q_8 = \left(\frac{1}{9}, \frac{t}{9}\right), q_9 = \left(\frac{1}{18}, \frac{t}{18}\right) \end{cases} \quad (S7)$$

Here, $t = \sqrt{3}$ and the unit of the coordinates is $2\pi/L$. The weighting factors $\{w_i\}$ are [19]:

$$w_{1,2,4,6,8,9} = \frac{1}{12}, w_{3,5,7} = \frac{1}{6}, \quad (S8)$$

Using Eqs. (S6-S8), Eq. (4) in the main text can be evaluated as follows:

$$\Gamma_{tot} = \frac{\pi \hbar^3}{2} \sum_{k=1}^{9} w_k \sum_\lambda \frac{e^{-\beta E_{q_k\lambda}} \rho(E_{q_k\lambda}) \left|V_{\lambda,\lambda+\frac{3r}{2}}(q_k)\right|^2}{Z \cdot E_{q_k\lambda}^2}. \quad (S9)$$

This equation was used to calculate the transition rate presented in Fig. 3 in the main text.

## 4  Derivation of Fermi's golden rule

The derivation of Fermi's golden rule is provided in a separate file (Fermi's Golden Rule for phonons.pdf) due to its extent.

## 5 Temperature dependence of interfacial thermal conductivity

In the main text, the target temperatures of the Langevin thermostats for the bottom and middle layers of graphene and *h*-BN were set to 225 K and 375 K, respectively. After reaching the steady-state, the average temperature of the system was found to be ~300 K. To check the effect of average temperature on our results, we calculated the cross-plane thermal conductivity ($\kappa_{\text{CP}}$) and the corresponding interfacial thermal resistance (ITR) at a different temperature gradient (325 K – 475 K), resulting the average steady-state temperature of ~400K. The protocol described in Section 1 above was used to perform these calculations, as well. Both ILP and Lennard-Jones (LJ) potential were tested for graphite whereas for the bulk *h*-BN simulations only the ILP was used. The results for graphite and bulk *h*-BN are illustrated in Fig. S6 and Fig. S7, respectively. For the ILP we find that the overall values of $\kappa_{\text{CP}}$ (ITR) decrease (increase) slightly with increasing average temperature, which is consistent with a recent experiment [10]. The LJ potential calculations, as well, exhibit very week dependence on average temperature within the range studied. Altogether, the layer dependences of both quantities remain mostly insensitive to the average temperature. The reason is that the thickness of the graphite and *h*-BN model systems was chosen to be much smaller than their phonon mean free path (~200 nm for graphite [1,10-13] and ~100 nm for bulk *h*-BN [14]) such that phonon transport is dominated by phonon-boundary scattering and the Umklapp process only makes marginal contribution [10].

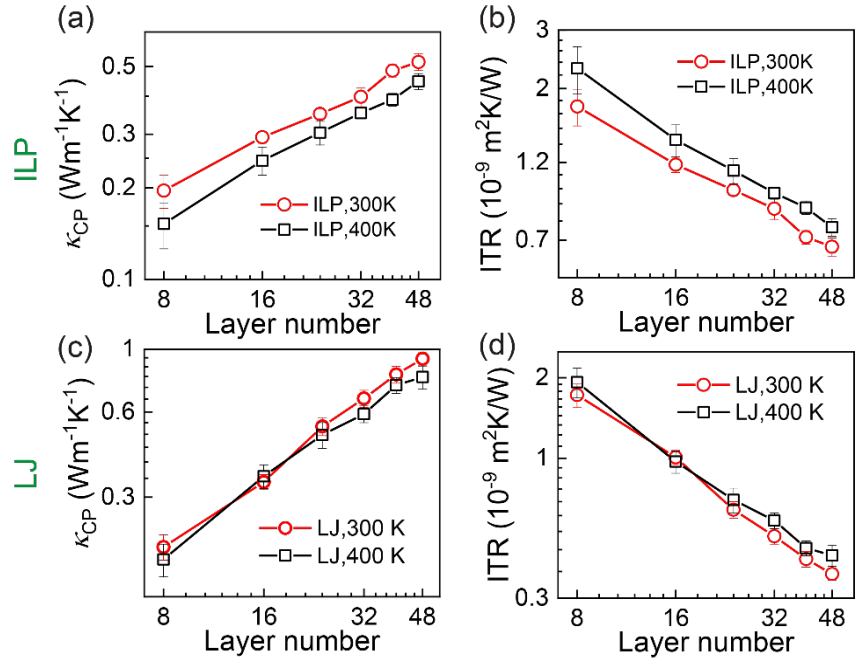

Fig. S6. Layer dependence of $\kappa_{CP}$ (a, c) and ITR (b, d) for Bernal-stacked graphite at average steady-state temperatures of 300 K (red circles) and 400 K (black squares). The left and right columns correspond to the $\kappa_{CP}$ and ITR calculated with ILP and LJ potential, respectively.

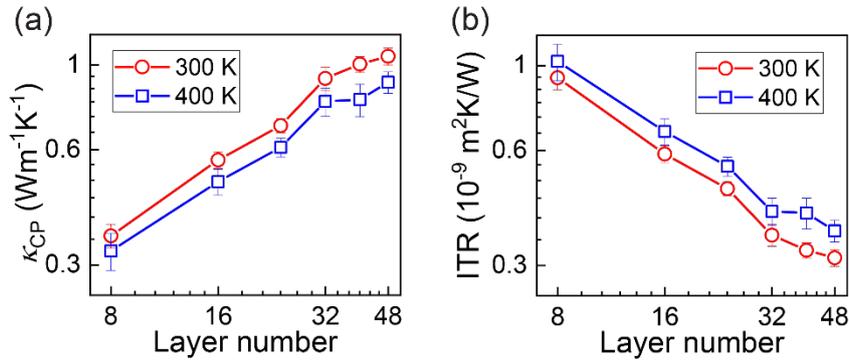

Fig. S7. Layer dependence of $\kappa_{CP}$ (a) and ITR (b) for AA'-stacked $h$-BN at average temperatures of 300 K (red circles) and 400 K (blue squares).

# Fermi's golden rule for phonons

## 1 Basic theory for phonons

### 1.1 Basic notations

Let us consider a 3D crystal with a total of $N = N_1 N_2 N_3$ unit cells and periodic boundary conditions. To be specific, let $\boldsymbol{a}_i, i = 1,2,3$ be the lattice vectors that define the unit cell. We index unit cells with $\boldsymbol{n} = (n_1, n_2, n_3)$ where each $n_i = 1, 2, \cdots, N_i$, and their locations are $\boldsymbol{R}_n = \sum_{i=1}^{3} n_i \boldsymbol{a}_i$. Assume that there are $r$ atoms in each unit cell, which are indexed with $s = 1, \cdots, r$. The mass and the equilibrium distance of the $s^{\text{th}}$ atom are notated as $M_s$ and $\boldsymbol{R}_s^0$, respectively. Then the location of the $s^{\text{th}}$ atom in the $n^{\text{th}}$ unit cell at time $t$ can be expressed as:

$$\boldsymbol{r}_{ns}(t) = \boldsymbol{R}_n + \boldsymbol{R}_s^0 + \boldsymbol{u}_{ns}(t), \tag{1.1}$$

where $\boldsymbol{u}_{ns}(t)$ is its displacement from its equilibrium position.

The Lagrangian for this classical problem can be written as

$$\mathcal{L} = \sum_{n=1}^{N} \sum_{s=1}^{r} \frac{M_s |\dot{\boldsymbol{r}}_{ns}|^2(t)}{2} - V, \tag{1.2}$$

where the second term is the sum of interactions between all pairs of atoms.

Under the harmonic approximation, i.e., expanding the total potential energy $V$ around the equilibrium positions, The Lagrangian can be simplified as

$$\mathcal{L} = \sum_{n=1}^{N} \sum_{s=1}^{r} \frac{M_s |\dot{\boldsymbol{u}}_{ns}|^2(t)}{2} - \frac{1}{2} \sum_{n,n'} \sum_{s,s'} \sum_{\alpha \alpha'} \phi_{\alpha \alpha'} \begin{pmatrix} n, n' \\ s, s' \end{pmatrix} u_{ns\alpha} u_{n's'\alpha'}, \tag{1.3}$$

where $u_{ns\alpha}, \alpha = 1,2,3$ are the Cartesian coordinates of the displacement $\boldsymbol{u}_{ns}(t)$ and

$$\phi_{\alpha \alpha'} \begin{pmatrix} n, n' \\ s, s' \end{pmatrix} = \left. \frac{\partial^2 V}{\partial r_{ns\alpha} r_{n's'\alpha'}} \right|_{\text{eq}} = \phi_{\alpha' \alpha} \begin{pmatrix} n', n \\ s', s \end{pmatrix} = \phi_{\alpha \alpha'} \begin{pmatrix} n - n' \\ s, \ s' \end{pmatrix}. \tag{1.4}$$

Note that the first order term vanishes because we are expanding around the equilibrium positions. $\phi_{\alpha \alpha'} \begin{pmatrix} n, n' \\ s, s' \end{pmatrix}$ represents the component of the force acted on the $s^{\text{th}}$ atom in the $n^{\text{th}}$ unit cell along $\alpha$ direction when the atom $s'$ in the unit cell $n'$ moves a unit displacement along $\alpha'$ direction. The symmetries of $\phi_{\alpha \alpha'} \begin{pmatrix} n, n' \\ s, s' \end{pmatrix}$ appearing in Eq. (1.4) arise from the intechangability of the second derivative and the translational invariance of the interactions.

### 1.2 Dynamical matrix

The equation of motion of the $s^{\text{th}}$ atom in the $n^{\text{th}}$ unit cell can be derived using the Euler–Lagrange equation as follows:

$$M_s \ddot{u}_{ns\alpha} = -\sum_{n's'\alpha'} \phi_{\alpha\alpha'} \begin{pmatrix} n - n' \\ s, & s' \end{pmatrix} u_{n's'\alpha'}, \tag{1.5}$$

If we displace all atoms equally, i.e. shifting $u_{n's'\alpha'}$ to $u_{n's'\alpha'} + \delta$, the total force on the $s^{th}$ atom in the $n^{th}$ unit cell does not change. From the above equation we have

$$\sum_{n'\alpha'} \phi_{\alpha\alpha'} \begin{pmatrix} n - n' \\ s, & s' \end{pmatrix} = 0, \tag{1.6}$$

We are looking for normal modes (because any general solution can be written as a linear combination of them); these are solutions where all atoms oscillate with the same frequency. Moreover, because of the lattice structure, we expect solutions to reflect this periodicity. So we guess solutions of the form

$$\tilde{u}_{ns\alpha}(t) = \frac{1}{\sqrt{M_s}} e_{s\alpha} e^{i(\mathbf{q} \cdot \mathbf{R}_n - \omega t)}, \tag{1.7}$$

where $e_{s\alpha}$ are real-space solutions that will be determined later and $\mathbf{q}$ is the wave-vector in reciprocal space. Substituting Eq. (1.7) into Eq. (1.5), we can get

$$\omega^2 e_{s\alpha}(\mathbf{q}) = \sum_{s'\alpha'} D_{s\alpha}^{s'\alpha'}(\mathbf{q}) e_{s'\alpha'}(\mathbf{q}), \tag{1.8}$$

where

$$D_{s\alpha}^{s'\alpha'}(\mathbf{q}) = \sum_{n'} \frac{1}{\sqrt{M_s M_{s'}}} \phi_{\alpha\alpha'} \begin{pmatrix} n - n' \\ s, & s' \end{pmatrix} e^{-i\mathbf{q} \cdot (\mathbf{R}_n - \mathbf{R}_{n'})} = \sum_l \frac{1}{\sqrt{M_s M_{s'}}} \phi_{\alpha\alpha'} \begin{pmatrix} l \\ s, & s' \end{pmatrix} e^{-i\mathbf{q} \cdot \mathbf{R}_l}, \tag{1.9}$$

is called dynamical matrix (dimension $3r \times 3r$). Note that we have defined the relative cell distance vector $\mathbf{R}_l \equiv \mathbf{R}_n - \mathbf{R}_{n'}$ and number index, $l \equiv n - n'$, where the infinite sum over $n'$ can be replaced by the sum over $l$ for any value of the index $n$. Note that dynamical matrix is Hermitian symmetric (i.e., $\left[D_{s\alpha}^{s'\alpha'}(\mathbf{q})\right]^* = D_{s'\alpha'}^{s\alpha}(-\mathbf{q})$) because $\phi$ is symmetric, so all the eigenvalues of Eq. (1.8) ($\omega_\lambda(\mathbf{q}), \lambda = 1,2 \cdots, 3r$) are real for each $\mathbf{q}$ in the Brillouin zone (BZ), which is determined by

$$\det \left| \omega_\lambda^2(\mathbf{q}) \delta_{s\alpha} \delta_{s'\alpha'} - D_{s\alpha}^{s'\alpha'}(\mathbf{q}) \right| = 0. \tag{1.10}$$

Taking the conjugate of this equation, we have

$$0 = \det \left| \left[\omega_\lambda^2(\mathbf{q})\right]^* \delta_{s\alpha} \delta_{s'\alpha'} - \left[D_{s\alpha}^{s'\alpha'}(\mathbf{q})\right]^* \right| = \det \left| \omega_\lambda^2(\mathbf{q}) \delta_{s\alpha} \delta_{s'\alpha'} - D_{s\alpha}^{s'\alpha'}(-\mathbf{q}) \right|, \tag{1.11}$$

while replacing $\mathbf{q}$ by $-\mathbf{q}$ in Eq. (1.10), we have $\det \left| \omega_\lambda^2(-\mathbf{q}) \delta_{s\alpha} \delta_{s'\alpha'} - D_{s\alpha}^{s'\alpha'}(-\mathbf{q}) \right| = 0$.

It's clear that $\omega_\lambda^2(\mathbf{q})$ and $\omega_\lambda^2(-\mathbf{q})$ obey the same equation, thus we have:

$$\omega_\lambda(-\mathbf{q}) = \omega_\lambda(\mathbf{q}). \tag{1.12}$$

The corresponding eigenvectors are orthonormal:

$$\sum_{s\alpha} \left(e_{s\alpha}^\lambda\right)^* e_{s\alpha}^{\lambda'} = \delta_{\lambda\lambda'}. \tag{1.13}$$

The complex conjugate of Eq. (1.8) gives

$$\omega^2 e_{s\alpha}^*(\mathbf{q}) = \sum_{s'\alpha'} \left[D_{s\alpha}^{s'\alpha'}(\mathbf{q})\right]^* e_{s'\alpha'}^*(\mathbf{q}) = \sum_{s'\alpha'} D_{s'\alpha'}^{s\alpha}(-\mathbf{q}) e_{s'\alpha'}^*(\mathbf{q}), \tag{1.14}$$

While replacing $\boldsymbol{q}$ by $-\boldsymbol{q}$ in Eq. (1.8) we get the following equation:

$$\omega^2 e_{s\alpha}(-\boldsymbol{q}) = \sum_{s'\alpha'} D_{s\alpha}^{s'\alpha'}(-\boldsymbol{q}) e_{s'\alpha'}(-\boldsymbol{q}), \tag{1.15}$$

From Eq. (1.14) and Eq. (1.15), we see that eigenvectors $[e_{s\alpha}^\lambda(-\boldsymbol{q})]^*$ and $e_{s\alpha}^\lambda(\boldsymbol{q})$ obey the same eigenvalue equation. Since the eigenvectors are normalized, we get the following property:

$$[e_{s\alpha}^\lambda(-\boldsymbol{q})]^* = e_{s\alpha}^\lambda(\boldsymbol{q}). \tag{1.16}$$

## 2 Second quantization

The general solution is a linear combination of all these normal modes, thus we have

$$u_{ns\alpha}(t) = \sum_{q\lambda} \frac{C_\lambda(\boldsymbol{q})}{\sqrt{NM_s}} [e_{s\alpha}^\lambda(\boldsymbol{q}) e^{-i\omega_{q\lambda} t}] e^{i\boldsymbol{q}\cdot\boldsymbol{R}_n} = \sum_{q\lambda} \frac{Q_\lambda(\boldsymbol{q},t)}{\sqrt{NM_s}} e_{s\alpha}^\lambda(\boldsymbol{q}) e^{i\boldsymbol{q}\cdot\boldsymbol{R}_n}, \tag{2.1}$$

where the we define the normal coordinates as $Q_\lambda(\boldsymbol{q},t) \equiv C_\lambda(\boldsymbol{q}) e^{-i\omega_\lambda(\boldsymbol{q})t}$ in the eigenvectors representation. To ensure that the displacements $u_{ns\alpha}(t)$ are real (namely, $u_{ns\alpha}^*(t) = u_{ns\alpha}(t)$), the following relation on $Q_\lambda(\boldsymbol{q},t)$ and $e_{s\alpha}^\lambda$ is enforced:

$$[Q_\lambda(\boldsymbol{q},t) e_{s\alpha}^\lambda(\boldsymbol{q})]^* = Q_\lambda(-\boldsymbol{q},t) e_{s\alpha}^\lambda(-\boldsymbol{q}), \tag{2.2}$$

where we used the fact that the sum over $\boldsymbol{q}$ runs symmetrically over both negative and positive values. Using Eq. (1.16), we have

$$[Q_\lambda(\boldsymbol{q},t)]^* = Q_\lambda(-\boldsymbol{q},t), \tag{2.3}$$

### 2.1 Kinetic energy term

Using Eq. (2.1), the kinetic energy of the system can be expressed as

$T = \frac{1}{2}\sum_{ns\alpha} M_s \dot{u}_{ns\alpha}^2(t) = \frac{1}{2}\sum_{s\alpha} \frac{M_s}{NM_s} \sum_{q\lambda} \sum_{q'\lambda'} [\dot{Q}_\lambda(\boldsymbol{q},t)\dot{Q}_{\lambda'}(\boldsymbol{q}',t)][e_{s\alpha}^\lambda(\boldsymbol{q}) e_{s\alpha}^{\lambda'}(\boldsymbol{q}')] \sum_n e^{i(\boldsymbol{q}+\boldsymbol{q}')\cdot\boldsymbol{R}_n} =$

$\frac{1}{2}\sum_{s\alpha}\sum_{qq'\lambda\lambda'} [\dot{Q}_\lambda(\boldsymbol{q},t)\dot{Q}_{\lambda'}(-\boldsymbol{q},t)][e_{s\alpha}^\lambda(\boldsymbol{q}) e_{s\alpha}^{\lambda'}(-\boldsymbol{q})] =$

$\frac{1}{2}\sum_{q\lambda\lambda'} [\dot{Q}_\lambda(\boldsymbol{q},t)\dot{Q}_{\lambda'}^*(\boldsymbol{q},t)] \sum_{s\alpha} \{e_{s\alpha}^\lambda(\boldsymbol{q}) [e_{s\alpha}^{\lambda'}(\boldsymbol{q})]^*\} = \frac{1}{2}\sum_{q\lambda} [\dot{Q}_\lambda(\boldsymbol{q},t)\dot{Q}_\lambda^*(\boldsymbol{q},t)] =$

$\frac{1}{2}\sum_{q\lambda} [\dot{Q}_\lambda(\boldsymbol{q},t)\dot{Q}_\lambda(-\boldsymbol{q},t)],$

i.e.,

$$T = \frac{1}{2}\sum_{q\lambda} [\dot{Q}_\lambda(\boldsymbol{q},t)\dot{Q}_\lambda(-\boldsymbol{q},t)], \tag{2.4}$$

To derive Eq. (2.4), we used Eqs. (1.13) and (1.16), as well as the following equations:

$$\frac{1}{N}\sum_n e^{i(\boldsymbol{q}+\boldsymbol{q}')\cdot\boldsymbol{R}_n} = \delta_{\boldsymbol{q}+\boldsymbol{q}',0}, \tag{2.5}$$

## 2.2 Potential energy term

Similarity, the potential energy can be rewritten in terms of the normal coordinates as follows:

$$U = \frac{1}{2}\sum_{n,n'}\sum_{s,s'}\sum_{\alpha\alpha'}\phi_{ns\alpha}^{n's'\alpha'}u_{ns\alpha}u_{n's'\alpha'} = \frac{1}{2}\sum_{s,s'}\sum_{\alpha\alpha'}\sum_{nn'}\phi_{\alpha\alpha'}\binom{n-n'}{s,s'}\frac{e^{i(q\cdot R_n + q'\cdot R_{n'})}}{N\sqrt{M_s M_s'}}\sum_{q\lambda}\sum_{q'\lambda'}[Q_\lambda(q,t)Q_{\lambda'}(q',t)]\left[e_{s\alpha}^\lambda(q)e_{s'\alpha'}^{\lambda'}(q')\right]$$

$$= \frac{1}{2}\sum_{s,s'}\sum_{\alpha\alpha'}\sum_{nl}\phi_{\alpha\alpha'}\binom{l}{s,s'}\frac{e^{i(q+q')\cdot R_n}e^{-iq'\cdot R_l}}{N\sqrt{M_s M_s'}}\sum_{q\lambda}\sum_{q'\lambda'}[Q_\lambda(q,t)Q_{\lambda'}(q',t)]\left[e_{s\alpha}^\lambda(q)e_{s'\alpha'}^{\lambda'}(q')\right]$$

$$= \frac{1}{2}\sum_{s,s'}\sum_{\alpha\alpha'}\sum_{l}\phi_{\alpha\alpha'}\binom{l}{s,s'}\frac{e^{-iq'\cdot R_l}}{\sqrt{M_s M_s'}}\sum_{q\lambda}\sum_{q'\lambda'}[Q_\lambda(q,t)Q_{\lambda'}(q',t)]\left[e_{s\alpha}^\lambda(q)e_{s'\alpha'}^{\lambda'}(q')\right]\underbrace{\sum_n\frac{1}{N}e^{i(q+q')\cdot R_n}}_{=\delta_{q+q'=0}}$$

$$= \frac{1}{2}\sum_{s,s'}\sum_{\alpha\alpha'}\sum_{l}\phi_{\alpha\alpha'}\binom{l}{s,s'}\frac{e^{iqR_l}}{\sqrt{M_s M_s'}}\sum_{q\lambda\lambda'}[Q_\lambda(q,t)Q_{\lambda'}(-q,t)]\left[e_{s\alpha}^\lambda(q)e_{s'\alpha'}^{\lambda'}(-q)\right]$$

$$= \frac{1}{2}\sum_{q\lambda\lambda'}[Q_\lambda(q,t)Q_{\lambda'}(-q,t)]\sum_{s\alpha}e_{s\alpha}^\lambda(q)\sum_{s'\alpha'}\left[D_{s\alpha}^{s'\alpha'}(-q)e_{s'\alpha'}^{\lambda'}(-q)\right]$$

$$= \frac{1}{2}\sum_{q\lambda\lambda'}[Q_\lambda(q,t)Q_{\lambda'}(-q,t)]\sum_{s\alpha}e_{s\alpha}^\lambda(q)\,\omega_{\lambda'}^2(-q)e_{s\alpha}^{\lambda'}(-q)$$

$$= \frac{1}{2}\sum_{q\lambda\lambda'}[Q_\lambda(q,t)Q_{\lambda'}(-q,t)]\omega_\lambda^2(q)\delta_{\lambda\lambda'} = \frac{1}{2}\sum_{q\lambda}\omega_\lambda^2(q)Q_\lambda(q,t)Q_\lambda(-q,t)$$

i.e.,

$$U = \frac{1}{2}\sum_{q\lambda}\omega_\lambda^2(q)Q_\lambda(q,t)Q_\lambda(-q,t), \tag{2.6}$$

where in above deriviation, the orthogonality of its eigenvectors and the symmetry of its eigenvalues are used. Thus the Lagrangian reads as:

$$\mathcal{L} = T - V = \frac{1}{2}\sum_{q\lambda}[\dot{Q}_\lambda(-q,t)\dot{Q}_\lambda(q,t) - \omega_\lambda^2(q)Q_\lambda(-q,t)Q_\lambda(q,t)], \tag{2.7}$$

Using the relation $P_\lambda(q,t) = \partial\mathcal{L}/\partial\dot{Q}_\lambda(q,t)$, the Hamiltonian of the system then can be written as:

$$H = T + V = \frac{1}{2}\sum_{q\lambda}[P_\lambda(-q,t)P_\lambda(q,t) + \omega_\lambda^2(q)Q_\lambda(-q,t)Q_\lambda(q,t)], \tag{2.8}$$

Now, we quantize $H$ by asking the momenta and coordinates to be operators:

$$\hat{H} = \frac{1}{2}\sum_{q\lambda}[\hat{P}_{-q\lambda}\hat{P}_{q\lambda} + \omega_{q\lambda}^2\hat{Q}_{-q\lambda}\hat{Q}_{q\lambda}], \tag{2.9}$$

here $\hat{P}_{q\lambda}$ and $\hat{Q}_{q\lambda}$ obey the following commutation relations:

$$\begin{cases} [\hat{Q}_{q\lambda},\hat{P}_{q'\lambda'}] = i\hbar\delta_{qq'}\delta_{\lambda\lambda'} \\ [\hat{Q}_{q\lambda},\hat{Q}_{q'\lambda'}] = 0,\ [\hat{P}_{q\lambda},\hat{P}_{q'\lambda'}] = 0 \end{cases}. \tag{2.10}$$

Similar to the case of ordinary quantum harmonic oscillators, it is convenient to define ladder operators for each mode as follows:

$$\hat{Q}_{q\lambda} = \sqrt{\frac{\hbar}{2\omega_{q\lambda}}}(\hat{b}_{q\lambda} + \hat{b}_{-q\lambda}^\dagger),\ \hat{P}_{q\lambda} = i\sqrt{\frac{\hbar\omega_{q\lambda}}{2}}(\hat{b}_{q\lambda}^\dagger - \hat{b}_{-q\lambda}), \tag{2.11}$$

where $\hat{b}_{q\lambda}^\dagger$ and $\hat{b}_{q\lambda}$ are Bosonic creation and annihilation operators for phonons with momentum $q$, branch index $\lambda$, and frequency $\omega_{q\lambda}$, which obeys the Bosonic commutation relation:

$$\begin{cases} [\hat{b}_{q\lambda}, \hat{b}^\dagger_{q'\lambda'}] = \delta_{qq'}\delta_{\lambda\lambda'} \\ [\hat{b}_{q\lambda}, \hat{b}_{q'\lambda'}] = \hat{0}, \ [\hat{b}^\dagger_{q\lambda}, \hat{b}^\dagger_{q'\lambda'}] = \hat{0} \end{cases} \quad (2.12)$$

Substituting Eqs. (2.11) into (2.9) and using the properties Eq. (2.10) and (2.12), we have

$$H = \frac{1}{2}\sum_{q\lambda}[\hat{P}_{-q\lambda}\hat{P}_{q\lambda} + \omega^2_{q\lambda}\hat{Q}_{-q\lambda}\hat{Q}_{q\lambda}]$$

$$= \frac{1}{2}\sum_{q\lambda}\left[-\frac{\hbar\omega_{q\lambda}}{2}(\hat{b}^\dagger_{-q\lambda} - \hat{b}_{q\lambda})(\hat{b}^\dagger_{q\lambda} - \hat{b}_{-q\lambda}) + \omega^2_{q\lambda}\frac{\hbar}{2\omega_{q\lambda}}(\hat{b}_{-q\lambda} + \hat{b}^\dagger_{q\lambda})(\hat{b}_{q\lambda} + \hat{b}^\dagger_{-q\lambda})\right]$$

$$= \frac{\hbar}{4}\sum_{q\lambda}\omega_{q\lambda}[-(\hat{b}^\dagger_{-q\lambda}\hat{b}^\dagger_{q\lambda} - \hat{b}^\dagger_{-q\lambda}\hat{b}_{-q\lambda} - \hat{b}_{q\lambda}\hat{b}^\dagger_{q\lambda} + \hat{b}_{q\lambda}\hat{b}_{-q\lambda})$$

$$+ (\hat{b}_{-q\lambda}\hat{b}_{q\lambda} + \hat{b}_{-q\lambda}\hat{b}^\dagger_{-q\lambda} + \hat{b}^\dagger_{q\lambda}\hat{b}_{q\lambda} + \hat{b}^\dagger_{q\lambda}\hat{b}^\dagger_{-q\lambda})]$$

$$= \frac{\hbar}{4}\sum_{q\lambda}\omega_{q\lambda}[(\hat{b}^\dagger_{-q\lambda}\hat{b}_{-q\lambda} + \hat{b}_{-q\lambda}\hat{b}^\dagger_{-q\lambda}) + (\hat{b}_{q\lambda}\hat{b}^\dagger_{q\lambda} + \hat{b}^\dagger_{q\lambda}\hat{b}_{q\lambda})]$$

$$= \frac{\hbar}{4}\sum_{q\lambda}\omega_{q\lambda}[(2\hat{b}^\dagger_{-q\lambda}\hat{b}_{-q\lambda} + 1) + (2\hat{b}^\dagger_{q\lambda}\hat{b}_{q\lambda} + 1)] = \sum_{q\lambda}\hbar\omega_{q\lambda}\left(\hat{b}^\dagger_{q\lambda}\hat{b}_{q\lambda} + \frac{1}{2}\right)$$

Therefore, we finally get the quantized representation of non-interacting phonons:

$$H = \sum_{q\lambda}\hbar\omega_{q\lambda}\left(\hat{b}^\dagger_{q\lambda}\hat{b}_{q\lambda} + \frac{1}{2}\right). \quad (2.13)$$

The operator of atom displacements (Eq. 1.14) is expressed in terms of the phonon operators by:

$$\hat{u}_{ns\alpha} = \sum_{q\lambda}\sqrt{\frac{\hbar}{2NM_s\omega_{q\lambda}}}e^\lambda_{s\alpha}e^{iq\cdot R_n}(\hat{b}_{q\lambda} + \hat{b}^\dagger_{-q\lambda}). \quad (2.14)$$

These equations will be used below.

## 3 Inter-phonon coupling within harmonic approximation

### 3.1 Hamiltonian with inter-phonon coupling

In this section, we consider systems that consist of two (or more) covalently bonded units that are weakly coupled between them. Unlike previous studies that considered phonon-phonon couplings resulting from anharmonicity effects [1], here all phonon mode considered are Harmonic and the couplings arise from the division of the entire system into subunits. The Hamiltonian of the whole system can be written as:

$$H = H_1 + H_2 + H_{12}, \quad (3.1)$$

To derive the expressions of $H_1$, $H_2$ and $H_{12}$ in Eq. (3.1), we consider the Hamiltonian of the whole system written as the function of the atomic displacements:

$$H = T + V = \sum_{ns\alpha}\frac{M_s\dot{u}^2_{ns\alpha}(t)}{2} + \frac{1}{2}\sum_{n,n'}\sum_{s,s'}\sum_{\alpha\alpha'}\phi_{\alpha\alpha'}\binom{n-n'}{s,\ s'}u_{ns\alpha}u_{n's'\alpha'}, \quad (3.2)$$

Let's assume that subsystem I and II contain atoms with indeices ranging from $s = 1, 2, \cdots, r/2$ and $s = \frac{r}{2}+1, \frac{r}{2}+2, \cdots, r$, respectively. Then the kinetic energy term in Eq. (3.2) can be rewritten as:

$$T = \sum_{n\alpha}\left[\sum_{s=1}^{r/2}\frac{M_s \dot{u}_{ns\alpha}^2(t)}{2} + \sum_{s=\frac{r}{2}+1}^{r}\frac{M_s \dot{u}_{ns\alpha}^2(t)}{2}\right]. \tag{3.3}$$

While the potential energy term is:

$$V = \frac{1}{2}\sum_{n,n'}\sum_{\alpha,\alpha'}\left\{\sum_{s,s'=1}^{r/2}\phi_{\alpha\alpha'}\binom{n-n'}{s,\ s'}u_{ns\alpha}u_{n's'\alpha'} + \sum_{s,s'=r/2+1}^{r}\phi_{\alpha\alpha'}\binom{n-n'}{s,\ s'}u_{ns\alpha}u_{n's'\alpha'} + \sum_{s=1}^{r/2}\sum_{s'=r/2+1}^{r}\phi_{\alpha\alpha'}\binom{n-n'}{s,\ s'}u_{ns\alpha}u_{n's'\alpha'}$$

$$+ \sum_{s=r/2+1}^{r}\sum_{s'=1}^{r/2}\phi_{\alpha\alpha'}\binom{n-n'}{s,\ s'}u_{ns\alpha}u_{n's'\alpha'}\right\} = V_{11} + V_{22} + V_{12} + V_{21}$$

Or equivalently,

$$\begin{cases} V_{11} = \frac{1}{2}\sum_{n,n'}\sum_{\alpha,\alpha'}\sum_{s,s'=1}^{\frac{r}{2}}\phi_{\alpha\alpha'}\binom{n-n'}{s,\ s'}u_{ns\alpha}u_{n's'\alpha'} \\ V_{22} = \frac{1}{2}\sum_{n,n'}\sum_{\alpha,\alpha'}\sum_{s,s'=r/2+1}^{r}\phi_{\alpha\alpha'}\binom{n-n'}{s,\ s'}u_{ns\alpha}u_{n's'\alpha'} \\ V_{12} = \frac{1}{2}\sum_{n,n'}\sum_{\alpha,\alpha'}\sum_{s=1}^{r/2}\sum_{s'=r/2+1}^{r}\phi_{\alpha\alpha'}\binom{n-n'}{s,\ s'}u_{ns\alpha}u_{n's'\alpha'} \\ V_{21} = \frac{1}{2}\sum_{n,n'}\sum_{\alpha,\alpha'}\sum_{s=r/2+1}^{r}\sum_{s'=1}^{r/2}\phi_{\alpha\alpha'}\binom{n-n'}{s,\ s'}u_{ns\alpha}u_{n's'\alpha'} \end{cases}, \tag{3.4}$$

## *3.2 Second quantization*

We may now quantize this Hamiltonian and write it in the basis of the eigenstates of the coupled subunits. In second quantization, the atomic displacement operators of the two subunits are given in the following form:

$$\hat{u}_{ns\alpha} = \begin{cases} \sum_{q\lambda}\sqrt{\frac{\hbar}{2NM_s\omega_{q\lambda}}}e_{s\alpha}^{\lambda}e^{iq\cdot R_n}(\hat{a}_{q\lambda} + \hat{a}_{-q\lambda}^{\dagger}), & s \in \left[1, \frac{r}{2}\right] \\ \sum_{q'\lambda'}\sqrt{\frac{\hbar}{2NM_s\widetilde{\omega}_{q'\lambda'}}}\widetilde{e}_{s\alpha}^{\lambda'}e^{iq'\cdot R_n}\left(\hat{\widetilde{a}}_{q'\lambda'} + \hat{\widetilde{a}}_{-q'\lambda'}^{\dagger}\right), & s \in \left[\frac{r}{2}+1, r\right] \end{cases}. \tag{3.5}$$

Here, we use different notations for the creation and annihilation operators ($\hat{a}_{q\lambda}, \hat{a}_{-q\lambda}^{\dagger}$ and $\hat{\widetilde{a}}_{q\lambda}, \hat{\widetilde{a}}_{-q\lambda}^{\dagger}$), eigenvalues ($\omega_{q\lambda}, \widetilde{\omega}_{q\lambda}$) and eigenvectors ($e_{s\alpha}^{\lambda}, \widetilde{e}_{s\alpha}^{\lambda}$) for the two subunits. Note that $e_{s\alpha}^{\lambda}$ and $\widetilde{e}_{s\alpha}^{\lambda}$ are of the dimensions of the whole system, however their non-zero elements appear only on the relevant subunits such that they obey the following relations: $\sum_{s\alpha}(e_{s\alpha}^{\lambda})^*e_{s\alpha}^{\lambda'} = \delta_{\lambda\lambda'}$ ; $\sum_{s\alpha}(\widetilde{e}_{s\alpha}^{\lambda})^*\widetilde{e}_{s\alpha}^{\lambda'} = \delta_{\lambda\lambda'}$ ; $\sum_{s\alpha}(\widetilde{e}_{s\alpha}^{\lambda})^*e_{s\alpha}^{\lambda'} = \sum_{s\alpha}(e_{s\alpha}^{\lambda})^*\widetilde{e}_{s\alpha}^{\lambda'} = 0$. Defining the normal coordinates of the two subunits as:

$$\hat{Q}_{q\lambda} \equiv \sqrt{\frac{\hbar}{2\omega_{q\lambda}}}(\hat{a}_{q\lambda} + \hat{a}_{-q\lambda}^{\dagger}) \ ; \ \hat{\widetilde{Q}}_{q\lambda} \equiv \sqrt{\frac{\hbar}{2\widetilde{\omega}_{q\lambda}}}(\hat{\widetilde{a}}_{q\lambda} + \hat{\widetilde{a}}_{-q\lambda}^{\dagger}), \tag{3.6}$$

and the corresponding momenta operators as:

$$\hat{P}_{q\lambda} = i\sqrt{\frac{\hbar\omega_{q\lambda}}{2}}\left(\hat{a}^\dagger_{q\lambda} - \hat{a}_{-q\lambda}\right) \;;\; \hat{\tilde{P}}_{q\lambda} = i\sqrt{\frac{\hbar\tilde{\omega}_{q\lambda}}{2}}\left(\hat{\tilde{a}}^\dagger_{q\lambda} - \hat{\tilde{a}}_{-q\lambda}\right), \tag{3.7}$$

Eq. (3.5) reads:

$$\hat{u}_{ns\alpha} = \begin{cases} \sum_{q\lambda} \dfrac{e^\lambda_{s\alpha}(q)}{\sqrt{NM_s}} e^{iq\cdot R_n} \hat{Q}_{q\lambda}, & s \in \left[1, \dfrac{r}{2}\right] \\ \sum_{q\lambda} \dfrac{\tilde{e}^\lambda_{s\alpha}(q)}{\sqrt{NM_s}} e^{iq\cdot R_n} \hat{\tilde{Q}}_{q\lambda}, & s \in \left[\dfrac{r}{2}+1, r\right] \end{cases}, \tag{3.8}$$

The second quantized kinetic energy operator is then written as:

$$\hat{T} = \hat{T}_1 + \hat{T}_2 = \frac{1}{2}\sum_{q\lambda} \hat{P}_{q\lambda}\hat{P}_{-q\lambda} + \frac{1}{2}\sum_{q'\lambda'} \hat{\tilde{P}}_{q'\lambda'}\hat{\tilde{P}}_{-q'\lambda'}, \tag{3.9}$$

Correspondingly, the various potential energy terms are obtained by substituting Eq. (3.8) in Eq. (3.4):

$$\hat{V}_{11} = \frac{1}{2}\sum_{\alpha,\alpha'} \sum_{s,s'=1}^{\frac{r}{2}} \sum_{n,n'} \phi_{\alpha\alpha'}\binom{n-n'}{s,\;s'} \sum_{q\lambda,q'\lambda'} \frac{e^\lambda_{s\alpha}(q)e^{\lambda'}_{s'\alpha'}(q')}{\sqrt{M_s M_{s'}}} \frac{1}{N} e^{iq\cdot R_n + iq'\cdot R_{n'}} \hat{Q}_{q\lambda}\hat{Q}_{q'\lambda'}$$

$$= \frac{1}{2}\sum_{\alpha,\alpha'} \sum_{s,s'=1}^{\frac{r}{2}} \sum_{qq'\lambda,\lambda'} \hat{Q}_{q\lambda}\hat{Q}_{q'\lambda'}\, e^\lambda_{s\alpha}(q)e^{\lambda'}_{s'\alpha'}(q') \sum_{n'}\frac{1}{N} e^{i(q+q')R_{n'}} \sum_n \phi_{\alpha\alpha'}\binom{n-n'}{s,\;s'}\frac{e^{iq\cdot(R_n-R_{n'})}}{\sqrt{M_s M_{s'}}}$$

$$= \frac{1}{2}\sum_{\alpha,\alpha'} \sum_{s,s'=1}^{\frac{r}{2}} \sum_{qq'\lambda,\lambda'} \hat{Q}_{q\lambda}\hat{Q}_{q'\lambda'}\, e^\lambda_{s\alpha}(q)e^{\lambda'}_{s'\alpha'}(q') \left[\sum_l \phi_{\alpha\alpha'}\binom{l}{s,\;s'}\frac{e^{iq\cdot R_l}}{\sqrt{M_s M_{s'}}}\right]\left[\sum_{n'}\frac{1}{N} e^{i(q+q')\cdot R_{n'}}\right]$$

$$= \frac{1}{2}\sum_{\alpha,\alpha'} \sum_{s,s'=1}^{\frac{r}{2}} \sum_{q\lambda,\lambda'} \hat{Q}_{q\lambda}\hat{Q}_{-q\lambda'}\, e^\lambda_{s\alpha}(q)e^{\lambda'}_{s'\alpha'}(-q) \sum_l \phi_{\alpha\alpha'}\binom{l}{s,\;s'}\frac{e^{iq\cdot R_l}}{\sqrt{M_s M_{s'}}}$$

$$= \frac{1}{2}\sum_\alpha \sum_{s=1}^{\frac{r}{2}} \sum_{q\lambda,\lambda'} \hat{Q}_{q\lambda}\hat{Q}_{-q\lambda'}\, e^\lambda_{s\alpha}(q) \sum_{s'=1}^{\frac{r}{2}} \sum_{\alpha'} D^{s'\alpha'}_{s\alpha}(-q) e^{\lambda'}_{s'\alpha'}(-q)$$

$$= \frac{1}{2}\sum_{q\lambda,\lambda'} \omega^2_{q\lambda'} \hat{Q}_{q\lambda}\hat{Q}_{-q\lambda'} \sum_\alpha \sum_{s=1}^{\frac{r}{2}} e^\lambda_{s\alpha}(q)[e^{\lambda'}_{s\alpha}(q)]^* = \frac{1}{2}\sum_{q\lambda} \omega^2_{q\lambda}\hat{Q}_{q\lambda}\hat{Q}_{-q\lambda}$$

i.e.,

$$\hat{V}_{11} = \frac{1}{2}\sum_{q\lambda}\omega^2_{q\lambda}\hat{Q}_{q\lambda}\hat{Q}_{-q\lambda'}, \tag{3.10}$$

Going from the second line to the third line we used the fact that the sum over $n'$ runs between $\pm\infty$ and the summand depends only on the difference between $n$ and $n'$, hence the sum is independent of the value of the index $n$. Therefore, we can replace the sum over $n'$ by a sum over $l \equiv n - n'$ amd define $R_l \equiv R_n - R_{n'}$. Following the same procedure, we can get the corresponding expressions for the second diagonal term:

$$\hat{V}_{22} = \frac{1}{2}\sum_{q\lambda}\tilde{\omega}^2_{q\lambda}\hat{\tilde{Q}}_{q\lambda}\hat{\tilde{Q}}_{-q\lambda}, \tag{3.11}$$

Similarly, for the off-diagonal terms we get:

$$\hat{V}_{12} = \frac{1}{2}\sum_{\alpha,\alpha'}\sum_{s=1}^{\frac{r}{2}}\sum_{s'=\frac{r}{2}+1}^{r}\sum_{n,n'}\phi_{\alpha\alpha'}\begin{pmatrix}n-n'\\s,\ s'\end{pmatrix}\sum_{q\lambda,q'\lambda'}\frac{e^{\lambda}_{s\alpha}(q)\tilde{e}^{\lambda'}_{s'\alpha'}(q')}{\sqrt{M_s M_{s'}}}\frac{1}{N}e^{iq\cdot R_n + iq'\cdot R_{n'}}\hat{Q}_{q\lambda}\hat{\tilde{Q}}_{q'\lambda'}$$

$$= \frac{1}{2}\sum_{\alpha,\alpha'}\sum_{s=1}^{\frac{r}{2}}\sum_{s'=\frac{r}{2}+1}^{r}\sum_{qq'\lambda,\lambda'}\hat{Q}_{q\lambda}\hat{\tilde{Q}}_{q'\lambda'}\,e^{\lambda}_{s\alpha}(q)\tilde{e}^{\lambda'}_{s'\alpha'}(q')\sum_{n'}\frac{1}{N}e^{i(q+q')\cdot R_{n'}}\sum_{n}\phi_{\alpha\alpha'}\begin{pmatrix}n-n'\\s,\ s'\end{pmatrix}\frac{e^{iq\cdot(R_n-R_{n'})}}{\sqrt{M_s M_{s'}}}$$

$$= \frac{1}{2}\sum_{\alpha,\alpha'}\sum_{s=1}^{\frac{r}{2}}\sum_{s'=\frac{r}{2}+1}^{r}\sum_{qq'\lambda,\lambda'}\hat{Q}_{q\lambda}\hat{\tilde{Q}}_{q'\lambda'}\,e^{\lambda}_{s\alpha}(q)\tilde{e}^{\lambda'}_{s'\alpha'}(q')\left[\sum_{l}\phi_{\alpha\alpha'}\begin{pmatrix}l\\s,\ s'\end{pmatrix}\frac{e^{iq\cdot R_l}}{\sqrt{M_s M_{s'}}}\right]\left[\sum_{n'}\frac{1}{N}e^{i(q+q')\cdot R_{n'}}\right]$$

$$= \frac{1}{2}\sum_{\alpha,\alpha'}\sum_{s=1}^{\frac{r}{2}}\sum_{s'=\frac{r}{2}+1}^{r}\sum_{q,\lambda,\lambda'}\hat{Q}_{q\lambda}\hat{\tilde{Q}}_{-q\lambda'}\,e^{\lambda}_{s\alpha}(q)\tilde{e}^{\lambda'}_{s'\alpha'}(-q)\sum_{l}\phi_{\alpha\alpha'}\begin{pmatrix}l\\s,\ s'\end{pmatrix}\frac{e^{iq\cdot R_l}}{\sqrt{M_s M_{s'}}}$$

$$= \frac{1}{2}\sum_{q}\sum_{\lambda=1}^{3r/2}\sum_{\lambda'=\frac{3r}{2}+1}^{3r}\hat{Q}_{q\lambda}\hat{\tilde{Q}}^{\dagger}_{q\lambda'}\left[\sum_{\alpha\alpha'}\sum_{s=1}^{\frac{r}{2}}\sum_{s'=\frac{r}{2}+1}^{r}\left(\tilde{e}^{\lambda'}_{s'\alpha'}(q)\right)^{*}\sum_{l}\phi_{\alpha\alpha'}\begin{pmatrix}l\\s,\ s'\end{pmatrix}\frac{e^{iq\cdot R_l}}{\sqrt{M_s M_{s'}}}e^{\lambda}_{s\alpha}(q)\right]$$

$$= \frac{1}{2}\sum_{q}\sum_{\lambda=1}^{3r/2}\sum_{\lambda'=\frac{3r}{2}+1}^{3r}V_{\lambda\lambda'}(q)\hat{Q}_{q\lambda}\hat{\tilde{Q}}^{\dagger}_{q\lambda'}$$

Where we have defined:

$$V_{\lambda\lambda'}(\boldsymbol{q}) \equiv \sum_{\alpha\alpha'}\sum_{s=1}^{\frac{r}{2}}\sum_{s'=\frac{r}{2}+1}^{r}\left(\tilde{e}^{\lambda'}_{s'\alpha'}(\boldsymbol{q})\right)^{*}V^{s'\alpha'}_{s\alpha}(\boldsymbol{q})e^{\lambda}_{s\alpha}(\boldsymbol{q}),\tag{3.12}$$

where

$$V^{s'\alpha'}_{s\alpha}(\boldsymbol{q}) \equiv \sum_{l}\frac{e^{iq\cdot R_l}}{\sqrt{M_s M_{s'}}}\phi_{\alpha\alpha'}\begin{pmatrix}l\\s,\ s'\end{pmatrix}.\tag{3.13}$$

It's easy to show that $V_{\lambda\lambda'}(\boldsymbol{q})$ has the following property:

$$V^{*}_{\lambda\lambda'}(\boldsymbol{q}) = V_{\lambda\lambda'}(-\boldsymbol{q}).\tag{3.14}$$

Following the same procedure, we have:

$$\hat{V}_{21} = \frac{1}{2}\sum_{\alpha,\alpha'}\sum_{s'=1}^{\frac{r}{2}}\sum_{s=\frac{r}{2}+1}^{r}\sum_{n,n'}\phi_{\alpha\alpha'}\binom{n-n'}{s,\ s'}\sum_{q\lambda,q'\lambda'}\frac{\tilde{e}_{s\alpha}^{\lambda}(q)e_{s'\alpha'}^{\lambda'}(q')}{\sqrt{M_s M_{s'}}}\frac{1}{N}e^{iq\cdot R_n + iq'\cdot R_{n'}}\hat{\tilde{Q}}_{q\lambda}\hat{Q}_{q'\lambda'}$$

$$= \frac{1}{2}\sum_{\alpha',\alpha}\sum_{s=1}^{\frac{r}{2}}\sum_{s'=\frac{r}{2}+1}^{r}\sum_{n',n}\phi_{\alpha'\alpha}\binom{n'-n}{s',s}\sum_{q'\lambda',q\lambda}\frac{\tilde{e}_{s'\alpha'}^{\lambda'}(q')e_{s\alpha}^{\lambda}(q)}{\sqrt{M_{s'}M_s}}\frac{1}{N}e^{iq'\cdot R_{n'}+iq\cdot R_n}\hat{\tilde{Q}}_{q'\lambda'}\hat{Q}_{q\lambda}$$

$$= \frac{1}{2}\sum_{\alpha,\alpha'}\sum_{s=1}^{\frac{r}{2}}\sum_{s'=\frac{r}{2}+1}^{r}\sum_{qq'\lambda,\lambda'}\hat{\tilde{Q}}_{q'\lambda'}\hat{Q}_{q\lambda}\tilde{e}_{s'\alpha'}^{\lambda'}(q')e_{s\alpha}^{\lambda}(q)\sum_{n'}\frac{1}{N}e^{i(q+q')\cdot R_{n'}}\sum_{n}\phi_{\alpha\alpha'}\binom{n-n'}{s,s'}\frac{e^{iq\cdot(R_n-R_{n'})}}{\sqrt{M_s M_{s'}}}$$

$$= \frac{1}{2}\sum_{\alpha,\alpha'}\sum_{s=1}^{\frac{r}{2}}\sum_{s'=\frac{r}{2}+1}^{r}\sum_{qq'\lambda,\lambda'}\hat{\tilde{Q}}_{q'\lambda'}\hat{Q}_{q\lambda}\tilde{e}_{s'\alpha'}^{\lambda'}(q')e_{s\alpha}^{\lambda}(q)\left[\sum_{l}\phi_{\alpha\alpha'}\binom{l}{s,s'}\frac{e^{iq\cdot R_l}}{\sqrt{M_s M_{s'}}}\right]\left[\sum_{n'}\frac{1}{N}e^{i(q+q')\cdot R_{n'}}\right]$$

$$= \frac{1}{2}\sum_{\alpha,\alpha'}\sum_{s=1}^{\frac{r}{2}}\sum_{s'=\frac{r}{2}+1}^{r}\sum_{q'\lambda,\lambda'}\hat{\tilde{Q}}_{q'\lambda'}\hat{Q}_{-q'\lambda}\tilde{e}_{s'\alpha'}^{\lambda'}(q')e_{s\alpha}^{\lambda}(-q')\sum_{l}\phi_{\alpha\alpha'}\binom{l}{s,s'}\frac{e^{-iq'\cdot R_l}}{\sqrt{M_s M_{s'}}}$$

$$= \frac{1}{2}\sum_{q'}\sum_{\lambda=1}^{3r/2}\sum_{\lambda'=\frac{3r}{2}+1}^{3r}\hat{\tilde{Q}}_{q'\lambda'}\hat{Q}_{q'\lambda}^{\dagger}\left[\sum_{\alpha\alpha'}\sum_{s=1}^{\frac{r}{2}}\sum_{s'=\frac{r}{2}+1}^{r}\left(e_{s\alpha}^{\lambda}(q')\right)^{*}\sum_{l}\phi_{\alpha\alpha'}\binom{l}{s,s'}\frac{e^{-iq'\cdot R_l}}{\sqrt{M_s M_{s'}}}\tilde{e}_{s'\alpha'}^{\lambda'}(q')\right]$$

$$= \frac{1}{2}\sum_{q}\sum_{\lambda=1}^{3r/2}\sum_{\lambda'=\frac{3r}{2}+1}^{3r}\hat{\tilde{Q}}_{q\lambda'}\hat{Q}_{q\lambda}^{\dagger}\left[\sum_{\alpha\alpha'}\sum_{s=1}^{\frac{r}{2}}\sum_{s'=\frac{r}{2}+1}^{r}\left(e_{s\alpha}^{\lambda}(q)\right)^{*}\sum_{l}\phi_{\alpha\alpha'}\binom{l}{s,\ s'}\frac{e^{-iq\cdot R_l}}{\sqrt{M_s M_{s'}}}\tilde{e}_{s'\alpha'}^{\lambda'}(q)\right]$$

$$= \frac{1}{2}\sum_{q}\sum_{\lambda=1}^{3r/2}\sum_{\lambda'=\frac{3r}{2}+1}^{3r}\left(\hat{Q}_{q\lambda}\hat{\tilde{Q}}_{q\lambda'}^{\dagger}\right)^{\dagger}\left[\sum_{\alpha\alpha'}\sum_{s=1}^{\frac{r}{2}}\sum_{s'=\frac{r}{2}+1}^{r}\left(\tilde{e}_{s'\alpha'}^{\lambda'}(q)\right)^{*}\sum_{l}\phi_{\alpha\alpha'}\binom{l}{s,\ s'}\frac{e^{iq\cdot R_l}}{\sqrt{M_s M_{s'}}}e_{s\alpha}^{\lambda}(q)\right]^{*}$$

$$= \left\{\frac{1}{2}\sum_{q}\sum_{\lambda=1}^{\frac{3r}{2}}\sum_{\lambda'=\frac{3r}{2}+1}^{3r}V_{\lambda\lambda'}(q)\hat{Q}_{q\lambda}\hat{\tilde{Q}}_{q\lambda'}^{\dagger}\right\}^{\dagger} = V_{12}^{\dagger}$$

Define:

$$\begin{cases}\hat{H}_1 = \hat{T}_{11} + \hat{V}_{11} \\ \hat{H}_2 = \hat{T}_{22} + \hat{V}_{22}, \\ \hat{H}_{12} = \hat{V}_{12} + \hat{V}_{21}\end{cases} \tag{3.15}$$

we finally get the expressions of $\hat{H}_1$, $\hat{H}_2$ and $\hat{H}_{12}$ in Eq. (3.1) as follows:

$$\begin{cases}\hat{H}_1 = \sum_q \sum_{\lambda=1}^{3r/2}\hbar\omega_{q\lambda}\left(\hat{a}_{q\lambda}^{\dagger}\hat{a}_{q\lambda}+\frac{1}{2}\right) \\ \hat{H}_2 = \sum_q \sum_{\lambda'=1+\frac{3r}{2}}^{3r}\hbar\tilde{\omega}_{q\lambda'}\left(\hat{\tilde{a}}_{q\lambda'}^{\dagger}\hat{\tilde{a}}_{q\lambda'}+\frac{1}{2}\right) \\ \hat{H}_{12} = \frac{1}{2}\sum_q\left\{\sum_{\lambda=1}^{3r/2}\sum_{\lambda'=\frac{3r}{2}+1}^{3r}V_{\lambda\lambda'}(q)\hat{Q}_{q\lambda}\hat{\tilde{Q}}_{q\lambda'}^{\dagger}+\text{h.c.}\right\}\end{cases} \tag{3.16}$$

Here h.c. means the Hermitian conjugate. Since the indices of $\hat{a}_{q\lambda}$, $\hat{\tilde{a}}_{q\lambda'}$ and $\hat{Q}_{q\lambda}$, $\hat{\tilde{Q}}_{q\lambda'}$ belong to the two system sections, we can define an abbreviated notation $\hat{a}_{q\lambda}$ and $\hat{Q}_{q\lambda}$ using the index $\lambda$ to identify which subsystem they belong to. In this case, the Hamiltonian of the coupled system can be

simplified as follows:
$$\hat{H} = \hat{H}_0 + \hat{H}_C, \tag{3.17}$$
where
$$\begin{cases} \hat{H}_0 = \sum_q \sum_{\lambda=1}^{3r} \hbar \omega_{q\lambda} \left( \hat{a}^\dagger_{q\lambda} \hat{a}_{q\lambda} + \frac{1}{2} \right) \\ \hat{H}_C = \frac{1}{2} \sum_q \left\{ \sum_{\lambda=1}^{3r/2} \sum_{\lambda'=\frac{3r}{2}+1}^{3r} V_{\lambda\lambda'}(q) \hat{Q}_{q\lambda} \hat{Q}^\dagger_{q\lambda'} + h.c. \right\} \end{cases} \tag{3.18}$$

# 4 Green's function

## 4.1 Dynamics of the ladder operators

In order to describe the dynamics of the ladder operators appearing in the Hamiltonian of Eq. (3.17) we express them in the Heisenberg picture as follows:
$$\hat{a}_p(\tau) = e^{\frac{i}{\hbar}\hat{H}t} \hat{a}_p e^{-\frac{i}{\hbar}\hat{H}t} = e^{\frac{\hat{H}\tau}{\hbar}} \hat{a}_p e^{-\frac{\hat{H}\tau}{\hbar}}, \tag{4.1}$$
where we define the imaginary time $\tau \equiv it$ and introduce the notation $p \equiv (q, \lambda)$ and $\bar{p} \equiv (-q, \lambda)$. The corresponding equation of motion for the ladder operators is given by:
$$\hbar \frac{\partial \hat{a}_p(\tau)}{\partial \tau} = [\hat{H}, \hat{a}_p(\tau)]. \tag{4.2}$$
For the uncoupled system ($\hat{H}_C = \hat{0}$) this gives:
$$\hbar \frac{\partial \hat{a}_p(\tau)}{\partial \tau} = [\hat{H}_0, \hat{a}_p(\tau)] = \left[ \hat{H}_0, e^{\frac{\hat{H}_0\tau}{\hbar}} \hat{a}_p e^{-\frac{\hat{H}_0\tau}{\hbar}} \right] = \hat{H}_0 e^{\frac{\hat{H}_0\tau}{\hbar}} \hat{a}_p e^{-\frac{\hat{H}_0\tau}{\hbar}} - e^{\frac{\hat{H}_0\tau}{\hbar}} \hat{a}_p e^{-\frac{\hat{H}_0\tau}{\hbar}} \hat{H}_0$$
$$= e^{\frac{\hat{H}_0\tau}{\hbar}} (\hat{H}_0 \hat{a}_p - \hat{a}_p \hat{H}_0) e^{-\frac{\hat{H}_0\tau}{\hbar}} = e^{\frac{\hat{H}_0\tau}{\hbar}} [\hat{H}_0, \hat{a}_p] e^{-\frac{\hat{H}_0\tau}{\hbar}}$$
i.e.,
$$\hbar \frac{\partial \hat{a}_p(\tau)}{\partial \tau} = e^{\frac{\hat{H}_0\tau}{\hbar}} [\hat{H}_0, \hat{a}_p] e^{-\frac{\hat{H}_0\tau}{\hbar}}. \tag{4.3}$$
The commutator on the right-hand-side of Eq. (4.3) can be evaluated as follows:
$$[\hat{H}_0, \hat{a}_{p'}] = \left[ \sum_p \hbar \omega_p \left( \hat{a}^\dagger_p \hat{a}_p + \frac{1}{2} \right), \hat{a}_{p'} \right] = \sum_p \hbar \omega_p [\hat{a}^\dagger_p \hat{a}_p, \hat{a}_{p'}] = \sum_p \hbar \omega_p (\hat{a}^\dagger_p \hat{a}_p \hat{a}_{p'} - \hat{a}_{p'} \hat{a}^\dagger_p \hat{a}_p) =$$
$$\sum_p \hbar \omega_p [\hat{a}^\dagger_p \hat{a}_p \hat{a}_{p'} - (\delta_{p'p} + \hat{a}^\dagger_p \hat{a}_{p'}) \hat{a}_p] = -\hbar \omega_{p'} \hat{a}_{p'} + \sum_p \hbar \omega_p [\hat{a}^\dagger_p \hat{a}_p \hat{a}_{p'} - \hat{a}^\dagger_p \hat{a}_p \hat{a}_{p'}] = -\hbar \omega_{p'} \hat{a}_{p'}.$$
i.e.,
$$[\hat{H}_0, \hat{a}_{p'}] = -\hbar \omega_{p'} \hat{a}_{p'}. \tag{4.4}$$
Therefore, we have:
$$\hbar \frac{\partial \hat{a}_p(\tau)}{\partial \tau} = e^{\frac{\hat{H}_0\tau}{\hbar}} [\hat{H}_0, \hat{a}_p] e^{-\frac{\hat{H}_0\tau}{\hbar}} = -\hbar \omega_p e^{\frac{\hat{H}_0\tau}{\hbar}} \hat{a}_p e^{-\frac{\hat{H}_0\tau}{\hbar}} = -\hbar \omega_p \hat{a}_p(\tau). \tag{4.5}$$

The solution of Eq. (4.5) is $\hat{a}_p(\tau) = \hat{a}_p(0)e^{-\omega_p \tau} = \hat{a}_p e^{-\omega_p \tau}$. For $\hat{a}_p^\dagger(\tau)$ we have

$$[\hat{H}_0, \hat{a}_{p'}^\dagger(0)] = [\hat{H}_0, \hat{a}_{p'}^\dagger] = \left[\sum_p \hbar\omega_p \left(\hat{a}_p^\dagger \hat{a}_p + \frac{1}{2}\right), \hat{a}_{p'}^\dagger\right] = \sum_p \hbar\omega_p [\hat{a}_p^\dagger \hat{a}_p, \hat{a}_{p'}^\dagger]$$

$$= \sum_p \hbar\omega_p \left(\hat{a}_p^\dagger \hat{a}_p \hat{a}_{p'}^\dagger - \hat{a}_{p'}^\dagger \hat{a}_p^\dagger \hat{a}_p\right) = \sum_p \hbar\omega_p \left[\hat{a}_p^\dagger \left(\delta_{pp'} + \hat{a}_{p'}^\dagger \hat{a}_p\right) - \hat{a}_{p'}^\dagger \hat{a}_p^\dagger \hat{a}_p\right]$$

$$= \hbar\omega_{p'} \hat{a}_{p'}^\dagger + \sum_p \hbar\omega_p \left[\hat{a}_p^\dagger \hat{a}_{p'}^\dagger \hat{a}_p - \hat{a}_{p'}^\dagger \hat{a}_p^\dagger \hat{a}_p\right] = \hbar\omega_{p'} \hat{a}_{p'}^\dagger$$

In summary, we have

$$\begin{cases} \hat{a}_p(\tau) = \hat{a}_p(0)e^{-\omega_p \tau} = \hat{a}_p e^{-\omega_p \tau} \\ \hat{a}_p^\dagger(\tau) = \hat{a}_p^\dagger(0)e^{\omega_p \tau} = \hat{a}_p^\dagger e^{\omega_p \tau} \end{cases} \quad (4.6)$$

## 4.2 Green's function

To describe thermal transport properties between different system sections, we use the formalism of thermal (or imaginary time) phonon Green's function [2]. To this end, we define the thermal Green's function for phonons as:

$$\hat{G}_{pp'}(\tau, \tau') = -\langle \hat{T}_\tau \hat{a}_p(\tau) \hat{a}_{p'}^\dagger(\tau') \rangle = -\text{Tr}\left\{\hat{\rho}_H \hat{T}_\tau \left[\hat{a}_p(\tau) \hat{a}_{p'}^\dagger(\tau')\right]\right\}, \quad (4.7)$$

where $\hat{T}_\tau$ is the time ordering operator and $\hat{\rho}_H$ is the statistical operator for the grand canonical ensemble (note that the chemical potential for phonons is zero):

$$\hat{\rho}_H = e^{-\beta \hat{H}}/Z_H. \quad (4.8)$$

where the partition function is given by $Z_H \equiv \text{Tr}(e^{-\beta \hat{H}})$, and $\beta = 1/k_B T$, with $k_B$ being Boltzmann's constant and $T$ the temperature. For time independent Hamiltonians, $G_{pp'}(\tau, \tau')$ depends only on $\tau - \tau'$, i.e.,

$$G_{pp'}(\tau, \tau') = G_{pp'}(\tau - \tau', 0). \quad (4.9)$$

To show this we shall first assume that $\tau > \tau'$ such that

$$G_{pp'}(\tau, \tau') = -\langle \hat{T}_\tau \hat{a}_p(\tau) \hat{a}_{p'}^\dagger(\tau') \rangle = -\text{Tr}\left\{\hat{\rho}_H \hat{T}_\tau \left[\hat{a}_p(\tau) \hat{a}_{p'}^\dagger(\tau')\right]\right\} = -\text{Tr}\left\{\hat{\rho}_H \hat{a}_p(\tau) \hat{a}_{p'}^\dagger(\tau')\right\}$$

$$= -\text{Tr}\left\{\hat{\rho}_H e^{\frac{\hat{H}\tau}{\hbar}} \hat{a}_p e^{-\frac{\hat{H}\tau}{\hbar}} e^{\frac{\hat{H}\tau'}{\hbar}} \hat{a}_{p'}^\dagger e^{-\frac{\hat{H}\tau'}{\hbar}}\right\} = -\text{Tr}\left\{\hat{\rho}_H e^{-\frac{\hat{H}\tau'}{\hbar}} e^{\frac{\hat{H}\tau}{\hbar}} \hat{a}_p e^{-\frac{\hat{H}(\tau-\tau')}{\hbar}} \hat{a}_{p'}^\dagger\right\}$$

$$= -\text{Tr}\left\{\hat{\rho}_H e^{\frac{\hat{H}(\tau-\tau')}{\hbar}} \hat{a}_p e^{-\frac{\hat{H}(\tau-\tau')}{\hbar}} \hat{a}_{p'}^\dagger\right\} = -\text{Tr}\left\{\hat{\rho}_H \hat{a}_p(\tau - \tau') \hat{a}_{p'}^\dagger(0)\right\}$$

$$= -\text{Tr}\left\{\hat{\rho}_H \hat{T}_\tau \left[\hat{a}_p(\tau - \tau') \hat{a}_{p'}^\dagger(0)\right]\right\} = -\langle \hat{T}_\tau \hat{a}_p(\tau - \tau') \hat{a}_{p'}^\dagger(0) \rangle = G_{pp'}(\tau - \tau', 0)$$

Where we have used the commutativity of $\hat{\rho}_H$ and $e^{\pm \frac{\hat{H}\tau'}{\hbar}}$ and the invariance of the trace operation towards cyclic permutations. Similarly, for $\tau < \tau'$ we have:

$$G_{pp'}(\tau, \tau') = -\langle \hat{T}_\tau \hat{a}_p(\tau) \hat{a}_{p'}^\dagger(\tau') \rangle = -\text{Tr}\left\{\hat{\rho}_H \hat{T}_\tau \left[\hat{a}_p(\tau) \hat{a}_{p'}^\dagger(\tau')\right]\right\} = -\text{Tr}\left\{\hat{\rho}_H \hat{a}_{p'}^\dagger(\tau') \hat{a}_p(\tau)\right\}$$

$$= -\text{Tr}\left\{\hat{\rho}_H e^{\frac{\hat{H}\tau'}{\hbar}} \hat{a}_{p'}^\dagger e^{-\frac{\hat{H}\tau'}{\hbar}} e^{\frac{\hat{H}\tau}{\hbar}} \hat{a}_p e^{-\frac{\hat{H}\tau}{\hbar}}\right\} = -\text{Tr}\left\{\hat{\rho}_H \hat{a}_{p'}^\dagger e^{\frac{\hat{H}(\tau-\tau')}{\hbar}} \hat{a}_p e^{-\frac{\hat{H}\tau}{\hbar}} e^{\frac{\hat{H}\tau'}{\hbar}}\right\}$$

$$= -\text{Tr}\left\{\hat{\rho}_H \hat{a}_{p'}^\dagger e^{\frac{\hat{H}(\tau-\tau')}{\hbar}} \hat{a}_p e^{-\frac{\hat{H}(\tau-\tau')}{\hbar}}\right\} = -\text{Tr}\left\{\hat{\rho}_H \hat{a}_{p'}^\dagger(0) \hat{a}_p(\tau - \tau')\right\}$$

$$= -\text{Tr}\left\{\hat{\rho}_H \hat{T}_\tau \left[\hat{a}_{p'}^\dagger(0) \hat{a}_p(\tau - \tau')\right]\right\} = -\text{Tr}\left\{\hat{\rho}_H \hat{T}_\tau \left[\hat{a}_p(\tau - \tau') \hat{a}_{p'}^\dagger(0)\right]\right\}$$

$$= -\langle \hat{T}_\tau \hat{a}_p(\tau - \tau') \hat{a}_{p'}^\dagger(0) \rangle = G_{pp'}(\tau - \tau', 0)$$

For simplicity, we will introduce the notation $G_{pp'}(\tau, 0) \equiv G_{pp'}(\tau)$. We further note that when using imaginary time $G_{pp'}(\tau)$ is a periodic functions in the domain $[-\beta\hbar, \beta\hbar]$ with a period of $\beta\hbar$ (see Page 236 of Ref. [2]):

$$\begin{cases} G_{pp'}(\tau + \beta\hbar, 0) = G_{pp'}(\tau, 0), \tau < 0 \\ G_{pp'}(\tau - \beta\hbar, 0) = G_{pp'}(\tau, 0), \tau > 0 \end{cases} \quad (4.10)$$

To show this, we shall again assume first that $-\beta\hbar < \tau < 0$ to write, that is

$$G_{pp'}(\tau + \beta\hbar, 0) = -\langle \hat{T}_\tau \hat{a}_p(\tau + \beta\hbar) \hat{a}_{p'}^\dagger(0) \rangle = -\langle \hat{a}_p(\tau + \beta\hbar) \hat{a}_{p'}^\dagger(0) \rangle$$

$$= -\text{Tr}\left\{\hat{\rho}_H e^{\frac{\hat{H}(\tau+\beta\hbar)}{\hbar}} \hat{a}_p e^{-\frac{\hat{H}(\tau+\beta\hbar)}{\hbar}} \hat{a}_{p'}\right\} = -\text{Tr}\left\{\hat{\rho}_H e^{\beta\hat{H}} e^{\frac{\hat{H}\tau}{\hbar}} \hat{a}_p e^{-\frac{\hat{H}\tau}{\hbar}} e^{-\beta\hat{H}} \hat{a}_{p'}\right\}$$

$$= -\text{Tr}\{\hat{\rho}_H e^{\beta\hat{H}} \hat{a}_p(\tau) e^{-\beta\hat{H}} \hat{a}_{p'}(0)\} = -\frac{1}{Z_H}\text{Tr}\{e^{-\beta\hat{H}} e^{\beta\hat{H}} \hat{a}_p(\tau) e^{-\beta\hat{H}} \hat{a}_{p'}(0)\}$$

$$= -\frac{1}{Z_H}\text{Tr}\{\hat{a}_p(\tau) e^{-\beta\hat{H}} \hat{a}_{p'}(0)\} = -\frac{1}{Z_H}\text{Tr}\{e^{-\beta\hat{H}} \hat{a}_{p'}(0) \hat{a}_p(\tau)\}$$

$$= -\frac{1}{Z_H}\text{Tr}\{e^{-\beta\hat{H}} \hat{T}_\tau [\hat{a}_p(\tau) \hat{a}_{p'}(0)]\} = -\text{Tr}\{\hat{\rho}_H \hat{T}_\tau [\hat{a}_p(\tau) \hat{a}_{p'}]\} = G_{pp'}(\tau, 0)$$

Similarly, for $\beta\hbar > \tau > 0$ we have:

$$G_{pp'}(\tau - \beta\hbar, 0) = -\langle \hat{T}_\tau \hat{a}_p(\tau - \beta\hbar) \hat{a}_{p'}^\dagger(0) \rangle = -\langle \hat{a}_{p'}^\dagger(0) \hat{a}_p(\tau - \beta\hbar) \rangle$$

$$= -\text{Tr}\left\{\hat{\rho}_H \hat{a}_{p'} e^{\frac{\hat{H}(\tau-\beta\hbar)}{\hbar}} \hat{a}_p e^{-\frac{\hat{H}(\tau-\beta\hbar)}{\hbar}}\right\} = -\text{Tr}\left\{\hat{\rho}_H \hat{a}_{p'} e^{-\beta\hat{H}} e^{\frac{\hat{H}\tau}{\hbar}} \hat{a}_p e^{-\frac{\hat{H}\tau}{\hbar}} e^{\beta\hat{H}}\right\}$$

$$= -\text{Tr}\{\hat{\rho}_H \hat{a}_{p'}(0) e^{-\beta\hat{H}} \hat{a}_p(\tau) e^{\beta\hat{H}}\} = -\frac{1}{Z_H}\text{Tr}\{e^{-\beta\hat{H}} \hat{a}_{p'}(0) e^{-\beta\hat{H}} \hat{a}_p(\tau) e^{\beta\hat{H}}\}$$

$$= -\frac{1}{Z_H}\text{Tr}\{\hat{a}_{p'}(0) e^{-\beta\hat{H}} \hat{a}_p(\tau)\} = -\frac{1}{Z_H}\text{Tr}\{e^{-\beta\hat{H}} \hat{a}_p(\tau) \hat{a}_{p'}(0)\}$$

$$= -\frac{1}{Z_H}\text{Tr}\{e^{-\beta\hat{H}} \hat{T}_\tau [\hat{a}_p(\tau) \hat{a}_{p'}(0)]\} = -\text{Tr}\{\hat{\rho}_H \hat{T}_\tau [\hat{a}_p(\tau) \hat{a}_{p'}]\} = G_{pp'}(\tau, 0)$$

Therefore, $G_{pp'}(\tau)$ can be expanded as a Fourier series in the domain $[0, \beta\hbar]$ as follows:

$$G_{pp'}(\tau) = \frac{1}{\beta\hbar}\sum_{-\infty}^{\infty} e^{-i\omega_n \tau} G_{pp'}(i\omega_n). \tag{4.11}$$

where $\omega_n = \frac{2\pi n}{\beta\hbar}$ and the associated Fourier coefficient is given by

$$G_{pp'}(i\omega_n) = \int_0^{\beta\hbar} d\tau\, e^{i\omega_n \tau} G_{pp'}(\tau), \quad \omega_n = \frac{2n\pi}{\beta\hbar}. \tag{4.12}$$

Having proven the translational time invariance and the periodicity of the Green's functios we can now calculate it for the uncoupled system:

$$G^0_{pp'}(\tau, \tau') = -\langle \hat{T}_\tau \hat{a}_p(\tau) \hat{a}^\dagger_{p'}(\tau')\rangle_0 = \begin{cases} -\langle \hat{a}_p(0) e^{-\omega_p \tau} \hat{a}^\dagger_{p'}(0) e^{\omega_{p'} \tau'}\rangle, & \tau - \tau' > 0 \\ -\langle \hat{a}^\dagger_{p'}(0) e^{\omega_{p'} \tau'} \hat{a}_p(0) e^{-\omega_p \tau}\rangle, & \tau - \tau' < 0 \end{cases}$$

$$= \begin{cases} -e^{-\omega_p \tau} e^{\omega_{p'} \tau'} \langle \hat{a}_p(0) \hat{a}^\dagger_{p'}(0)\rangle, & \tau - \tau' > 0 \\ -e^{-\omega_p \tau} e^{\omega_{p'} \tau'} \langle \hat{a}^\dagger_{p'}(0) \hat{a}_p(0)\rangle, & \tau - \tau' < 0 \end{cases}$$

$$= \begin{cases} -e^{-\omega_p \tau} e^{\omega_{p'} \tau'} \langle \hat{a}^\dagger_{p'}(0)\hat{a}_p(0) + [\hat{a}_p(0), \hat{a}^\dagger_{p'}(0)]\rangle, & \tau - \tau' > 0 \\ -e^{-\omega_p \tau} e^{\omega_{p'} \tau'} \langle \hat{a}^\dagger_{p'}(0) \hat{a}_p(0)\rangle, & \tau - \tau' < 0 \end{cases}$$

$$= \begin{cases} -e^{-\omega_p \tau} e^{\omega_{p'} \tau'} \left[\langle \hat{a}^\dagger_{p'}(0)\hat{a}_p(0)\rangle + \delta_{pp'}\right], & \tau - \tau' > 0 \\ -e^{-\omega_p \tau} e^{\omega_{p'} \tau'} \langle \hat{a}^\dagger_{p'}(0) \hat{a}_p(0)\rangle, & \tau - \tau' < 0 \end{cases}$$

i.e.,

$$G^0_{pp'}(\tau, \tau') = \begin{cases} -e^{-\omega_p \tau} e^{\omega_{p'} \tau'} \left[\langle \hat{a}^\dagger_{p'}(0)\hat{a}_p(0)\rangle + \delta_{pp'}\right], & \tau - \tau' > 0 \\ -e^{-\omega_p \tau} e^{\omega_{p'} \tau'} \langle \hat{a}^\dagger_{p'}(0) \hat{a}_p(0)\rangle, & \tau - \tau' < 0 \end{cases}, \tag{4.13}$$

where we have used Eq. (4.6) for $\hat{a}_p(\tau), \hat{a}^\dagger_{p'}(\tau')$ and Commutation relation $[\hat{a}_p(0), \hat{a}^\dagger_{p'}(0)] = \delta_{pp'}$.

To calculate $\langle \hat{a}^\dagger_{p'}(0)\hat{a}_p(0)\rangle$, we first prove following equation:

$$e^{\hat{A}}\hat{B}e^{-\hat{A}} = \sum_{n=0}^{\infty} \frac{1}{n!}[\hat{A}^{(n)}, \hat{B}], \tag{4.14}$$

where $[\hat{A}^{(k)}, \hat{B}] \equiv [\hat{A}, [\hat{A}^{(k-1)}, \hat{B}]]$. To prove this identity, we define the operator $\hat{f}(t) = e^{t\hat{A}}\hat{B}e^{-t\hat{A}}$ and taylor expand it around $t = 0$:

$$\hat{f}(t) = \hat{f}(0) + t\hat{f}'(0) + \frac{t^2}{2}\hat{f}''(0) + \cdots = \sum_{n=0}^{\infty} \frac{t^n}{n!}\frac{d^n \hat{f}}{dt^n}\bigg|_{t=0}, \tag{4.15}$$

The corresponding derivatives are given by:

$$\begin{cases} \hat{f}'(t) = e^{t\hat{A}}\hat{A}\hat{B}e^{-t\hat{A}} - e^{t\hat{A}}\hat{B}\hat{A}e^{-t\hat{A}} = e^{t\hat{A}}[\hat{A}, \hat{B}]e^{-t\hat{A}} \\ \hat{f}''(t) = e^{t\hat{A}}(\hat{A}[\hat{A}, \hat{B}] - [\hat{A}, \hat{B}]\hat{A})e^{-t\hat{A}} = e^{t\hat{A}}[\hat{A}^{(2)}, \hat{B}]e^{-t\hat{A}} \\ \vdots \\ \hat{f}^{(n)}(t) = e^{t\hat{A}}[\hat{A}^{(n)}, \hat{B}]e^{-t\hat{A}} \end{cases}, \tag{4.16}$$

Substituting Eq. (4.16) into Eq. (4.15), we have:

$$e^{t\hat{A}}\hat{B}e^{-t\hat{A}} = \sum_{n=0}^{\infty} \frac{t^n}{n!}[\hat{A}^{(n)}, \hat{B}]. \tag{4.17}$$

It's clear that Eq. (4.13) is a spectial case of Eq. (4.16) with $t = 1$. With this we can proceed as follows:

$$\langle \hat{a}_{p'}^{\dagger}(0)\hat{a}_{p}(0)\rangle = \frac{1}{Z_{H_0}}\text{Tr}\{e^{-\beta H_0}\hat{a}_{p'}^{\dagger}(0)\hat{a}_{p}(0)\} = \frac{1}{Z_{H_0}}\text{Tr}\{e^{-\beta H_0}\hat{a}_{p'}^{\dagger}(0)e^{\beta H_0}e^{-\beta H_0}\hat{a}_{p}(0)\}$$

$$= \frac{1}{Z_{H_0}}\text{Tr}\left\{\sum_{n=0}^{\infty}\frac{(-\beta)^n}{n!}\left[\hat{H}_0^{(n)}, \hat{a}_{p'}^{\dagger}(0)\right]e^{-\beta H_0}\hat{a}_{p}(0)\right\}$$

Note that $[\hat{H}_0, \hat{a}_{p'}^{\dagger}(0)] = \hbar\omega_{p'}\hat{a}_{p'}^{\dagger}(0)$, we have

$$\left[\hat{H}_0^{(n)}, \hat{a}_{p'}^{\dagger}(0)\right] = (\hbar\omega_{p'})^n \hat{a}_{p'}^{\dagger}(0), \tag{4.18}$$

such that:

$$\langle \hat{a}_{p'}^{\dagger}(0)\hat{a}_{p}(0)\rangle = \frac{1}{Z_{H_0}}\text{Tr}\left\{\sum_{n=0}^{\infty}\frac{(-\beta)^n}{n!}\left[\hat{H}_0^{(n)}, \hat{a}_{p'}^{\dagger}(0)\right]e^{-\beta H_0}\hat{a}_{p}(0)\right\}$$

$$= \frac{1}{Z_{H_0}}\text{Tr}\left\{\sum_{n=0}^{\infty}\frac{(-\beta\hbar\omega_{p'})^n}{n!}\hat{a}_{p'}^{\dagger}(0)\,e^{-\beta H_0}\hat{a}_{p}(0)\right\}$$

$$= e^{-\beta\hbar\omega_{p'}}\frac{1}{Z_{H_0}}\text{Tr}\{e^{-\beta H_0}\hat{a}_{p}(0)\hat{a}_{p'}^{\dagger}(0)\} = e^{-\beta\hbar\omega_{p'}}\langle \hat{a}_{p}(0)\hat{a}_{p'}^{\dagger}(0)\rangle$$

$$= e^{-\beta\hbar\omega_{p'}}\left(\langle \hat{a}_{p'}^{\dagger}(0)\hat{a}_{p}(0)\rangle + \delta_{pp'}\right)$$

therefore, we obtain $\langle \hat{a}_{p'}^{\dagger}(0)\hat{a}_{p}(0)\rangle\left(e^{\beta\hbar\omega_{p'}} - 1\right) = \delta_{pp'}$. For $\boldsymbol{p}' \neq \boldsymbol{p}$ and general $e^{\beta\hbar\omega_{p'}}$ we have $\langle \hat{a}_{p'}^{\dagger}(0)\hat{a}_{p}(0)\rangle = 0$. For $\boldsymbol{p}' = \boldsymbol{p}$ we obtain $\langle \hat{a}_{p}^{\dagger}(0)\hat{a}_{p}(0)\rangle = \left[e^{\beta\hbar\omega_{p}} - 1\right]^{-1}$. Hence, we have:

$$\langle \hat{a}_{p'}^{\dagger}(0)\hat{a}_{p}(0)\rangle = \delta_{p'p}\frac{1}{e^{\beta\hbar\omega_{p}}-1} \equiv \delta_{p'p}n_B(\omega_{p}), \tag{4.19}$$

where $n_B(\omega_{p})$ is the Bose-Einstein distribution for phonons. Substituting Eq. (4.18) in Eq. (4.13) we obtain:

$$G_{pp'}^{0}(\tau, \tau') = \begin{cases} -\delta_{pp'}[1 + n_B(\omega_{p})]e^{-\omega_{p}(\tau-\tau')}, & \tau - \tau' > 0 \\ -\delta_{pp'}n_B(\omega_{p})e^{-\omega_{p}(\tau-\tau')}, & \tau - \tau' < 0 \end{cases}. \tag{4.20}$$

Note that only those Green's functions of the form $G_{pp'}^{0}(\tau, \tau') = G_{p}^{0}(\tau - \tau')\delta_{pp'}$ are non-zero due to the orthogonality of normal modes. Looking at the non-vanishing terms and setting $\tau' = 0$, without loss of generality, we can calculate the Fourier transform of $G_p^0(\tau)$ from Eq. (4.12):

$$G_p^0(i\omega_n) = \int_0^{\beta\hbar} d\tau e^{i\omega_n\tau}G_p^0(\tau) = -\int_0^{\beta\hbar}d\tau e^{i\omega_n\tau}\left[1 + n_B(\omega_p)\right]e^{-\omega_p\tau} = -\frac{1+n_B(\omega_p)}{i\omega_n-\omega_p}\left[e^{(i\omega_n-\omega_p)\beta\hbar} - \right.$$

$$1] = -\frac{1+\left(\frac{1}{e^{\beta\hbar\omega_p}-1}\right)}{i\omega_n-\omega_p}\left(e^{i\frac{2\pi n}{\beta\hbar}\beta\hbar}e^{-\beta\hbar\omega_p}-1\right) = -\frac{e^{\beta\hbar\omega_p}\left(e^{-\beta\hbar\omega_p}-1\right)}{i\omega_n-\omega_p\ e^{\beta\hbar\omega_p}-1} = -\frac{1}{i\omega_n-\omega_p}\frac{1-e^{\beta\hbar\omega_p}}{e^{\beta\hbar\omega_p}-1} = \frac{1}{i\omega_n-\omega_p},$$

i.e.,

$$G_p^0(i\omega_n) == \frac{1}{i\omega_n-\omega_p}. \tag{4.21}$$

# 5 Fermi's golden rule

## 5.1 Interaction picture

Next, we can proceed with calculating the Green's function of the coupled system. To this end, we define the coupling Hamiltonian operator term in the interaction picture as:

$$\widehat{H}_C^I(\tau) = e^{\frac{\widehat{H}_0\tau}{\hbar}}\widehat{H}_C e^{-\frac{\widehat{H}_0\tau}{\hbar}}. \tag{5.1}$$

The time evolution of $\widehat{H}_C^I(\tau)$ is then given by:

$$\hbar\frac{\partial\widehat{H}_C^I(\tau)}{\partial\tau} = [\widehat{H}_0,\widehat{H}_C^I(\tau)]. \tag{5.2}$$

Note that the Green's function defined above was given in the Heisenberg picture. To proceed, we need to transform it to the interaction picture:

$$G_{pp'}(\tau,0) = -\langle\widehat{T}_\tau\widehat{a}_p(\tau)\widehat{a}_{p'}^\dagger(0)\rangle = -\mathrm{Tr}\left\{\widehat{\rho}_H\widehat{T}_\tau\left[e^{\frac{\widehat{H}\tau}{\hbar}}\widehat{a}_p(0)e^{-\frac{\widehat{H}\tau}{\hbar}}\widehat{a}_{p'}^\dagger(0)\right]\right\}$$

$$= -\mathrm{Tr}\left\{\widehat{\rho}_H\widehat{T}_\tau\left[e^{\frac{\widehat{H}\tau}{\hbar}}e^{-\frac{\widehat{H}_0\tau}{\hbar}}\left(e^{\frac{\widehat{H}_0\tau}{\hbar}}\widehat{a}_p(0)e^{-\frac{\widehat{H}_0\tau}{\hbar}}\right)e^{\frac{\widehat{H}_0\tau}{\hbar}}e^{-\frac{\widehat{H}\tau}{\hbar}}\widehat{a}_{p'}^\dagger(0)\right]\right\}$$

$$= -\mathrm{Tr}\left\{\widehat{\rho}_H\widehat{T}_\tau\left[\widehat{U}(0,\tau)\widehat{a}_p^I(\tau)\widehat{U}(\tau,0)\widehat{a}_{p'}^\dagger(0)\right]\right\}$$

where,

$$\widehat{a}_p^I(\tau) \equiv e^{\frac{\widehat{H}_0\tau}{\hbar}}\widehat{a}_p(0)e^{-\frac{\widehat{H}_0\tau}{\hbar}}, \tag{5.3}$$

is the operator in interaction picture and the operator $\widehat{U}$ is defined by:

$$\widehat{U}(\tau_1,\tau_2) \equiv e^{\frac{\widehat{H}_0\tau_1}{\hbar}}e^{-\frac{\widehat{H}(\tau_1-\tau_2)}{\hbar}}e^{-\frac{\widehat{H}_0\tau_2}{\hbar}}, \tag{5.4}$$

Note that while $\widehat{U}$ is not unitary, it satisfies the following group property:

$$\widehat{U}(\tau_1,\tau_2)\widehat{U}(\tau_2,\tau_3) = \widehat{U}(\tau_1,\tau_3), \tag{5.5}$$

and the boundary condition $\widehat{U}(\tau_1,\tau_1) = \widehat{1}$. In addition, the $\tau$ derivative of $\widehat{U}$ is simply:

$$\hbar\frac{\partial\widehat{U}(\tau,\tau')}{\partial\tau} = \left\{\widehat{H}_0 e^{\frac{\widehat{H}_0\tau}{\hbar}}e^{-\frac{\widehat{H}(\tau-\tau')}{\hbar}} - e^{\frac{\widehat{H}_0\tau}{\hbar}}\widehat{H}e^{-\frac{\widehat{H}(\tau-\tau')}{\hbar}}\right\}e^{-\frac{\widehat{H}_0\tau'}{\hbar}}$$

$$= e^{\frac{\widehat{H}_0\tau}{\hbar}}(\widehat{H}_0-\widehat{H})e^{-\frac{\widehat{H}_0\tau}{\hbar}}e^{\frac{\widehat{H}_0\tau}{\hbar}}e^{-\frac{\widehat{H}(\tau-\tau')}{\hbar}}e^{-\frac{\widehat{H}_0\tau'}{\hbar}} = e^{\frac{\widehat{H}_0\tau}{\hbar}}(-\widehat{H}_C)e^{-\frac{\widehat{H}_0\tau}{\hbar}}\widehat{U}(\tau,\tau')$$

$$= -\widehat{H}_C^I(\tau)\widehat{U}(\tau,\tau')$$

i.e.,

$$\hbar \frac{\partial \hat{U}(\tau,\tau')}{\partial \tau} = -\hat{H}_C^I(\tau)\hat{U}(\tau,\tau'), \tag{5.6}$$

The solution of Eq. (5.6) is (see page 235 of Ref. [2]):

$$\hat{U}(\tau,\tau') = \sum_{n=0}^{\infty} \frac{\left(-\frac{1}{\hbar}\right)^n}{n!} \int_{\tau'}^{\tau} d\tau_1 \cdots \int_{\tau'}^{\tau} d\tau_n \hat{T}_\tau[\hat{H}_C^I(\tau_1) \cdots \hat{H}_C^I(\tau_n)], \tag{5.7}$$

The exact thermal Green's function now may be rewritten in the interaction picture as:

$$\hat{G}_{pp'}(\tau,0) = -\frac{1}{Z_H} \text{Tr}\left\{e^{-\beta\hat{H}} \hat{T}_\tau[\hat{U}(0,\tau)\hat{a}_p^I(\tau)\hat{U}(\tau,0)]\hat{a}_{p'}^\dagger(0)\right\}$$

$$= -\frac{\text{Tr}\left\{e^{-\beta\hat{H}} \hat{T}_\tau\left[\hat{U}(0,\tau)\hat{a}_p^I(\tau)\hat{U}(\tau,0)\hat{a}_{p'}^\dagger(0)\right]\right\}}{\text{Tr}\{e^{-\beta\hat{H}}\}}$$

$$= -\frac{\text{Tr}\left\{e^{-\beta\hat{H}_0}\hat{U}(\beta\hbar,0)\hat{T}_\tau\left[\hat{a}_p^I(\tau)\hat{U}(0,\tau)\hat{U}(\tau,0)\hat{a}_{p'}^\dagger(0)\right]\right\}}{\text{Tr}\{e^{-\beta\hat{H}_0}\hat{U}(\beta\hbar,0)\}}$$

$$= -\frac{\text{Tr}\left\{e^{-\beta\hat{H}_0}\hat{U}(\beta\hbar,0)\hat{T}_\tau\left[\hat{a}_p^I(\tau)\hat{U}(0,0)\hat{a}_{p'}^\dagger(0)\right]\right\}}{\text{Tr}\{e^{-\beta\hat{H}_0}\hat{U}(\beta\hbar,0)\}} =$$

$$= -\frac{\text{Tr}\left\{e^{-\beta\hat{H}_0}\hat{T}_\tau\left[\hat{U}(\beta\hbar,0)\hat{a}_p^I(\tau)\hat{a}_{p'}^\dagger(0)\right]\right\}}{\text{Tr}\{e^{-\beta\hat{H}_0}\hat{U}(\beta\hbar,0)\}}$$

Where we used the fact that we are free to change the order of the operators within the time ordering operation (see pages 241-242 of Ref. [2]).

## 5.2 Wick's theorem

Thus the Green's function can be expanded as [2]

$$G_{pp'}(\tau,0) = -\frac{\sum_{m=0}^{\infty}\frac{1}{m!}\left(-\frac{1}{\hbar}\right)^m \int_0^{\beta\hbar} d\tau_1 \cdots \int_0^{\beta\hbar} d\tau_m \langle T_\tau \hat{H}_C^I(\tau_1) \cdots \hat{H}_C^I(\tau_m)\hat{a}_p^I(\tau)\hat{a}_{p'}^\dagger(0)\rangle_0}{\sum_{m=0}^{\infty}\frac{1}{m!}\left(-\frac{1}{\hbar}\right)^m \int_0^{\beta\hbar} d\tau_1 \cdots \int_0^{\beta\hbar} d\tau_m \langle T_\tau \hat{H}_C^I(\tau_1) \cdots \hat{H}_C^I(\tau_m)\rangle_0}, \tag{5.8}$$

where $\langle \cdots \rangle_0$ represents the ensemble average with respect to the non-interacting basis $\text{Tr}\{e^{-\beta\hat{H}_0}(\cdots)\}$. Or explicitly,

$$G_{pp'}(\tau,0) = -\frac{G_{pp'}^0(\tau) - \frac{1}{\hbar}\int_0^{\beta\hbar} d\tau_1 \langle T_\tau \hat{H}_C^I(\tau_1)\hat{a}_p(\tau)\hat{a}_{p'}^\dagger(0)\rangle_0 + \frac{1}{2\hbar^2}\int_0^{\beta\hbar} d\tau_1 \int_0^{\beta\hbar} d\tau_2 \langle T_\tau \hat{H}_C^I(\tau_1)\hat{H}_C^I(\tau_2)\hat{a}_p(\tau)\hat{a}_{p'}^\dagger(0)\rangle_0 + \cdots}{1 - \frac{1}{\hbar}\int_0^{\beta\hbar} d\tau_1 \langle T_\tau \hat{H}_C^I(\tau_1)\rangle_0 + \frac{1}{2\hbar^2}\int_0^{\beta\hbar} d\tau_1 \int_0^{\beta\hbar} d\tau_2 \langle T_\tau \hat{H}_C^I(\tau_1)H_C^I(\tau_2)\rangle_0 + \cdots}, \tag{5.9}$$

For consiceness we introduce the following notation:

$$D_m \equiv \frac{1}{m!}\left(-\frac{1}{\hbar}\right)^m \int_0^{\beta\hbar} d\tau_1 \cdots \int_0^{\beta\hbar} d\tau_m \langle T_\tau \hat{H}_C^I(\tau_1) \cdots \hat{H}_C^I(\tau_m)\rangle_0, \tag{5.10}$$

To simplify the calculation in Eq. (5.8), we adopt Wick's theorem (see pages 237-241 of Ref. [2]),which can be expressed as follows:

$$\langle T_\tau[\hat{A}\hat{B}\hat{C}\hat{D} \cdots \hat{Y}\hat{Z}]\rangle_0 = \langle T_\tau[\hat{A}\hat{B}]\rangle_0 \langle T_\tau[\hat{C}\hat{D}]\rangle_0 \cdots \langle T_\tau[\hat{Y}\hat{Z}]\rangle_0 + \langle T_\tau[\hat{A}\hat{C}]\rangle_0 \langle T_\tau[\hat{B}\hat{D}]\rangle_0 \cdots \langle T_\tau[\hat{X}\hat{Z}]\rangle_0 + \cdots, \tag{5.11}$$

Here $\hat{A}, \hat{B}, \ldots, \hat{Y}, \hat{Z}$ represent $\hat{a}_p(\tau)$ or $\hat{a}_p^\dagger(\tau)$. In Eq. (5.11), each term corresponds to a particular pairing of the operators $\hat{A}\hat{B}\hat{C}\hat{D}\cdots\hat{Y}\hat{Z}$ and all possible pairings are taken into account. Here, the only non-vanishing propagators will have the form [see Eq. (4.20)]:

$$\langle T_\tau[\hat{a}_{p'}^\dagger(\tau')\hat{a}_p(\tau)]\rangle_0 = -\langle T_\tau[\hat{a}_p(\tau)\hat{a}_{p'}^\dagger(\tau')]\rangle_0 = G^0_{pp'}(\tau,\tau') = G^0_p(\tau-\tau')\delta_{pp'}. \quad (5.12)$$

Therefore, the second term in the numerator of Eq. (5.9) can be calculated as:

$$I_2 = -\frac{1}{\hbar}\int_0^{\beta\hbar} d\tau_1 \langle T_\tau \hat{H}_C^I(\tau_1)\hat{a}_p(\tau)\hat{a}_{p'}^\dagger(0)\rangle_0$$

$$= -\frac{1}{\hbar}\int_0^{\beta\hbar} d\tau_1 \langle T_\tau \hat{H}_C^I(\tau_1)\rangle_0 \langle T_\tau \hat{a}_{p'}^\dagger(0)\hat{a}_p(\tau)\rangle_0 - \frac{1}{\hbar}\int_0^{\beta\hbar} d\tau_1 \langle T_\tau \hat{H}_C^I(\tau_1)\hat{a}_p(\tau)\hat{a}_{p'}^\dagger(0)\rangle_{0,c}$$

$$= G^0_{pp'}(\tau,0)\left[-\frac{1}{\hbar}\int_0^{\beta\hbar} d\tau_1 \langle T_\tau \hat{H}_C^I(\tau_1)\rangle_0\right] - \frac{1}{\hbar}\int_0^{\beta\hbar} d\tau_1 \langle T_\tau \hat{H}_C^I(\tau_1)\hat{a}_p(\tau)\hat{a}_{p'}^\dagger(0)\rangle_{0,c}$$

The first term in above equation is called disconnected part since the pairing is performed separately on $\hat{H}_C(\tau_1)$ and $\hat{a}_p(\tau)\hat{a}_{p'}^\dagger(0)$. All other terms have pairs that mix creation and annihilation operators of the Hamiltonian with $\hat{a}_p(\tau)$ or $\hat{a}_{p'}^\dagger(0)$ and are said to have connected party. For simplicity we include all these terms in the notation $\langle T_\tau \hat{H}_C(\tau_1)\hat{a}_p(\tau)\hat{a}_{p'}^\dagger(0)\rangle_{0,c}$. Furthermore, we define $G^{(1)}_{pp'}(\tau) \equiv -\frac{1}{\hbar}\int_0^{\beta\hbar} d\tau_1 \langle T_\tau \hat{H}_C(\tau_1)\hat{a}_p(\tau)\hat{a}_{p'}^\dagger(0)\rangle_{0,c}$. Then we have,

$$I_3 = -\frac{1}{\hbar}\int_0^{\beta\hbar} d\tau_1 \langle T_\tau \hat{H}_C^I(\tau_1)\hat{a}_p(\tau)\hat{a}_{p'}^\dagger(0)\rangle_0 = G^0_{pp'}(\tau,0)D_1 + G^{(1)}_{pp'}(\tau), \quad (5.13)$$

Using this method, the third term in the numerator of Eq. (5.9) can be calculated as:

$I_3 = \frac{1}{2\hbar^2}\int_0^{\beta\hbar} d\tau_1 \int_0^{\beta\hbar} d\tau_2 \langle T_\tau \hat{H}_C^I(\tau_1)\hat{H}_C^I(\tau_2)\hat{a}_p(\tau)\hat{a}_{p'}^\dagger(0)\rangle_0 = \frac{1}{2\hbar^2}\int_0^{\beta\hbar} d\tau_1 \int_0^{\beta\hbar} d\tau_2 \langle T_\tau \hat{H}_C^I(\tau_1)\hat{H}_C^I(\tau_2)\rangle_0 \langle T_\tau \hat{a}_{p'}^\dagger(0)\hat{a}_p(\tau)\rangle_0 +$

$\frac{1}{2\hbar^2}\int_0^{\beta\hbar} d\tau_1 \int_0^{\beta\hbar} d\tau_2 \langle T_\tau \hat{H}_C^I(\tau_2)\rangle_0 \langle T_\tau \hat{H}_C^I(\tau_1)\hat{a}_p(\tau)\hat{a}_{p'}^\dagger(0)\rangle_{0,c} + \frac{1}{2\hbar^2}\int_0^{\beta\hbar} d\tau_1 \int_0^{\beta\hbar} d\tau_2 \langle T_\tau \hat{H}_C^I(\tau_1)\rangle_0 \langle T_\tau \hat{H}_C^I(\tau_2)\hat{a}_p(\tau)\hat{a}_{p'}^\dagger(0)\rangle_{0,c} +$

$\frac{1}{2\hbar^2}\int_0^{\beta\hbar} d\tau_1 \int_0^{\beta\hbar} d\tau_2 \langle T_\tau \hat{H}_C^I(\tau_1)\hat{H}_C^I(\tau_2)\hat{a}_p(\tau)\hat{a}_{p'}^\dagger(0)\rangle_{0,c} = G^0_{pp'}(\tau,0)\left[\frac{1}{2\hbar^2}\int_0^{\beta\hbar} d\tau_1 \int_0^{\beta\hbar} d\tau_2 \langle T_\tau \hat{H}_C^I(\tau_1)\hat{H}_C^I(\tau_2)\rangle_0\right] +$

$\frac{1}{2\hbar^2}\int_0^{\beta\hbar} d\tau_2 \langle T_\tau \hat{H}_C^I(\tau_2)\rangle_0 \int_0^{\beta\hbar} d\tau_1 \langle T_\tau \hat{H}_C^I(\tau_1)\hat{a}_p(\tau)\hat{a}_{p'}^\dagger(0)\rangle_{0,c} + \frac{1}{2\hbar^2}\int_0^{\beta\hbar} d\tau_1 \langle T_\tau \hat{H}_C^I(\tau_1)\rangle_0 \int_0^{\beta\hbar} d\tau_2 \langle T_\tau \hat{H}_C^I(\tau_2)\hat{a}_p(\tau)\hat{a}_{p'}^\dagger(0)\rangle_{0,c} +$

$\frac{1}{2\hbar^2}\int_0^{\beta\hbar} d\tau_1 \int_0^{\beta\hbar} d\tau_2 \langle T_\tau \hat{H}_C^I(\tau_1)\hat{H}_C^I(\tau_2)\hat{a}_p(\tau)\hat{a}_{p'}^\dagger(0)\rangle_{0,c}$.

Here, the last term is denoted as $G^{(2)}_{pp'}(\tau)$.

Using Eq. (5.13), the second and third terms on the right hand above equation can be simplified as follows:

$$\frac{1}{2\hbar^2}\int_0^{\beta\hbar} d\tau_2 \langle T_\tau \hat{H}_C^I(\tau_2)\rangle_0 \int_0^{\beta\hbar} d\tau_1 \langle T_\tau \hat{H}_C^I(\tau_1)\hat{a}_{\boldsymbol{p}}(\tau)\hat{a}_{\boldsymbol{p}'}^\dagger(0)\rangle_{0,c} + \frac{1}{2\hbar^2}\int_0^{\beta\hbar} d\tau_1 \langle T_\tau \hat{H}_C^I(\tau_1)\rangle_0 \int_0^{\beta\hbar} d\tau_2 \langle T_\tau \hat{H}_C^I(\tau_2)\hat{a}_{\boldsymbol{p}}(\tau)\hat{a}_{\boldsymbol{p}'}^\dagger(0)\rangle_{0,c}$$

$$= \frac{1}{2}\left[\frac{-1}{\hbar}\int_0^{\beta\hbar} d\tau_2 \langle T_\tau \hat{H}_C^I(\tau_2)\rangle_0\right]\left[\frac{-1}{\hbar}\int_0^{\beta\hbar} d\tau_1 \langle T_\tau \hat{H}_C^I(\tau_1)\hat{a}_{\boldsymbol{p}}(\tau)\hat{a}_{\boldsymbol{p}'}^\dagger(0)\rangle_{0,c}\right]$$

$$+ \frac{1}{2}\left[\frac{-1}{\hbar}\int_0^{\beta\hbar} d\tau_1 \langle T_\tau \hat{H}_C^I(\tau_1)\rangle_0\right]\left[\frac{-1}{\hbar}\int_0^{\beta\hbar} d\tau_2 \langle T_\tau \hat{H}_C^I(\tau_2)\hat{a}_{\boldsymbol{p}}(\tau)\hat{a}_{\boldsymbol{p}'}^\dagger(0)\rangle_{0,c}\right]$$

$$= \frac{1}{2}\left[\frac{-1}{\hbar}\int_0^{\beta\hbar} d\tau_2 \langle T_\tau \hat{H}_C^I(\tau_2)\rangle_0\right]\left[\frac{-1}{\hbar}\int_0^{\beta\hbar} d\tau_1 \langle T_\tau \hat{H}_C^I(\tau_1)\hat{a}_{\boldsymbol{p}}(\tau)\hat{a}_{\boldsymbol{p}'}^\dagger(0)\rangle_{0,c}\right]$$

$$+ \frac{1}{2}\left[\frac{-1}{\hbar}\int_0^{\beta\hbar} d\tau_2 \langle T_\tau \hat{H}_C^I(\tau_2)\rangle_0\right]\left[\frac{-1}{\hbar}\int_0^{\beta\hbar} d\tau_1 \langle T_\tau \hat{H}_C^I(\tau_1)\hat{a}_{\boldsymbol{p}}(\tau)\hat{a}_{\boldsymbol{p}'}^\dagger(0)\rangle_{0,c}\right]$$

$$= \left[\frac{-1}{\hbar}\int_0^{\beta\hbar} d\tau_2 \langle T_\tau \hat{H}_C^I(\tau_2)\rangle_0\right]\left[\frac{-1}{\hbar}\int_0^{\beta\hbar} d\tau_1 \langle T_\tau \hat{H}_C^I(\tau_1)\hat{a}_{\boldsymbol{p}}(\tau)\hat{a}_{\boldsymbol{p}'}^\dagger(0)\rangle_{0,c}\right] = D_1 G_{\boldsymbol{p}\boldsymbol{p}'}^{(1)}(\tau)$$

Where, we performed the following integration variables interchange $\tau_1 \leftrightarrow \tau_2$. Thus we have

$$I_3 = G_{\boldsymbol{p}\boldsymbol{p}'}^{(2)}(\tau) + G_{\boldsymbol{p}\boldsymbol{p}'}^{(1)}(\tau)D_1 + G_{\boldsymbol{p}\boldsymbol{p}'}^0(\tau,0)D_2, \tag{5.14}$$

Similarity, we can calculate the fourth term of Eq. (5.9) as:

$$I_4 = \frac{-1}{6\hbar^3}\int_0^{\beta\hbar} d\tau_1 \int_0^{\beta\hbar} d\tau_2 \int_0^{\beta\hbar} d\tau_3 \langle T_\tau \hat{H}_C^I(\tau_1)\hat{H}_C^I(\tau_2)\hat{H}_C^I(\tau_3)\hat{a}_{\boldsymbol{p}}(\tau)\hat{a}_{\boldsymbol{p}'}^\dagger(0)\rangle_0 =$$

$$\frac{-1}{6\hbar^3}\int_0^{\beta\hbar} d\tau_1 \int_0^{\beta\hbar} d\tau_2 \int_0^{\beta\hbar} d\tau_3 \langle T_\tau \hat{H}_C^I(\tau_1)\hat{H}_C^I(\tau_2)\hat{H}_C^I(\tau_3)\rangle_0 \langle T_\tau \hat{a}_{\boldsymbol{p}}(\tau)\hat{a}_{\boldsymbol{p}'}^\dagger(0)\rangle_0 + 3 \times$$

$$\frac{-1}{6\hbar^3}\int_0^{\beta\hbar} d\tau_1 \int_0^{\beta\hbar} d\tau_2 \int_0^{\beta\hbar} d\tau_3 \langle T_\tau \hat{H}_C^I(\tau_2)\hat{H}_C^I(\tau_3)\rangle_0 \langle T_\tau \hat{H}_C^I(\tau_1)\hat{a}_{\boldsymbol{p}}(\tau)\hat{a}_{\boldsymbol{p}'}^\dagger(0)\rangle_{0,c} + 3 \times$$

$$\frac{-1}{6\hbar^3}\int_0^{\beta\hbar} d\tau_1 \int_0^{\beta\hbar} d\tau_2 \int_0^{\beta\hbar} d\tau_3 \langle T_\tau \hat{H}_C^I(\tau_1)\rangle_0 \langle T_\tau \hat{H}_C^I(\tau_2)\hat{H}_C^I(\tau_3)\hat{a}_{\boldsymbol{p}}(\tau)\hat{a}_{\boldsymbol{p}'}^\dagger(0)\rangle_{0,c} +$$

$$\frac{-1}{6\hbar^3}\int_0^{\beta\hbar} d\tau_1 \int_0^{\beta\hbar} d\tau_2 \int_0^{\beta\hbar} d\tau_3 \langle T_\tau \hat{H}_C^I(\tau_1)\hat{H}_C^I(\tau_2)\hat{H}_C^I(\tau_3)\hat{a}_{\boldsymbol{p}}(\tau)\hat{a}_{\boldsymbol{p}'}^\dagger(0)\rangle_{0,c} =$$

$$G_{\boldsymbol{p}\boldsymbol{p}'}^0(\tau,0)\left[\frac{-1}{6\hbar^3}\int_0^{\beta\hbar} d\tau_1 \int_0^{\beta\hbar} d\tau_2 \int_0^{\beta\hbar} d\tau_3 \langle T_\tau \hat{H}_C^I(\tau_1)\hat{H}_C^I(\tau_2)\hat{H}_C^I(\tau_3)\rangle_0\right] +$$

$$\frac{1}{2\hbar^2}\int_0^{\beta\hbar} d\tau_2 \int_0^{\beta\hbar} d\tau_3 \langle T_\tau \hat{H}_C^I(\tau_2)\hat{H}_C^I(\tau_3)\rangle_0 \left[\frac{-1}{\hbar}\int_0^{\beta\hbar} d\tau_1 \langle T_\tau \hat{H}_C^I(\tau_1)\hat{a}_{\boldsymbol{p}}(\tau)\hat{a}_{\boldsymbol{p}'}^\dagger(0)\rangle_{0,c}\right] +$$

$$\frac{1}{2\hbar^2}\int_0^{\beta\hbar} d\tau_2 \int_0^{\beta\hbar} d\tau_3 \langle T_\tau \hat{H}_C^I(\tau_2)\hat{H}_C^I(\tau_3)\hat{a}_{\boldsymbol{p}}(\tau)\hat{a}_{\boldsymbol{p}'}^\dagger(0)\rangle_{0,c}\left[\frac{-1}{\hbar}\int_0^{\beta\hbar} d\tau_1 \langle T_\tau \hat{H}_C^I(\tau_1)\rangle_0\right] + G_{\boldsymbol{p}\boldsymbol{p}'}^{(3)}(\tau) = G_{\boldsymbol{p}\boldsymbol{p}'}^{(3)}(\tau) +$$

$$G_{\boldsymbol{p}\boldsymbol{p}'}^{(2)}(\tau)D_1 + G_{\boldsymbol{p}\boldsymbol{p}'}^{(1)}(\tau)D_2 + G_{\boldsymbol{p}\boldsymbol{p}'}^0(\tau,0)D_3.$$

Higher order terms can be treated in the same manner (see pages 95-96 of Ref. [2]). Then when substituting $I_2, I_3, I_4$ and all higher order terms into Eq. (5.9), we can simplify the numerator as follows:

$$G_{\boldsymbol{p}\boldsymbol{p}'}^0(\tau) - \frac{1}{\hbar}\int_0^{\beta\hbar} d\tau_1 \langle T_\tau \hat{H}_C^I(\tau_1)\hat{a}_{\boldsymbol{p}}(\tau)\hat{a}_{\boldsymbol{p}'}^\dagger(0)\rangle_0$$

$$+ \frac{1}{2\hbar^2}\int_0^{\beta\hbar} d\tau_1 \int_0^{\beta\hbar} d\tau_2 \langle T_\tau \hat{H}_C^I(\tau_1)\hat{H}_C^I(\tau_2)\hat{a}_{\boldsymbol{p}}(\tau)\hat{a}_{\boldsymbol{p}'}^\dagger(0)\rangle_0 + \cdots$$

$$= \left[G_{\boldsymbol{p}\boldsymbol{p}'}^0(\tau) + G_{\boldsymbol{p}\boldsymbol{p}'}^{(1)}(\tau) + G_{\boldsymbol{p}\boldsymbol{p}'}^{(2)}(\tau) + \cdots\right](1 + D_1 + D_2 + \cdots)$$

Noting that the expression $(1 + D_1 + D_2 + \cdots)$ is exactly canceled with the denominator, the Green's function can be simplified as:

$$G_{pp'}(\tau) = G_{pp'}^0(\tau) + G_{pp'}^{(1)}(\tau) + G_{pp'}^{(2)}(\tau) + \cdots, \tag{5.15}$$

where,

$$\begin{cases} G_{pp'}^{(1)}(\tau) = -\frac{1}{\hbar}\int_0^{\beta\hbar} d\tau_1 \langle T_\tau \hat{H}_C^I(\tau_1)\hat{a}_{q\lambda}(\tau)\hat{a}_{q'\lambda'}^\dagger(0)\rangle_{0,c} \\ G_{pp'}^{(2)}(\tau) = \frac{1}{2\hbar^2}\int_0^{\beta\hbar} d\tau_1 \int_0^{\beta\hbar} d\tau_2 \langle T_\tau \hat{H}_C^I(\tau_1)\hat{H}_C^I(\tau_2)\hat{a}_{q\lambda}(\tau)\hat{a}_{q'\lambda'}^\dagger(0)\rangle_{0,c} \end{cases}. \tag{5.16}$$

## *5.3 Calculation of Green's function*

### *5.3.1 First order appriximation*

In Eq. (5.16), we go back to the full notation $\hat{a}_{q\lambda}$ to distinguish phonons of different branches, then $G_{pp'}^{(1)}(\tau)$ is calculated as:

$$G_{pp'}^{(1)}(\tau) = -\frac{1}{\hbar}\frac{1}{2}\int_0^{\beta\hbar} d\tau_1 \langle T_\tau \left[ \sum_{k}\sum_{j=1}^{3r/2}\sum_{j'=\frac{3r}{2}+1}^{3r} V_{jj'}(k)\hat{Q}_{kj}(\tau_1)\hat{\tilde{Q}}_{kj'}^\dagger(\tau_1) \right.$$

$$\left. + \sum_{k}\sum_{j=1}^{3r/2}\sum_{j'=\frac{3r}{2}+1}^{3r} V_{jj'}^*(k)\hat{\tilde{Q}}_{kj'}(\tau_1)\hat{Q}_{kj}^\dagger(\tau_1) \right]\hat{a}_{q\lambda}(\tau)\hat{a}_{q'\lambda'}^\dagger(0)\rangle_{0,c}$$

$$= -\frac{1}{2\hbar}\int_0^{\beta\hbar} d\tau_1 \langle T_\tau \left\{ \sum_{k}\sum_{j=1}^{3r/2}\sum_{j'=\frac{3r}{2}+1}^{3r} \frac{\hbar V_{jj'}(k)}{2\sqrt{\omega_{kj}\widetilde{\omega}_{kj'}}}[\hat{a}_{kj}(\tau_1) + \hat{a}_{-kj}^\dagger(\tau_1)][\hat{\tilde{a}}_{-kj'}(\tau_1) + \hat{\tilde{a}}_{kj'}^\dagger(\tau_1)] \right.$$

$$\left. + \sum_{k}\sum_{j=1}^{3r/2}\sum_{j'=\frac{3r}{2}+1}^{3r} \frac{\hbar V_{jj'}^*(k)}{2\sqrt{\omega_{kj}\widetilde{\omega}_{kj'}}}[\hat{\tilde{a}}_{kj'}(\tau_1) + \hat{\tilde{a}}_{-kj'}^\dagger(\tau_1)][\hat{a}_{-kj}(\tau_1) + \hat{a}_{kj}^\dagger(\tau_1)] \right\}\hat{a}_{q\lambda}(\tau)\hat{a}_{q'\lambda'}^\dagger(0)\rangle_{0,c}$$

According to Wick's theorem [2], the terms that contain $a_{kj}a_{-kj'}$ and $a_{-kj}^\dagger a_{kj'}^\dagger$ are equal to zero, thus the first term of above equation are calculated as:

$$g_{11} = -\frac{1}{4}\sum_{k}\sum_{j=1}^{\frac{3r}{2}}\sum_{j'=\frac{3r}{2}+1}^{3r} \frac{V_{jj'}(k)}{\sqrt{\omega_{kj}\widetilde{\omega}_{kj'}}}\int_0^{\beta\hbar} d\tau_1 \left\{ \langle T_\tau \hat{a}_{-kj}^\dagger(\tau_1)\hat{\tilde{a}}_{-kj'}(\tau_1)\hat{a}_{q\lambda}(\tau)\hat{a}_{q'\lambda'}^\dagger(0)\rangle_{0,c} \right.$$

$$\left. + \langle T_\tau \hat{a}_{kj}(\tau_1)\hat{\tilde{a}}_{kj'}^\dagger(\tau_1)\hat{a}_{q\lambda}(\tau)\hat{a}_{q'\lambda'}^\dagger(0)\rangle_{0,c} \right\}$$

$$= -\frac{1}{4}\sum_{k}\sum_{j=1}^{3r/2}\sum_{j'=\frac{3r}{2}+1}^{3r} \frac{V_{jj'}(k)}{\sqrt{\omega_{kj}\widetilde{\omega}_{kj'}}}\int_0^{\beta\hbar} d\tau_1 \left\{ \langle T_\tau \hat{a}_{-kj}^\dagger(\tau_1)\hat{a}_{q\lambda}(\tau)\rangle_0 \langle T_\tau \hat{a}_{q'\lambda'}^\dagger(0)\hat{\tilde{a}}_{-kj'}(\tau_1)\rangle_0 \right.$$

$$\left. + \langle T_\tau \hat{\tilde{a}}_{kj'}^\dagger(\tau_1)\hat{a}_{q\lambda}(\tau)\rangle_0 \langle T_\tau \hat{a}_{q'\lambda'}^\dagger(0)\hat{a}_{kj}(\tau_1)\rangle_0 \right\}$$

$$= -\frac{1}{4}\sum_{k}\sum_{j=1}^{3r/2}\sum_{j'=\frac{3r}{2}+1}^{3r} \frac{V_{jj'}(k)}{\sqrt{\omega_{kj}\widetilde{\omega}_{kj'}}}\int_0^{\beta\hbar} d\tau_1 \left\{ G_{q\lambda}^0(\tau,\tau_1)\delta_{q,-k}\delta_{\lambda j}G_{q'\lambda'}^0(\tau_1,0)\delta_{q',-k}\delta_{\lambda'j'} \right.$$

$$\left. + G_{q\lambda}^0(\tau,\tau_1)\delta_{q,k}\delta_{\lambda j'}G_{q'\lambda'}^0(\tau_1,0)\delta_{q',k}\delta_{\lambda'j} \right\}$$

Simplification of above equation gives:

$$g_{11} = \begin{cases} \frac{-V_{\lambda\lambda'}(-q)}{4\sqrt{\omega_{q\lambda}\widetilde{\omega}_{q\lambda'}}} \int_0^{\beta\hbar} d\tau_1\, G^0_{q\lambda}(\tau,\tau_1)G^0_{q'\lambda'}(\tau_1,0)\delta_{qq'}, & 1\le \lambda \le \frac{3r}{2} < \lambda' \le 3r \\ \frac{-V_{\lambda'\lambda}(q)}{4\sqrt{\widetilde{\omega}_{q\lambda}\omega_{q\lambda'}}} \int_0^{\beta\hbar} d\tau_1\, G^0_{q\lambda}(\tau,\tau_1)G^0_{q'\lambda'}(\tau_1,0)\delta_{qq'}, & 1\le \lambda' \le \frac{3r}{2} < \lambda \le 3r \\ 0 & \text{else} \end{cases} \quad (5.17)$$

Similarity, the second term of $G^{(1)}_{pp'}(\tau)$ is calculated as:

$$g_{12} = -\frac{1}{4}\sum_{k}\sum_{j=1}^{\frac{3r}{2}}\sum_{j'=\frac{3r}{2}+1}^{3r} \frac{-V^*_{jj'}(k)}{4\sqrt{\omega_{kj}\widetilde{\omega}_{kj'}}} \int_0^{\beta\hbar} d\tau_1 \Big\{ \langle T_\tau \hat{\tilde{a}}_{kj'}(\tau_1)\hat{a}^\dagger_{kj}(\tau_1)\hat{a}_{q\lambda}(\tau)\hat{a}^\dagger_{q'\lambda'}(0)\rangle_{0,c}$$

$$+ \langle T_\tau \hat{\tilde{a}}^\dagger_{-kj'}(\tau_1)\hat{a}_{-kj}(\tau_1)\hat{a}_{q\lambda}(\tau)\hat{a}^\dagger_{q'\lambda'}(0)\rangle_{0,c} \Big\}$$

$$= -\frac{1}{4}\sum_{k}\sum_{j=1}^{\frac{3r}{2}}\sum_{j'=\frac{3r}{2}+1}^{3r} \frac{-V^*_{jj'}(k)}{4\sqrt{\omega_{kj}\widetilde{\omega}_{kj'}}} \int_0^{\beta\hbar} d\tau_1 \Big\{ \langle T_\tau \hat{a}^\dagger_{q'\lambda'}(0)\hat{\tilde{a}}_{kj'}(\tau_1)\rangle_0 \langle T_\tau \hat{a}^\dagger_{kj}(\tau_1)\hat{a}_{q\lambda}(\tau)\rangle_0$$

$$+ \langle T_\tau \hat{a}^\dagger_{q'\lambda'}(0)\hat{a}_{-kj}(\tau_1)\rangle_0 \langle T_\tau \hat{\tilde{a}}^\dagger_{-kj'}(\tau_1)\hat{a}_{q\lambda}(\tau)\rangle_0 \Big\}$$

i.e.,

$$g_{12} = \begin{cases} \frac{-V^*_{\lambda\lambda'}(q)}{4\sqrt{\omega_{q\lambda}\widetilde{\omega}_{q\lambda'}}} \int_0^{\beta\hbar} d\tau_1\, G^0_{q'\lambda'}(\tau_1,0)G^0_{q\lambda}(\tau,\tau_1)\delta_{qq'}, & 1\le \lambda \le \frac{3r}{2} < \lambda' \le 3r \\ \frac{-V^*_{\lambda'\lambda}(-q)}{4\sqrt{\widetilde{\omega}_{q\lambda}\omega_{q\lambda'}}} \int_0^{\beta\hbar} d\tau_1\, G^0_{q'\lambda'}(\tau_1,0)G^0_{q\lambda}(\tau,\tau_1)\delta_{qq'}, & 1\le \lambda' \le \frac{3r}{2} < \lambda \le 3r \\ 0 & \text{else} \end{cases} \quad (5.18)$$

Note that since $V^*_{\lambda\lambda'}(q) = V_{\lambda\lambda'}(-q)$ [Eq. (3.14)], the first order approximation of Green's function can be written as

$$G^{(1)}_{q\lambda,q'\lambda'}(\tau) = \begin{cases} \frac{-V_{\lambda\lambda'}(-q)}{2\sqrt{\omega_{q\lambda}\widetilde{\omega}_{q\lambda'}}} \int_0^{\beta\hbar} d\tau_1\, G^0_{q\lambda}(\tau,\tau_1)G^0_{q'\lambda'}(\tau_1,0)\delta_{qq'}, & 1\le \lambda \le \frac{3r}{2} < \lambda' \le 3r \\ \frac{-V_{\lambda'\lambda}(q)}{2\sqrt{\widetilde{\omega}_{q\lambda}\omega_{q\lambda'}}} \int_0^{\beta\hbar} d\tau_1\, G^0_{q\lambda}(\tau,\tau_1)G^0_{q'\lambda'}(\tau_1,0)\delta_{qq'}, & 1\le \lambda' \le \frac{3r}{2} < \lambda \le 3r \\ 0 & \text{else} \end{cases} \quad (5.19)$$

To derive the Green's function in frequency domain, we use the expression for Fourier series, then we have:

$$\int_0^{\beta\hbar} d\tau_1\, G^0_{q\lambda}(\tau,\tau_1)G^0_{q'\lambda'}(\tau_1,0) = \int_0^{\beta\hbar} d\tau_1 \left[\frac{1}{\beta\hbar}\sum_n e^{-i\omega_n(\tau-\tau_1)}G^0_{q\lambda}(i\omega_n)\right]\left[\frac{1}{\beta\hbar}\sum_{n'} e^{-i\omega_{n'}\tau_1}G^0_{q'\lambda'}(i\omega_{n'})\right]$$

$$= \frac{1}{(\beta\hbar)^2}\sum_{n,n'} e^{-i\omega_n\tau}\left[\int_0^{\beta\hbar} d\tau_1\, e^{i(\omega_n-\omega_{n'})\tau_1}\right] G^0_{q\lambda}(i\omega_n)G^0_{q'\lambda'}(i\omega_{n'}) =$$

$$= \frac{1}{(\beta\hbar)^2}\sum_{n,n'} e^{-i\omega_n\tau}[\beta\hbar\delta_{nn'}]G^0_{q\lambda}(i\omega_n)G^0_{q'\lambda'}(i\omega_{n'}) = \frac{1}{\beta\hbar}\sum_n e^{-i\omega_n\tau}G^0_{q\lambda}(i\omega_n)G^0_{q'\lambda'}(i\omega_n)$$

According to the definition of Fourier series [see Eq. (4.12)], we get the Fourier coefficient of

$G^{(1)}_{q\lambda,q'\lambda'}(\tau)$:

$$G^{(1)}_{q\lambda,q'\lambda'}(i\omega_n) = \begin{cases} \dfrac{-V_{\lambda\lambda'}(-q)}{2\sqrt{\omega_{q\lambda}\widetilde{\omega}_{q\lambda'}}} G^0_{q\lambda}(i\omega_n)G^0_{q'\lambda'}(i\omega_n)\delta_{qq'}, & 1 \leq \lambda \leq \dfrac{3r}{2} < \lambda' \leq 3r \\ \dfrac{-V_{\lambda'\lambda}(q)}{2\sqrt{\widetilde{\omega}_{q\lambda}\omega_{q\lambda'}}} G^0_{q\lambda}(i\omega_n)G^0_{q'\lambda'}(i\omega_n)\delta_{qq'}, & 1 \leq \lambda' \leq \dfrac{3r}{2} < \lambda \leq 3r \\ 0, & \text{else} \end{cases} \quad (5.20)$$

where $G^{(1)}_{q\lambda,q'\lambda'}(i\omega_n) = G^0_{q\lambda}(i\omega_n) \times \Sigma^{(1)}_{q\lambda,q'\lambda'}(i\omega_n) \times G^0_{q'\lambda'}(i\omega_n)$ and $\Sigma^{(1)}_{q\lambda,q'\lambda'}(i\omega_n)$ is defined as

$$\Sigma^{(1)}_{q\lambda,q'\lambda'}(i\omega_n) = \begin{cases} \dfrac{-V_{\lambda\lambda'}(-q)}{2\sqrt{\omega_{q\lambda}\widetilde{\omega}_{q\lambda'}}}\delta_{qq'}, & 1 \leq \lambda \leq \dfrac{3r}{2} < \lambda' \leq 3r \\ \dfrac{-V_{\lambda'\lambda}(q)}{2\sqrt{\widetilde{\omega}_{q\lambda}\omega_{q\lambda'}}}\delta_{qq'}, & 1 \leq \lambda' \leq \dfrac{3r}{2} < \lambda \leq 3r \\ 0, & \text{else} \end{cases} \quad (5.21)$$

### 5.3.2 second order approximation

The second order approximation of the Green's function in Eq. (5.16), $G^{(2)}_{q\lambda,q'\lambda'}(\tau)$, is calculated as:

$$G^{(2)}_{q\lambda,q'\lambda'}(\tau) = \frac{1}{8\hbar^2} \int_0^{\beta\hbar} d\tau_1 \int_0^{\beta\hbar} d\tau_2 \langle T_\tau \Bigg[ \sum_{k_1} \sum_{j_1=1}^{\frac{3r}{2}} \sum_{j_1'=\frac{3r}{2}+1}^{3r} V_{j_1 j_1'}(k_1) \hat{Q}_{k_1 j_1}(\tau_1) \hat{\widetilde{Q}}^\dagger_{k_1 j_1'}(\tau_1)$$

$$+ \sum_{k_1} \sum_{i_1=1}^{\frac{3r}{2}} \sum_{i_1'=\frac{3r}{2}+1}^{3r} V^*_{i_1 i_1'}(k_1) \hat{\widetilde{Q}}_{k_1 i_1'}(\tau_1) \hat{Q}^\dagger_{k_1 i_1}(\tau_1) \Bigg] \Bigg[ \sum_{k_2} \sum_{j_2=1}^{\frac{3r}{2}} \sum_{j_2'=\frac{3r}{2}+1}^{3r} V_{j_2 j_2'}(k_2) \hat{Q}_{k_2 j_2}(\tau_2) \hat{\widetilde{Q}}^\dagger_{k_2 j_2'}(\tau_2)$$

$$+ \sum_{k_2} \sum_{i_2=1}^{\frac{3r}{2}} \sum_{i_2'=\frac{3r}{2}+1}^{3r} V^*_{i_2 i_2'}(k_2) \hat{\widetilde{Q}}_{k_2 i_2'}(\tau_2) \hat{Q}^\dagger_{k_2 i_2}(\tau_2) \Bigg] \hat{a}_{q\lambda}(\tau) \hat{a}^\dagger_{q'\lambda'}(0) \rangle_{0,c} = g_1 + g_2 + g_3 + g_4$$

where,

$$g_1 = \frac{1}{8\hbar^2} \int_0^{\beta\hbar} d\tau_1 \int_0^{\beta\hbar} d\tau_2 \begin{Bmatrix} \sum_{k_1 j_1 j_1'}^{1\leq j_1 \leq \frac{3r}{2} < j_1' \leq 3r} \sum_{k_2 j_2 j_2'}^{1\leq j_2 \leq \frac{3r}{2} < j_2' \leq 3r} V_{j_1 j_1'}(k_1) V_{j_2 j_2'}(k_2) \\ \times \langle T_\tau \hat{Q}_{k_1 j_1}(\tau_1) \hat{\widetilde{Q}}^\dagger_{k_1 j_1'}(\tau_1) \hat{Q}_{k_2 j_2}(\tau_2) \hat{\widetilde{Q}}^\dagger_{k_2 j_2'}(\tau_2) \hat{a}_{q\lambda}(\tau) \hat{a}^\dagger_{q'\lambda'}(0) \rangle_{0,c} \end{Bmatrix} \quad (5.22)$$

$$g_2 = \frac{1}{8\hbar^2} \int_0^{\beta\hbar} d\tau_1 \int_0^{\beta\hbar} d\tau_2 \begin{Bmatrix} \sum_{k_1 j_1 j_1'}^{1\leq j_1 \leq \frac{3r}{2} < j_1' \leq 3r} \sum_{k_2 i_2 i_2'}^{1\leq i_2 \leq \frac{3r}{2} < i_2' \leq 3r} V_{j_1 j_1'}(k_1) V^*_{i_2 i_2'}(k_2) \\ \times \langle T_\tau \hat{Q}_{k_1 j_1}(\tau_1) \hat{\widetilde{Q}}^\dagger_{k_1 j_1'}(\tau_1) \hat{\widetilde{Q}}_{k_2 i_2'}(\tau_2) \hat{Q}^\dagger_{k_2 i_2}(\tau_2) \hat{a}_{q\lambda}(\tau) \hat{a}^\dagger_{q'\lambda'}(0) \rangle_{0,c} \end{Bmatrix} \quad (5.23)$$

$$g_3 = \frac{1}{8\hbar^2} \int_0^{\beta\hbar} d\tau_1 \int_0^{\beta\hbar} d\tau_2 \begin{Bmatrix} \sum_{k_1 i_1 i_1'}^{1\leq i_1 \leq \frac{3r}{2} < i_1' \leq 3r} \sum_{k_2 j_2 j_2'}^{1\leq j_2 \leq \frac{3r}{2} < j_2' \leq 3r} V^*_{i_1 i_1'}(k_1) V_{j_2 j_2'}(k_2) \\ \times \langle T_\tau \hat{\widetilde{Q}}_{k_1 i_1'}(\tau_1) \hat{Q}^\dagger_{k_1 i_1}(\tau_1) \hat{Q}_{k_2 j_2}(\tau_2) \hat{\widetilde{Q}}^\dagger_{k_2 j_2'}(\tau_2) \hat{a}_{q\lambda}(\tau) \hat{a}^\dagger_{q'\lambda'}(0) \rangle_{0,c} \end{Bmatrix} \quad (5.24)$$

$$g_4 = \frac{1}{8\hbar^2} \int_0^{\beta\hbar} d\tau_1 \int_0^{\beta\hbar} d\tau_2 \left\{ \begin{array}{l} \sum_{k_1 i_1 i_1'}^{1 \leq i_1 \leq \frac{3r}{2} < i_1' \leq 3r} \sum_{k_2 i_2 i_2'}^{1 \leq i_2 \leq \frac{3r}{2} < i_2' \leq 3r} V_{i_1 i_1'}^*(k_1) V_{i_2 i_2'}^*(k_2) \\ \times \langle T_\tau \hat{\tilde{Q}}_{k_1 i_1'}(\tau_1) \hat{Q}_{k_1 i_1}^\dagger(\tau_1) \hat{\tilde{Q}}_{k_2 i_2'}(\tau_2) \hat{Q}_{k_2 i_2}^\dagger(\tau_2) \hat{a}_{q\lambda}(\tau) \hat{a}_{q'\lambda'}^\dagger(0) \rangle_{0,c} \end{array} \right\} \quad (5.25)$$

In what follows we will expand the term $g_2$. The expansion of all other terms follows the same lines and eventually results in the same expression. Expanding the product $\hat{Q}_{k_1 j_1} \hat{\tilde{Q}}_{k_1 j_1'}^\dagger \hat{\tilde{Q}}_{k_2 i_2'} \hat{Q}_{k_2 i_2}^\dagger$, we can rewrite Eq. (5.23) as follows

$$g_2 = \frac{1}{32} \int_0^{\beta\hbar} d\tau_1 \int_0^{\beta\hbar} d\tau_2 \left\{ \sum_{k_1 j_1 j_1'}^{1 \leq j_1 \leq \frac{3r}{2} < j_1' \leq 3r} \sum_{k_2 i_2 i_2'}^{1 \leq i_2 \leq \frac{3r}{2} < i_2' \leq 3r} \frac{V_{j_1 j_1'}(k_1) V_{i_2 i_2'}^*(k_2)}{\sqrt{\omega_{k_1 j_1} \tilde{\omega}_{k_1 j_1'} \omega_{k_2 i_2} \tilde{\omega}_{k_2 i_2'}}} \times I \right\} \quad (5.26)$$

Where

$$I = \langle T_\tau [\hat{a}_{k_1 j_1}(\tau_1) + \hat{a}_{-k_1 j_1}^\dagger(\tau_1)] [\hat{\tilde{a}}_{-k_1 j_1'}(\tau_1) + \hat{\tilde{a}}_{k_1 j_1'}^\dagger(\tau_1)] [\hat{\tilde{a}}_{k_2 i_2'}(\tau_2) + \hat{\tilde{a}}_{-k_2 i_2'}^\dagger(\tau_2)] [\hat{a}_{-k_2 i_2}(\tau_2) + \hat{a}_{k_2 i_2}^\dagger(\tau_2)] \hat{a}_{q\lambda}(\tau) \hat{a}_{q'\lambda'}^\dagger(0) \rangle_{0,c}$$

$$= \langle T_\tau [\hat{a}_{k_1 j_1} \hat{\tilde{a}}_{-k_1 j_1'} + \hat{a}_{k_1 j_1} \hat{\tilde{a}}_{k_1 j_1'}^\dagger + \hat{a}_{-k_1 j_1}^\dagger \hat{\tilde{a}}_{-k_1 j_1'} + \hat{a}_{-k_1 j_1}^\dagger \hat{\tilde{a}}_{k_1 j_1'}^\dagger]_{\tau_1} [\hat{\tilde{a}}_{k_2 i_2'} \hat{a}_{-k_2 i_2} + \hat{\tilde{a}}_{k_2 i_2'} \hat{a}_{k_2 i_2}^\dagger + \hat{\tilde{a}}_{-k_2 i_2'}^\dagger \hat{a}_{-k_2 i_2}$$

$$+ \hat{\tilde{a}}_{-k_2 i_2'}^\dagger \hat{a}_{k_2 i_2}^\dagger]_{\tau_2} \hat{a}_{q\lambda}(\tau) \hat{a}_{q'\lambda'}^\dagger(0) \rangle_{0,c}$$

$$= \langle T_\tau [\hat{a}_{k_1 j_1} \hat{\tilde{a}}_{k_1 j_1'}^\dagger + \hat{a}_{-k_1 j_1}^\dagger \hat{\tilde{a}}_{-k_1 j_1'}]_{\tau_1} [\hat{\tilde{a}}_{k_2 i_2'} \hat{a}_{k_2 i_2}^\dagger + \hat{\tilde{a}}_{-k_2 i_2'}^\dagger \hat{a}_{-k_2 i_2}]_{\tau_2} \hat{a}_{q\lambda}(\tau) \hat{a}_{q'\lambda'}^\dagger(0) \rangle_{0,c}$$

Where the symbol $[\ldots]_{\tau_1}$ signifies that the operators in the brackets are given in the interaction picture. The last equality in above equation comes from the fact that the contractions of the product of the operators are equal to zero when the number of creation and annihilation operators are not the same (see Wick's theorem in Ref. [2]). Using Wick's theorem [2], we are able to calculate the above equation term by term as follows:

$$\begin{cases} I_1 = \langle T_\tau \hat{a}_{k_1 j_1}(\tau_1) \hat{\tilde{a}}_{k_1 j_1'}^\dagger(\tau_1) \hat{\tilde{a}}_{k_2 i_2'}(\tau_2) \hat{a}_{k_2 i_2}^\dagger(\tau_2) \hat{a}_{q\lambda}(\tau) \hat{a}_{q'\lambda'}^\dagger(0) \rangle_{0,c} \\ I_2 = \langle T_\tau \hat{a}_{k_1 j_1}(\tau_1) \hat{\tilde{a}}_{k_1 j_1'}^\dagger(\tau_1) \hat{\tilde{a}}_{-k_2 i_2'}^\dagger(\tau_2) \hat{a}_{-k_2 i_2}(\tau_2) \hat{a}_{q\lambda}(\tau) \hat{a}_{q'\lambda'}^\dagger(0) \rangle_{0,c} \\ I_3 = \langle T_\tau \hat{a}_{-k_1 j_1}^\dagger(\tau_1) \hat{\tilde{a}}_{-k_1 j_1'}(\tau_1) \hat{\tilde{a}}_{k_2 i_2'}(\tau_2) \hat{a}_{k_2 i_2}^\dagger(\tau_2) \hat{a}_{q\lambda}(\tau) \hat{a}_{q'\lambda'}^\dagger(0) \rangle_{0,c} \\ I_4 = \langle T_\tau \hat{a}_{-k_1 j_1}^\dagger(\tau_1) \hat{\tilde{a}}_{-k_1 j_1'}(\tau_1) \hat{\tilde{a}}_{-k_2 i_2'}^\dagger(\tau_2) \hat{a}_{-k_2 i_2}(\tau_2) \hat{a}_{q\lambda}(\tau) \hat{a}_{q'\lambda'}^\dagger(0) \rangle_{0,c} \end{cases} \quad (5.27)$$

We shall now calculate them term by term:

$$I_1 = \langle T_\tau \hat{a}_{k_1 j_1}(\tau_1) \hat{\tilde{a}}^\dagger_{k_1 j'_1}(\tau_1) \hat{\tilde{a}}_{k_2 i'_2}(\tau_2) \hat{a}^\dagger_{k_2 i_2}(\tau_2) \hat{a}_{q\lambda}(\tau) a^\dagger_{q'\lambda'}(0) \rangle_{0,c}$$

$$= \langle T_\tau \hat{a}^\dagger_{q'\lambda'}(0) \hat{a}_{k_1 j_1}(\tau_1) \rangle_0 \langle T_\tau \hat{a}^\dagger_{k_2 i_2}(\tau_2) \hat{\tilde{a}}_{k_2 i'_2}(\tau_2) \rangle_0 \langle T_\tau \hat{\tilde{a}}^\dagger_{k_1 j'_1}(\tau_1) \hat{a}_{q\lambda}(\tau) \rangle_0$$

$$+ \langle T_\tau \hat{a}^\dagger_{k_2 i_2}(\tau_2) \hat{a}_{k_1 j_1}(\tau_1) \rangle_0 \langle T_\tau \hat{a}^\dagger_{q'\lambda'}(0) \hat{\tilde{a}}_{k_2 i'_2}(\tau_2) \rangle_0 \langle T_\tau \hat{\tilde{a}}^\dagger_{k_1 j'_1}(\tau_1) \hat{a}_{q\lambda}(\tau) \rangle_0$$

$$+ \langle T_\tau \hat{a}^\dagger_{q'\lambda'}(0) \hat{\tilde{a}}_{k_2 i'_2}(\tau_2) \rangle_0 \langle T_\tau \hat{\tilde{a}}^\dagger_{k_1 j'_1}(\tau_1) \hat{a}_{k_1 j_1}(\tau_1) \rangle_0 \langle T_\tau \hat{a}^\dagger_{k_2 i_2}(\tau_2) \hat{a}_{q\lambda}(\tau) \rangle_0$$

$$+ \langle T_\tau \hat{\tilde{a}}^\dagger_{k_1 j'_1}(\tau_1) \hat{\tilde{a}}_{k_2 i'_2}(\tau_2) \rangle_0 \langle T_\tau \hat{a}^\dagger_{q'\lambda'}(0) \hat{a}_{k_1 j_1}(\tau_1) \rangle_0 \langle T_\tau \hat{a}^\dagger_{k_2 i_2}(\tau_2) \hat{a}_{q\lambda}(\tau) \rangle_{0,c}$$

$$= \left[ G^0_{q'\lambda'}(\tau_1, 0) \delta_{k_1 q'} \delta_{j_1 \lambda'} \cdot G^0_{k_2 j_2}(\tau_2, \tau_2) \delta_{i_2 i'_2} \cdot G^0_{q\lambda}(\tau, \tau_1) \delta_{k_1 q} \delta_{j'_1 \lambda} \right]$$

$$+ \left[ G^0_{k_1 j_1}(\tau_1, \tau_2) \delta_{k_1 k_2} \delta_{j_1 i_2} \cdot G^0_{q'\lambda'}(\tau_2, 0) \delta_{k_2 q'} \delta_{i'_2 \lambda'} \cdot G^0_{q\lambda}(\tau, \tau_1) \delta_{k_1 q} \delta_{j'_1 \lambda} \right]$$

$$+ \left[ G^0_{q'\lambda'}(\tau_2, 0) \delta_{k_2 q'} \delta_{i'_2 \lambda'} \cdot G^0_{k_1 j_1}(\tau_1, \tau_1) \delta_{j_1 j'_1} \cdot G^0_{q\lambda}(\tau, \tau_2) \delta_{k_2 q} \delta_{i_2 \lambda} \right]$$

$$+ \left[ G^0_{k_1 j'_1}(\tau_2, \tau_1) \delta_{k_1 k_2} \delta_{i'_2 j'_1} \cdot G^0_{q'\lambda'}(\tau_1, 0) \delta_{k_1 q'} \delta_{j_1 \lambda'} \cdot G^0_{q\lambda}(\tau, \tau_2) \delta_{k_2 q} \delta_{i_2 \lambda} \right]$$

$$= \left[ G^0_{k_1 j_1}(\tau_1, \tau_2) \delta_{k_1 k_2} \delta_{j_1 i_2} \cdot G^0_{q'\lambda'}(\tau_2, 0) \delta_{k_2 q'} \delta_{i'_2 \lambda'} \cdot G^0_{q\lambda}(\tau, \tau_1) \delta_{k_1 q} \delta_{j'_1 \lambda} \right]$$

$$+ \left[ G^0_{k_1 j'_1}(\tau_2, \tau_1) \delta_{k_1 k_2} \delta_{i'_2 j'_1} \cdot G^0_{q'\lambda'}(\tau_1, 0) \delta_{k_1 q'} \delta_{j_1 \lambda'} \cdot G^0_{q\lambda}(\tau, \tau_2) \delta_{k_2 q} \delta_{i_2 \lambda} \right]$$

The last equality results from the fact that $j_1$ and $j'_1$ are indices of different ranges (belonging to different subsystems) and the same holds for $i_2$ and $i'_2$. Similarity,

$$I_4 = \langle T_\tau \hat{a}^\dagger_{-k_1 j_1}(\tau_1) \hat{\tilde{a}}_{-k_1 j'_1}(\tau_1) \hat{\tilde{a}}^\dagger_{-k_2 i'_2}(\tau_2) \hat{a}_{-k_2 i_2}(\tau_2) \hat{a}_{q\lambda}(\tau) \hat{\tilde{a}}^\dagger_{q'\lambda'}(0) \rangle_{0,c}$$

$$= \langle T_\tau \hat{a}^\dagger_{q'\lambda'}(0) \hat{\tilde{a}}_{-k_1 j'_1}(\tau_1) \rangle_0 \langle T_\tau \hat{a}^\dagger_{-k_1 j_1}(\tau_1) \hat{a}_{q\lambda}(\tau) \rangle_0 \langle T_\tau \hat{\tilde{a}}^\dagger_{-k_2 i'_2}(\tau_2) \hat{a}_{-k_2 i_2}(\tau_2) \rangle_0$$

$$+ \langle T_\tau \hat{a}^\dagger_{q'\lambda'}(0) \hat{a}_{-k_2 i_2}(\tau_2) \rangle_0 \langle T_\tau \hat{a}^\dagger_{-k_1 j_1}(\tau_1) \hat{a}_{q\lambda}(\tau) \rangle_0 \langle T_\tau \hat{\tilde{a}}^\dagger_{-k_2 i'_2}(\tau_2) \hat{\tilde{a}}_{-k_1 j'_1}(\tau_1) \rangle_0$$

$$+ \langle T_\tau \hat{a}^\dagger_{q'\lambda'}(0) \hat{a}_{-k_2 i_2}(\tau_2) \rangle_0 \langle T_\tau \hat{\tilde{a}}^\dagger_{-k_2 i'_2}(\tau_2) \hat{a}_{q\lambda}(\tau) \rangle_0 \langle T_\tau \hat{a}^\dagger_{-k_1 j_1}(\tau_1) \hat{\tilde{a}}_{-k_1 j'_1}(\tau_1) \rangle_0$$

$$+ \langle T_\tau \hat{a}^\dagger_{q'\lambda'}(0) \hat{\tilde{a}}_{-k_1 j'_1}(\tau_1) \rangle_0 \langle T_\tau \hat{\tilde{a}}^\dagger_{-k_2 i'_2}(\tau_2) \hat{a}_{q\lambda}(\tau) \rangle_0 \langle T_\tau \hat{a}^\dagger_{-k_1 j_1}(\tau_1) \hat{a}_{-k_2 i_2}(\tau_2) \rangle_0$$

$$= \left[ G^0_{q'\lambda'}(\tau_1, 0) \delta_{-k_1 q'} \delta_{j'_1 \lambda'} \cdot G^0_{q\lambda}(\tau, \tau_1) \delta_{-k_1 q} \delta_{j_1 \lambda} \cdot G^0_{-k_2 i_2}(\tau_2, \tau_2) \delta_{i_2 i'_2} \right]$$

$$+ \left[ G^0_{q'\lambda'}(\tau_2, 0) \delta_{-k_2 q'} \delta_{i_2 \lambda'} \cdot G^0_{q\lambda}(\tau, \tau_1) \delta_{-k_1 q} \delta_{j_1 \lambda} \cdot G^0_{-k_1 j'_1}(\tau_1, \tau_2) \delta_{k_1 k_2} \delta_{j'_1 i'_2} \right]$$

$$+ \left[ G^0_{q'\lambda'}(\tau_2, 0) \delta_{-k_2 q'} \delta_{i_2 \lambda'} \cdot G^0_{q\lambda}(\tau, \tau_2) \delta_{-k_2 q} \delta_{i'_2 \lambda} \cdot G^0_{-k_1 j_1}(\tau_1, \tau_1) \delta_{j_1 j'_1} \right]$$

$$+ \left[ G^0_{q'\lambda'}(\tau_1, 0) \delta_{-k_1 q'} \delta_{j'_1 \lambda'} \cdot G^0_{q\lambda}(\tau, \tau_2) \delta_{-k_2 q} \delta_{i'_2 \lambda} \cdot G^0_{-k_1 j_1}(\tau_2, \tau_1) \delta_{k_1 k_2} \delta_{j_1 i_2} \right]$$

$$= \left[ G^0_{q'\lambda'}(\tau_2, 0) \delta_{-k_2 q'} \delta_{i_2 \lambda'} \cdot G^0_{q\lambda}(\tau, \tau_1) \delta_{-k_1 q} \delta_{j_1 \lambda} \cdot G^0_{-k_1 j'_1}(\tau_1, \tau_2) \delta_{k_1 k_2} \delta_{j'_1 i'_2} \right]$$

$$+ \left[ G^0_{q'\lambda'}(\tau_1, 0) \delta_{-k_1 q'} \delta_{j'_1 \lambda'} \cdot G^0_{q\lambda}(\tau, \tau_2) \delta_{-k_2 q} \delta_{i'_2 \lambda} \cdot G^0_{-k_1 j_1}(\tau_2, \tau_1) \delta_{k_1 k_2} \delta_{j_1 i_2} \right].$$

Where, again, the last equality results from the fact that $j_1$ and $j'_1$ are indices of different ranges (belonging to different subsystems) and the same holds for $i_2$ and $i'_2$. Similarly, we obtain $I_2 = I_3 = 0$ for the same reason, as shown below:

$$I_2 = \langle T_\tau \hat{a}_{\boldsymbol{k}_1 j_1}(\tau_1) \hat{\tilde{a}}^\dagger_{\boldsymbol{k}_1 j'_1}(\tau_1) \hat{\tilde{a}}^\dagger_{-\boldsymbol{k}_2 i'_2}(\tau_2) \hat{a}_{-\boldsymbol{k}_2 i_2}(\tau_2) \hat{a}_{\boldsymbol{q}\lambda}(\tau) \hat{a}^\dagger_{\boldsymbol{q}'\lambda'}(0) \rangle_{0,c}$$

$$= \langle T_\tau \hat{a}^\dagger_{\boldsymbol{q}'\lambda'}(0) \hat{a}_{\boldsymbol{k}_1 j_1}(\tau_1) \rangle_0 \langle T_\tau \hat{\tilde{a}}^\dagger_{\boldsymbol{k}_1 j'_1}(\tau_1) \hat{a}_{\boldsymbol{q}\lambda}(\tau) \rangle_0 \langle T_\tau \hat{\tilde{a}}^\dagger_{-\boldsymbol{k}_2 i'_2}(\tau_2) \hat{a}_{-\boldsymbol{k}_2 i_2}(\tau_2) \rangle_0$$

$$+ \langle T_\tau \hat{\tilde{a}}^\dagger_{-\boldsymbol{k}_2 i'_2}(\tau_2) \hat{a}_{\boldsymbol{k}_1 j_1}(\tau_1) \rangle_0 \langle T_\tau \hat{a}^\dagger_{\boldsymbol{q}'\lambda'}(0) \hat{a}_{-\boldsymbol{k}_2 i_2}(\tau_2) \rangle_0 \langle T_\tau \hat{\tilde{a}}^\dagger_{\boldsymbol{k}_1 j'_1}(\tau_1) \hat{a}_{\boldsymbol{q}\lambda}(\tau) \rangle_0$$

$$+ \langle T_\tau \hat{a}^\dagger_{\boldsymbol{q}'\lambda'}(0) \hat{a}_{-\boldsymbol{k}_2 i_2}(\tau_2) \rangle_0 \langle T_\tau \hat{\tilde{a}}^\dagger_{-\boldsymbol{k}_2 i'_2}(\tau_2) \hat{a}_{\boldsymbol{q}\lambda}(\tau) \rangle_0 \langle T_\tau \hat{\tilde{a}}^\dagger_{\boldsymbol{k}_1 j'_1}(\tau_1) \hat{a}_{\boldsymbol{k}_1 j_1}(\tau_1) \rangle_0$$

$$+ \langle T_\tau \hat{\tilde{a}}^\dagger_{\boldsymbol{k}_1 j'_1}(\tau_1) \hat{a}_{-\boldsymbol{k}_2 i_2}(\tau_2) \rangle_0 \langle T_\tau \hat{a}^\dagger_{\boldsymbol{q}'\lambda'}(0) \hat{a}_{\boldsymbol{k}_1 j_1}(\tau_1) \rangle_0 \langle T_\tau \hat{\tilde{a}}^\dagger_{-\boldsymbol{k}_2 i'_2}(\tau_2) \hat{a}_{\boldsymbol{q}\lambda}(\tau) \rangle_0$$

$$= \left[ G^0_{\boldsymbol{q}'\lambda'}(\tau_1, 0) \delta_{\boldsymbol{k}_1 \boldsymbol{q}'} \delta_{j_1 \lambda'} \cdot G^0_{\boldsymbol{q}\lambda}(\tau, \tau_1) \delta_{\boldsymbol{k}_1 \boldsymbol{q}} \delta_{j'_1 \lambda} \cdot G^0_{-\boldsymbol{k}_2 i_2}(\tau_2, \tau_2) \delta_{i_2 i'_2} \right]$$

$$+ \left[ G^0_{\boldsymbol{k}_1 j_1}(\tau_1, \tau_2) \delta_{-\boldsymbol{k}_2 \boldsymbol{k}_1} \delta_{j_1 i'_2} \cdot G^0_{\boldsymbol{q}'\lambda'}(\tau_2, 0) \delta_{-\boldsymbol{k}_2 \boldsymbol{q}'} \delta_{i_2 \lambda'} \cdot G^0_{\boldsymbol{q}\lambda}(\tau, \tau_1) \delta_{\boldsymbol{k}_1 \boldsymbol{q}} \delta_{j'_1 \lambda} \right]$$

$$+ \left[ G^0_{\boldsymbol{q}'\lambda'}(\tau_2, 0) \delta_{-\boldsymbol{k}_2 \boldsymbol{q}'} \delta_{i_2 \lambda'} \cdot G^0_{\boldsymbol{q}\lambda}(\tau, \tau_2) \delta_{-\boldsymbol{k}_2 \boldsymbol{q}} \delta_{i'_2 \lambda} \cdot G^0_{\boldsymbol{k}_1 j_1}(\tau_1, \tau_1) \delta_{j_1 j'_1} \right]$$

$$+ \left[ G^0_{\boldsymbol{k}_1 j'_1}(\tau_2, \tau_1) \delta_{-\boldsymbol{k}_2 \boldsymbol{k}_1} \delta_{i_2 j'_1} \cdot G^0_{\boldsymbol{q}'\lambda'}(\tau_1, 0) \delta_{\boldsymbol{k}_1 \boldsymbol{q}'} \delta_{j_1 \lambda'} \cdot G^0_{\boldsymbol{q}\lambda}(\tau, \tau_2) \delta_{-\boldsymbol{k}_2 \boldsymbol{q}} \delta_{i'_2 \lambda} \right] = 0.$$

$$I_3 = \langle T_\tau \hat{a}^\dagger_{-\boldsymbol{k}_1 j_1}(\tau_1) \hat{\tilde{a}}_{-\boldsymbol{k}_1 j'_1}(\tau_1) \hat{\tilde{a}}_{\boldsymbol{k}_2 i'_2}(\tau_2) \hat{a}^\dagger_{\boldsymbol{k}_2 i_2}(\tau_2) \hat{a}_{\boldsymbol{q}\lambda}(\tau) \hat{a}^\dagger_{\boldsymbol{q}'\lambda'}(0) \rangle_{0,c}$$

$$= \langle T_\tau \hat{a}^\dagger_{\boldsymbol{q}'\lambda'}(0) \hat{\tilde{a}}_{-\boldsymbol{k}_1 j'_1}(\tau_1) \rangle_0 \langle T_\tau \hat{a}^\dagger_{-\boldsymbol{k}_1 j_1}(\tau_1) \hat{a}_{\boldsymbol{q}\lambda}(\tau) \rangle_0 \langle T_\tau \hat{a}^\dagger_{\boldsymbol{k}_2 i_2}(\tau_2) \hat{\tilde{a}}_{\boldsymbol{k}_2 i'_2}(\tau_2) \rangle_0$$

$$+ \langle T_\tau \hat{a}^\dagger_{\boldsymbol{k}_2 i_2}(\tau_2) \hat{\tilde{a}}_{-\boldsymbol{k}_1 j'_1}(\tau_1) \rangle_0 \langle T_\tau \hat{a}^\dagger_{\boldsymbol{q}'\lambda'}(0) \hat{\tilde{a}}_{\boldsymbol{k}_2 i'_2}(\tau_2) \rangle_0 \langle T_\tau \hat{a}^\dagger_{-\boldsymbol{k}_1 j_1}(\tau_1) \hat{a}_{\boldsymbol{q}\lambda}(\tau) \rangle_0$$

$$+ \langle T_\tau \hat{a}^\dagger_{\boldsymbol{q}'\lambda'}(0) \hat{\tilde{a}}_{\boldsymbol{k}_2 i'_2}(\tau_2) \rangle_0 \langle T_\tau \hat{a}^\dagger_{\boldsymbol{k}_2 i_2}(\tau_2) \hat{a}_{\boldsymbol{q}\lambda}(\tau) \rangle_0 \langle T_\tau \hat{a}^\dagger_{-\boldsymbol{k}_1 j_1}(\tau_1) \hat{\tilde{a}}_{-\boldsymbol{k}_1 j'_1}(\tau_1) \rangle_0$$

$$+ \langle T_\tau \hat{a}^\dagger_{-\boldsymbol{k}_1 j_1}(\tau_1) \hat{\tilde{a}}_{\boldsymbol{k}_2 i'_2}(\tau_2) \rangle_0 \langle T_\tau \hat{a}^\dagger_{\boldsymbol{q}'\lambda'}(0) \hat{\tilde{a}}_{-\boldsymbol{k}_1 j'_1}(\tau_1) \rangle_0 \langle T_\tau \hat{a}^\dagger_{\boldsymbol{k}_2 i_2}(\tau_2) \hat{a}_{\boldsymbol{q}\lambda}(\tau) \rangle_0$$

$$= \left[ G^0_{\boldsymbol{q}'\lambda'}(\tau_1, 0) \delta_{-\boldsymbol{k}_1 \boldsymbol{q}'} \delta_{j'_1 \lambda'} \cdot G^0_{\boldsymbol{q}\lambda}(\tau, \tau_1) \delta_{-\boldsymbol{k}_1 \boldsymbol{q}} \delta_{j_1 \lambda} \cdot G^0_{\boldsymbol{k}_2 i_2}(\tau_2, \tau_2) \delta_{i_2 i'_2} \right]$$

$$+ \left[ G^0_{\boldsymbol{k}_2 i_2}(\tau_1, \tau_2) \delta_{-\boldsymbol{k}_1 \boldsymbol{k}_2} \delta_{j'_1 i_2} \cdot G^0_{\boldsymbol{q}'\lambda'}(\tau_2, 0) \delta_{\boldsymbol{k}_2 \boldsymbol{q}'} \delta_{i'_2 \lambda'} \cdot G^0_{\boldsymbol{q}\lambda}(\tau, \tau_1) \delta_{-\boldsymbol{k}_1 \boldsymbol{q}} \delta_{j_1 \lambda} \right]$$

$$+ \left[ G^0_{\boldsymbol{q}'\lambda'}(\tau_2, 0) \delta_{\boldsymbol{k}_2 \boldsymbol{q}'} \delta_{i'_2 \lambda'} \cdot G^0_{\boldsymbol{q}\lambda}(\tau, \tau_2) \delta_{\boldsymbol{k}_2 \boldsymbol{q}} \delta_{i_2 \lambda} \cdot G^0_{-\boldsymbol{k}_1 j_1}(\tau_1, \tau_1) \delta_{j_1 j'_1} \right]$$

$$+ \left[ G^0_{\boldsymbol{k}_2 i'_2}(\tau_2, \tau_1) \delta_{-\boldsymbol{k}_1 \boldsymbol{k}_2} \delta_{i'_2 j_1} \cdot G^0_{\boldsymbol{q}'\lambda'}(\tau_1, 0) \delta_{-\boldsymbol{k}_1 \boldsymbol{q}'} \delta_{j'_1 \lambda'} \cdot G^0_{\boldsymbol{q}\lambda}(\tau, \tau_2) \delta_{\boldsymbol{k}_2 \boldsymbol{q}} \delta_{i_2 \lambda} \right] = 0.$$

The two non-vanishing terms ($I_1$ and $I_4$) produce the following contributions to $g_2$ of Eq. (5.26):

$$g_{21} = \frac{1}{32} \int_0^{\beta\hbar} d\tau_1 \int_0^{\beta\hbar} d\tau_2 \left\{ \sum_{\boldsymbol{k}_1 j_1 j'_1}^{1 \leq j_1 \leq \frac{3r}{2} < j'_1 \leq 3r} \sum_{\boldsymbol{k}_2 i_2 i'_2}^{1 \leq i_2 \leq \frac{3r}{2} < i'_2 \leq 3r} \frac{V_{j_1 j'_1}(\boldsymbol{k}_1) V^*_{i_2 i'_2}(\boldsymbol{k}_2)}{\sqrt{\omega_{\boldsymbol{k}_1 j_1} \widetilde{\omega}_{\boldsymbol{k}_1 j'_1} \omega_{\boldsymbol{k}_2 i_2} \widetilde{\omega}_{\boldsymbol{k}_2 i'_2}}} I_1 \right\} =$$

$$\frac{1}{32} \int_0^{\beta\hbar} d\tau_1 \int_0^{\beta\hbar} d\tau_2 \sum_{\boldsymbol{k}_1 j_1 j'_1}^{1 \leq j_1 \leq \frac{3r}{2} < j'_1 \leq 3r} \sum_{\boldsymbol{k}_2 i_2 i'_2}^{1 \leq i_2 \leq \frac{3r}{2} < i'_2 \leq 3r} \frac{V_{j_1 j'_1}(\boldsymbol{k}_1) V^*_{i_2 i'_2}(\boldsymbol{k}_2)}{\sqrt{\omega_{\boldsymbol{k}_1 j_1} \widetilde{\omega}_{\boldsymbol{k}_1 j'_1} \omega_{\boldsymbol{k}_2 i_2} \widetilde{\omega}_{\boldsymbol{k}_2 i'_2}}} \left[ G^0_{\boldsymbol{k}_1 j_1}(\tau_1, \tau_2) \delta_{\boldsymbol{k}_1 \boldsymbol{k}_2} \delta_{j_1 i_2} \cdot G^0_{\boldsymbol{q}'\lambda'}(\tau_2, 0) \delta_{\boldsymbol{k}_2 \boldsymbol{q}'} \delta_{i'_2 \lambda'} \cdot \right.$$

$$\left. G^0_{\boldsymbol{q}\lambda}(\tau, \tau_1) \delta_{\boldsymbol{k}_1 \boldsymbol{q}} \delta_{j'_1 \lambda} \right] + \frac{1}{32} \int_0^{\beta\hbar} d\tau_1 \int_0^{\beta\hbar} d\tau_2 \sum_{\boldsymbol{k}_1 j_1 j'_1}^{1 \leq j_1 \leq \frac{3r}{2} < j'_1 \leq 3r} \sum_{\boldsymbol{k}_2 i_2 i'_2}^{1 \leq i_2 \leq \frac{3r}{2} < i'_2 \leq 3r} \frac{V_{j_1 j'_1}(\boldsymbol{k}_1) V^*_{i_2 i'_2}(\boldsymbol{k}_2)}{\sqrt{\omega_{\boldsymbol{k}_1 j_1} \widetilde{\omega}_{\boldsymbol{k}_1 j'_1} \omega_{\boldsymbol{k}_2 i_2} \widetilde{\omega}_{\boldsymbol{k}_2 i'_2}}} \left[ G^0_{\boldsymbol{k}_1 j'_1}(\tau_2, \tau_1) \delta_{\boldsymbol{k}_1 \boldsymbol{k}_2} \delta_{i'_2 j'_1} \cdot \right.$$

$$\left. G^0_{\boldsymbol{q}'\lambda'}(\tau_1, 0) \delta_{\boldsymbol{k}_1 \boldsymbol{q}'} \delta_{j_1 \lambda'} \cdot G^0_{\boldsymbol{q}\lambda}(\tau, \tau_2) \delta_{\boldsymbol{k}_2 \boldsymbol{q}} \delta_{i_2 \lambda} \right] =$$

$$\begin{cases} \frac{\delta_{\boldsymbol{q}\boldsymbol{q}'}}{32} \int_0^{\beta\hbar} d\tau_1 \int_0^{\beta\hbar} d\tau_2 \sum_{j_1 = 1}^{\frac{3r}{2}} \frac{V_{j_1 \lambda}(\boldsymbol{q}) V^*_{j_1 \lambda'}(\boldsymbol{q}')}{\omega_{\boldsymbol{q} j_1} \sqrt{\omega_{\boldsymbol{q}\lambda} \omega_{\boldsymbol{q}'\lambda'}}} \left[ G^0_{\boldsymbol{q} j_1}(\tau_1, \tau_2) \cdot G^0_{\boldsymbol{q}'\lambda'}(\tau_2, 0) G^0_{\boldsymbol{q}\lambda}(\tau, \tau_1) \right], & \lambda, \lambda' \in \left[ \frac{3r}{2} + 1, 3r \right] \\ \frac{\delta_{\boldsymbol{q}\boldsymbol{q}'}}{32} \int_0^{\beta\hbar} d\tau_1 \int_0^{\beta\hbar} d\tau_2 \sum_{j'_1 = \frac{3r}{2} + 1}^{3r} \frac{V_{\lambda' j'_1}(\boldsymbol{q}') V^*_{\lambda j'_1}(\boldsymbol{q})}{\omega_{\boldsymbol{q} j'_1} \sqrt{\omega_{\boldsymbol{q}\lambda} \omega_{\boldsymbol{q}'\lambda'}}} \left[ G^0_{\boldsymbol{q} j'_1}(\tau_2, \tau_1) \cdot G^0_{\boldsymbol{q}'\lambda'}(\tau_1, 0) G^0_{\boldsymbol{q}\lambda}(\tau, \tau_2) \right], & \lambda, \lambda' \in \left[ 1, \frac{3r}{2} \right] \\ 0 & \text{else} \end{cases}.$$

and

$$g_{24} = \frac{1}{32}\int_0^{\beta\hbar} d\tau_1 \int_0^{\beta\hbar} d\tau_2 \left\{ \sum_{k_1 j_1 j_1'}^{1\le j_1 \le \frac{3r}{2}<j_1'\le 3r} \sum_{k_2 i_2 i_2'}^{1\le i_2 \le \frac{3r}{2}<i_2'\le 3r} \frac{V_{j_1 j_1'}(k_1)V_{i_2 i_2'}^*(k_2)}{\sqrt{\omega_{k_1 j_1}\widetilde{\omega}_{k_1 j_1'}\omega_{k_2 i_2}\widetilde{\omega}_{k_2 i_2'}}} I_4 \right\} =$$

$$\frac{1}{32}\int_0^{\beta\hbar} d\tau_1 \int_0^{\beta\hbar} d\tau_2 \sum_{k_1 j_1 j_1'}^{1\le j_1 \le \frac{3r}{2}<j_1'\le 3r} \sum_{k_2 i_2 i_2'}^{1\le i_2 \le \frac{3r}{2}<i_2'\le 3r} \frac{V_{j_1 j_1'}(k_1)V_{i_2 i_2'}^*(k_2)}{\sqrt{\omega_{k_1 j_1}\widetilde{\omega}_{k_1 j_1'}\omega_{k_2 i_2}\widetilde{\omega}_{k_2 i_2'}}} \left[ G^0_{q'\lambda'}(\tau_2,0)\delta_{-k_2 q'}\delta_{i_2\lambda'} \cdot G^0_{q\lambda}(\tau,\tau_1)\delta_{-k_1 q}\delta_{j_1\lambda} \cdot \right.$$

$$\left. G^0_{-k_1 j_1'}(\tau_1,\tau_2)\delta_{k_1 k_2}\delta_{j_1' i_2'} \right] + \frac{1}{32}\int_0^{\beta\hbar} d\tau_1 \int_0^{\beta\hbar} d\tau_2 \sum_{k_1 j_1 j_1'}^{1\le j_1 \le \frac{3r}{2}<j_1'\le 3r} \sum_{k_2 i_2 i_2'}^{1\le i_2 \le \frac{3r}{2}<i_2'\le 3r} \frac{V_{j_1 j_1'}(k_1)V_{i_2 i_2'}^*(k_2)}{\sqrt{\omega_{k_1 j_1}\widetilde{\omega}_{k_1 j_1'}\omega_{k_2 i_2}\widetilde{\omega}_{k_2 i_2'}}} \left[ G^0_{q'\lambda'}(\tau_1,0)\delta_{-k_1 q'}\delta_{j_1'\lambda'} \cdot \right.$$

$$\left. G^0_{q\lambda}(\tau,\tau_2)\delta_{-k_2 q}\delta_{i_2'\lambda} \cdot G^0_{-k_1 j_1}(\tau_2,\tau_1)\delta_{k_1 k_2}\delta_{j_1 i_2} \right] =$$

$$\begin{cases} \frac{\delta_{qq'}}{32}\int_0^{\beta\hbar} d\tau_1 \int_0^{\beta\hbar} d\tau_2 \sum_{j_1'=\frac{3r}{2}+1}^{3r} \frac{V_{\lambda' j_1'}(q')V_{\lambda j_1'}^*(q)}{\omega_{q j_1'}\sqrt{\omega_{q\lambda}\omega_{q'\lambda'}}} \left[ G^0_{q j_1'}(\tau_1,\tau_2)G^0_{q'\lambda'}(\tau_2,0)G^0_{q\lambda}(\tau,\tau_1) \right], & \lambda,\lambda' \in \left[1,\frac{3r}{2}\right] \\ \frac{\delta_{qq'}}{32}\int_0^{\beta\hbar} d\tau_1 \int_0^{\beta\hbar} d\tau_2 \sum_{j_1=1}^{\frac{3r}{2}} \frac{V_{j_1\lambda}(q)V_{j_1\lambda'}^*(q')}{\omega_{q j_1}\sqrt{\omega_{q\lambda}\omega_{q'\lambda'}}} \left[ G^0_{q j_1}(\tau_2,\tau_1)G^0_{q'\lambda'}(\tau_1,0)G^0_{q\lambda}(\tau,\tau_2) \right], & \lambda,\lambda' \in \left[\frac{3r}{2}+1,3r\right] \\ 0 & \text{else} \end{cases}$$

Here we used the following properties: $\omega_{-q\lambda} = \omega_{q\lambda}$ and $V_{\lambda j_1'}(-q) = V_{\lambda j_1'}^*(q)$ [see Eqs. (1.12) and (3.14)].

The equivalence of $g_{21}$ and $g_{24}$ can be seen by changing the integration variables $\tau_2 \leftrightarrow \tau_1$. Substituting $g_{21}$ and $g_{24}$ into Eq. (5.26), $g_2$ is given by:

$$g_2 = \begin{cases} \frac{\delta_{qq'}}{16}\int_0^{\beta\hbar} d\tau_1 \int_0^{\beta\hbar} d\tau_2 \sum_{j_1=1}^{\frac{3r}{2}} \frac{V_{j_1\lambda}(q)V_{j_1\lambda'}^*(q')}{\omega_{q j_1}\sqrt{\omega_{q\lambda}\omega_{q'\lambda'}}} \left[ G^0_{q j_1}(\tau_1,\tau_2)G^0_{q'\lambda'}(\tau_2,0)G^0_{q\lambda}(\tau,\tau_1) \right], & \lambda,\lambda' \in \left[\frac{3r}{2}+1,3r\right] \\ \frac{\delta_{qq'}}{16}\int_0^{\beta\hbar} d\tau_1 \int_0^{\beta\hbar} d\tau_2 \sum_{j_1'=\frac{3r}{2}+1}^{3r} \frac{V_{\lambda' j_1'}(q')V_{\lambda j_1'}^*(q)}{\omega_{q j_1'}\sqrt{\omega_{q\lambda}\omega_{q'\lambda'}}} \left[ G^0_{q j_1'}(\tau_2,\tau_1)G^0_{q'\lambda'}(\tau_1,0)G^0_{q\lambda}(\tau,\tau_2) \right], & \lambda,\lambda' \in \left[1,\frac{3r}{2}\right] \\ 0 & \text{else} \end{cases} \quad (5.28)$$

By changing the indices and integration variables, it's straightforward to show that the other three terms ($g_1, g_3, g_4$) are identical to $g_2$, thus the final expression of $G^{(2)}_{q\lambda,q'\lambda'}(\tau)$ [Eq. (5.16))] can be calculated as:

$$G^{(2)}_{q\lambda,q'\lambda'}(\tau) = 4g_2 = \begin{cases} \frac{\delta_{qq'}}{4}\int_0^{\beta\hbar} d\tau_1 \int_0^{\beta\hbar} d\tau_2 \sum_{j_1'=\frac{3r}{2}+1}^{3r} \frac{V_{\lambda' j_1'}(q')V_{\lambda j_1'}^*(q)}{\omega_{q j_1'}\sqrt{\omega_{q\lambda}\omega_{q'\lambda'}}} \left[ G^0_{q j_1'}(\tau_2,\tau_1)G^0_{q'\lambda'}(\tau_1,0)G^0_{q\lambda}(\tau,\tau_2) \right], & \lambda,\lambda' \in \left[1,\frac{3r}{2}\right] \\ \frac{\delta_{qq'}}{4}\int_0^{\beta\hbar} d\tau_1 \int_0^{\beta\hbar} d\tau_2 \sum_{j_1=1}^{\frac{3r}{2}} \frac{V_{j_1\lambda}(q)V_{j_1\lambda'}^*(q')}{\omega_{q j_1}\sqrt{\omega_{q\lambda}\omega_{q'\lambda'}}} \left[ G^0_{q j_1}(\tau_1,\tau_2)G^0_{q'\lambda'}(\tau_2,0)G^0_{q\lambda}(\tau,\tau_1) \right], & \lambda,\lambda' \in \left[\frac{3r}{2}+1,3r\right] \\ 0 & \text{else} \end{cases} \quad (5.29)$$

To calculate $G^{(2)}_{q\lambda,q'\lambda'}$ in frequency space, we expand $G^{(2)}_{q\lambda,q'\lambda'}(\tau)$ in a Fourier series:

$$\begin{cases} G^{(2)}_{q\lambda,q'\lambda'}(\tau) = \frac{1}{\beta\hbar}\sum_n e^{-i\omega_n\tau}G^{(2)}_{q\lambda,q'\lambda'}(i\omega_n) \\ G^{(2)}_{q\lambda,q'\lambda'}(i\omega_n) = \int_0^{\beta\hbar} d\tau e^{i\omega_n\tau}G^{(2)}_{q\lambda,q'\lambda'}(\tau) \end{cases} \quad (5.30)$$

Using Eq. (4.21) we get for $\lambda,\lambda' \in \left[1,\frac{3r}{2}\right]$,

$$G^{(2)}_{q\lambda,q'\lambda'}(\tau) = \frac{\delta_{qq'}}{4}\int_0^{\beta\hbar}d\tau_1\int_0^{\beta\hbar}d\tau_2\sum_{j_1'=\frac{3r}{2}+1}^{3r}\frac{V_{\lambda'j_1'}(q')V^*_{\lambda j_1'}(q)}{\omega_{qj_1'}\sqrt{\omega_{q\lambda}\omega_{q'\lambda'}}}\left[G^0_{qj_1'}(\tau_2,\tau_1)G^0_{q\lambda'}(\tau_1,0)G^0_{q\lambda}(\tau,\tau_2)\right]$$

$$= \frac{\delta_{qq'}}{4}\sum_{j_1'=\frac{3r}{2}+1}^{3r}\frac{V_{\lambda'j_1'}(q')V^*_{\lambda j_1'}(q)}{\omega_{qj_1'}\sqrt{\omega_{q\lambda}\omega_{q'\lambda'}}}\int_0^{\beta\hbar}d\tau_1\int_0^{\beta\hbar}d\tau_2\left[\frac{1}{\beta\hbar}\sum_{n_2}e^{-i\omega_{n_2}(\tau_2-\tau_1)}G^0_{qj_1'}(i\omega_{n_2})\cdot\frac{1}{\beta\hbar}\sum_{n_1}e^{-i\omega_{n_1}\tau_1}G^0_{q\lambda'}(i\omega_{n_1})\right.$$

$$\left.\cdot\frac{1}{\beta\hbar}\sum_n e^{-i\omega_n(\tau-\tau_2)}G^0_{q\lambda}(i\omega_n)\right]$$

$$= \frac{\delta_{qq'}}{4}\sum_{j_1'=\frac{3r}{2}+1}^{3r}\frac{V_{\lambda'j_1'}(q')V^*_{\lambda j_1'}(q)}{\omega_{qj_1'}\sqrt{\omega_{q\lambda}\omega_{q'\lambda'}}}\frac{1}{\beta\hbar}\sum_{n,n_1,n_2}e^{-i\omega_n\tau}G^0_{q\lambda}(i\omega_n)G^0_{q\lambda'}(i\omega_{n_1})G^0_{qj_1'}(i\omega_{n_2})\begin{Bmatrix}\frac{1}{\beta\hbar}\int_0^{\beta\hbar}d\tau_1\,e^{-i(\omega_{n_1}-\omega_{n_2})\tau_1}\\\times\frac{1}{\beta\hbar}\int_0^{\beta\hbar}d\tau_2\,e^{-i(\omega_{n_2}-\omega_n)\tau_2}\end{Bmatrix}$$

$$= \frac{\delta_{qq'}}{4}\sum_{j_1'=\frac{3r}{2}+1}^{3r}\frac{V_{\lambda'j_1'}(q')V^*_{\lambda j_1'}(q)}{\omega_{qj_1'}\sqrt{\omega_{q\lambda}\omega_{q'\lambda'}}}\frac{1}{\beta\hbar}\sum_{n,n_1,n_2}e^{-i\omega_n\tau}G^0_{q\lambda}(i\omega_n)G^0_{q\lambda'}(i\omega_{n_1})G^0_{qj_1'}(i\omega_{n_2})\delta_{n_1n_2}\delta_{nn_2}$$

$$= \frac{\delta_{qq'}}{\beta\hbar}\sum_n e^{-i\omega_n\tau}\left\{G^0_{q\lambda}(i\omega_n)\left[\frac{1}{4}\sum_{j_1'=\frac{3r}{2}+1}^{3r}\frac{V_{\lambda'j_1'}(q')V^*_{\lambda j_1'}(q)}{\omega_{qj_1'}\sqrt{\omega_{q\lambda}\omega_{q'\lambda'}}}G^0_{qj_1'}(i\omega_n)\right]G^0_{q\lambda'}(i\omega_n)\right\}$$

Comparing above expression with Eq. (5.30), the Fourier transform of $G^{(2)}_{q\lambda,q'\lambda'}(\tau)$ reads as

$$G^{(2)}_{q\lambda,q'\lambda'}(i\omega_n) = G^0_{q\lambda}(i\omega_n)\left[\frac{\delta_{qq'}}{4}\sum_{j_1'=\frac{3r}{2}+1}^{3r}\frac{V_{\lambda'j_1'}(q')V^*_{\lambda j_1'}(q)}{\omega_{qj_1'}\sqrt{\omega_{q\lambda}\omega_{q'\lambda'}}}G^0_{qj_1'}(i\omega_n)\right]G^0_{q\lambda'}(i\omega_n) \qquad (5.31)$$

where we have introduced the notation:

$$\Sigma^{(2)}_{q\lambda,q'\lambda'}(i\omega_n) \equiv \frac{\delta_{qq'}}{4}\sum_{j_1'=\frac{3r}{2}+1}^{3r}\frac{V_{\lambda'j_1'}(q')V^*_{\lambda j_1'}(q)}{\omega_{qj_1'}\sqrt{\omega_{q\lambda}\omega_{q'\lambda'}}}G^0_{qj_1'}(i\omega_n) \qquad (5.32)$$

and $G^{(2)}_{q\lambda,q'\lambda'}(i\omega_n) = G^0_{q\lambda}(i\omega_n)\cdot\Sigma^{(2)}_{q\lambda,q'\lambda'}(i\omega_n)\cdot G^0_{q\lambda'}(i\omega_n)$. Similarly, for $\lambda,\lambda' \in \left[\frac{3r}{2}+1,3r\right]$ we have:

$$\Sigma^{(2)}_{q\lambda,q'\lambda'}(i\omega_n) = \frac{\delta_{qq'}}{4}\sum_{j_1=1}^{\frac{3r}{2}}\frac{V_{j_1\lambda}(q)V^*_{j_1\lambda'}(q')}{\omega_{qj_1}\sqrt{\omega_{q\lambda}\omega_{q'\lambda'}}}G^0_{qj_1}(i\omega_n) \qquad (5.33)$$

and all together we have for $\Sigma^{(2)}_{q\lambda,q'\lambda'}(i\omega_n)$:

$$\Sigma^{(2)}_{q\lambda,q'\lambda'}(i\omega_n) = \begin{cases}\frac{\delta_{qq'}}{4}\sum_{j_1'=\frac{3r}{2}+1}^{3r}\frac{V_{\lambda'j_1'}(q')V^*_{\lambda j_1'}(q)}{\omega_{qj_1'}\sqrt{\omega_{q\lambda}\omega_{q'\lambda'}}}G^0_{qj_1'}(i\omega_n), & \lambda,\lambda' \in \left[1,\frac{3r}{2}\right]\\ \frac{\delta_{qq'}}{4}\sum_{j_1=1}^{\frac{3r}{2}}\frac{V_{j_1\lambda}(q)V^*_{j_1\lambda'}(q')}{\omega_{qj_1}\sqrt{\omega_{q\lambda}\omega_{q'\lambda'}}}G^0_{qj_1}(i\omega_n), & \lambda,\lambda' \in \left[\frac{3r}{2}+1,3r\right]\\ 0 & \text{else}\end{cases} \qquad (5.34)$$

### 5.3.3 Dyson's equation

Substituting Eqs. (5.21) and (5.34) into Eq. (5.16) and performing a Fourier transform, we have:

$$G_{q\lambda,q'\lambda'}(i\omega_n) = G^0_{q\lambda,q'\lambda'}(i\omega_n) + G^{(1)}_{q\lambda,q'\lambda'}(i\omega_n) + G^{(2)}_{q\lambda,q'\lambda'}(i\omega_n) + \cdots$$

$$= G^0_{q\lambda}(i\omega_n)\delta_{qq'}\delta_{\lambda\lambda'} + G^0_{q\lambda}(i\omega_n) \cdot \Sigma^{(1)}_{q\lambda,q'\lambda'}(i\omega_n) \cdot G^0_{q'\lambda'}(i\omega_n) + G^0_{q\lambda}(i\omega_n) \cdot \Sigma^{(2)}_{q\lambda,q'\lambda'}(i\omega_n)$$

$$\cdot G^0_{q'\lambda'}(i\omega_n) + \cdots = G^0_{q\lambda}(i\omega_n)\delta_{qq'}\delta_{\lambda\lambda'} + G^0_{q\lambda}(i\omega_n) \cdot \Sigma_{q\lambda,q'\lambda'}(i\omega_n) \cdot G^0_{q'\lambda'}(i\omega_n)$$

or

$$G_{q\lambda,q'\lambda'}(i\omega_n) = G^0_{q\lambda}(i\omega_n)\delta_{qq'}\delta_{\lambda\lambda'} + G^0_{q\lambda}(i\omega_n) \cdot \Sigma_{q\lambda,q'\lambda'}(i\omega_n) \cdot G^0_{q'\lambda'}(i\omega_n) \quad (5.35)$$

Eq. (5.35) is the so-called Dyson's equation (see Ref. [2]), where

$$\Sigma_{q\lambda,q\lambda'}(i\omega_n) = \Sigma^{(1)}_{q\lambda,q'\lambda'}(i\omega_n) + \Sigma^{(2)}_{q\lambda,q'\lambda'}(i\omega_n) + \cdots$$

is called self-energy and $\Sigma^{(n)}_{q\lambda,q\lambda'}(i\omega_n), n = 1,2,\cdots$ is the self-energy of $n^{\text{th}}$ order approximation. Up to the second order approximation, the self-energy is written as

$$\Sigma_{q\lambda,q'\lambda'}(i\omega_n) = \begin{cases} -\dfrac{\delta_{qq'}}{2}\dfrac{V^*_{\lambda\lambda'}(q)}{\sqrt{\omega_{q\lambda}\widetilde{\omega}_{q\lambda'}}}, & 1 \le \lambda \le \dfrac{3r}{2} < \lambda' \le 3r \\[2mm] -\dfrac{\delta_{qq'}}{2}\dfrac{V_{\lambda'\lambda}(q)}{\sqrt{\widetilde{\omega}_{q\lambda}\omega_{q\lambda'}}}, & 1 \le \lambda' \le \dfrac{3r}{2} < \lambda \le 3r \\[2mm] \dfrac{\delta_{qq'}}{4}\sum_{j'_1=\frac{3r}{2}+1}^{3r}\dfrac{V_{\lambda'j'_1}(q')V^*_{\lambda j'_1}(q)}{\widetilde{\omega}_{qj'_1}\sqrt{\omega_{q\lambda}\omega_{q'\lambda'}}} G^0_{qj'_1}(i\omega_n), & \lambda,\lambda' \in \left[1,\dfrac{3r}{2}\right] \\[2mm] \dfrac{\delta_{qq'}}{4}\sum_{j_1=1}^{\frac{3r}{2}}\dfrac{V_{j_1\lambda}(q)V^*_{j_1\lambda'}(q')}{\omega_{qj_1}\sqrt{\widetilde{\omega}_{q\lambda}\widetilde{\omega}_{q'\lambda'}}} G^0_{qj_1}(i\omega_n), & \lambda,\lambda' \in \left[\dfrac{3r}{2}+1,3r\right] \end{cases} \quad (5.36)$$

### 5.4 Fermi's golden Rule

To get the Fermi's golden Rule, we need to calculate the retarded Green's function, which can be calculated from $G_{q\lambda,q\lambda'}(i\omega_n)$ by analytic continuation to the real axis via $i\omega_n \to \omega + i\eta$ with an infinitesimal positive $\eta$ [1,2]:

$$G^r_{q\lambda,q\lambda'}(\omega) \equiv G_{q\lambda,q\lambda'}(i\omega_n \to \omega + i\eta), \eta \to 0 \quad (5.37)$$

Similarity, the retarded self-energy is defined as

$$\Sigma^r_{q\lambda,q\lambda'}(\omega) \equiv \Sigma_{q\lambda,q\lambda'}(i\omega_n \to \omega + i\eta), \eta \to 0 \quad (5.38)$$

The transition rate of phonons of branch $\lambda$ at $q$, $\Gamma_{q\lambda}(\omega)$, is related to the retarded self energy $\Sigma^r_{q\lambda}(\omega)$ via [1,3]:

$$\Gamma_{q\lambda}(\omega) = -2\text{Im}\Sigma^r_{q\lambda,q\lambda}(\omega). \quad (5.39)$$

Using the expression for $G^0_{q\lambda}(i\omega_n)$ [Eq. (4.21)] and focused on the second order approximation [1,3] we have:

$$\Sigma_{q\lambda,q\lambda}^{r}(i\omega_n) = \begin{cases} \frac{1}{4}\sum_{j_1'=\frac{3r}{2}+1}^{3r} \frac{|V_{\lambda j_1'}(q)|^2}{\widetilde{\omega}_{qj_1'}\omega_{q\lambda}} \frac{1}{(\omega-\omega_{qj_1'})+i\eta}, & \lambda \in \left[1,\frac{3r}{2}\right] \\ \frac{1}{4}\sum_{j_1=1}^{\frac{3r}{2}} \frac{|V_{j_1\lambda}(q)|^2}{\omega_{qj_1}\widetilde{\omega}_{q\lambda}} \frac{1}{(\omega-\omega_{qj_1})+i\eta}, & \lambda \in \left[\frac{3r}{2}+1,3r\right] \end{cases} \quad (5.40)$$

Using the relation:

$$\text{Im}\left[\frac{1}{(\omega-\omega_{qj_1'})+i\eta}\right] = -\pi\delta(\omega-\omega_{qj_1'}). \quad (5.41)$$

The transition rate reads as

$$\Gamma_{q\lambda}(\omega) = \begin{cases} \frac{\pi}{2}\sum_{j_1'=\frac{3r}{2}+1}^{3r} \frac{|V_{\lambda j_1'}(q)|^2}{\widetilde{\omega}_{qj_1'}\omega_{q\lambda}} \delta(\omega-\omega_{qj_1'}), & \lambda \in \left[1,\frac{3r}{2}\right] \\ \frac{\pi}{2}\sum_{j_1=1}^{\frac{3r}{2}} \frac{|V_{j_1\lambda}(q)|^2}{\omega_{qj_1}\widetilde{\omega}_{q\lambda}} \delta(\omega-\omega_{qj_1}), & \lambda \in \left[\frac{3r}{2}+1,3r\right] \end{cases} \quad (5.42)$$

Or equivalently,

$$\Gamma_{q\lambda}(E=\hbar\omega) = \begin{cases} \frac{\pi\hbar^3}{2}\sum_{j_1'=\frac{3r}{2}+1}^{3r} \frac{|V_{\lambda j_1'}(q)|^2}{E_{qj_1'}E_{q\lambda}} \delta(E-E_{qj_1'}), & \lambda \in \left[1,\frac{3r}{2}\right] \\ \frac{\pi\hbar^3}{2}\sum_{j_1=1}^{\frac{3r}{2}} \frac{|V_{j_1\lambda}(q)|^2}{E_{qj_1}E_{q\lambda}} \delta(E-E_{qj_1}), & \lambda \in \left[\frac{3r}{2}+1,3r\right] \end{cases} \quad (5.43)$$

$\Gamma_{q\lambda}(E)$ represents the probability per unit time of a transition with energy $E$ from phonon branch $\lambda$ at wave number $q$ in subsystem I (II) to a set of phonon branches in subsystem II (I) with the same wave number. Here, state $q\lambda$ in one subsystem is coupled to state $qj_1'$ in the other subsystem via: $V_{\lambda j_1'}(q)$. Since the two expressions in Eq. (5.43) are completely equivalent just representing transitions of opposite directions we consider only the first one in what follows. Assuming that subsystem II sufficiently large such that its density of states is nearly continuous, the sum over its states in the above expression can be replaced by an integral over the energy $\sum_j(\cdots) \to \int(\cdots)\rho(E_q)dE_q$ giving:

$$\Gamma_{q\lambda}(E) = \frac{\pi\hbar^3}{2}\int dE_q'\rho(E_q')\frac{\{|V_{\lambda j'}(q)|^2\}_{E_{qj'}=E_q'}}{E_q'E_{q\lambda}}\delta(E-E_q') = \frac{\pi\hbar^3}{2}\rho(E)\frac{\{|V_{\lambda j'}(q)|^2\}_{E_{qj'}=E}}{EE_{q\lambda}}. \quad (5.44)$$

Eq. (5.44) represents the transition rate of the phonons with energy $E$ from branch $\lambda$ and wave number $q$ in subsystem I to the continuous manifold of states in subsystem II. The total transition rate between the two subsystems is then given by the sum of transition rates from all states in subsystem I, weighted by their phonon populations, to the manifold of states in subsystem II. Assuming that the two subsystems are weakly coupled such that the width of state $q\lambda$ in subsystem I is small and the probability to leave it at an energy other than $E_{q\lambda}$ is negligible we can now write total transition rate as:

$$\Gamma_{\text{tot}} = \sum_{q\lambda} \frac{1}{Z} e^{-\beta E_{q\lambda}} \Gamma_{q\lambda}(E_{q\lambda}) = \frac{\pi\hbar^3}{2} \sum_{q\lambda} \frac{e^{-\beta E_{q\lambda}}}{Z} \frac{\rho(E_{q\lambda}) \left\{ |V_{\lambda j'}(q)|^2 \right\}_{E_{qj'}=E_{q\lambda}}}{E_{q\lambda}^2}$$

$$= \frac{\pi\hbar^3}{2} \sum_{q\lambda} \frac{e^{-\beta E_{q\lambda}}}{Z} \frac{\rho(E_{q\lambda}) \left| V_{\lambda,\lambda+\frac{3r}{2}}(q) \right|^2}{E_{q\lambda}^2}.$$

Finally we get

$$\Gamma_{\text{tot}} = \Gamma_{q\lambda}(E) = \frac{\pi\hbar^3}{2} \sum_{q\lambda} \frac{e^{-\beta E_{q\lambda}}}{Z} \frac{\rho(E_{q\lambda}) \left| V_{\lambda,\lambda+\frac{3r}{2}}(q) \right|^2}{E_{q\lambda}^2}. \tag{5.45}$$

where $Z = \sum_{q\lambda} e^{-\beta E_{q\lambda}}$ is the partition function. Here, we choose the index of phonon branches $\lambda$ such that the phonon branch with lower energy has a smaller index, i.e., $E_{q\lambda_1} < E_{q\lambda_2}$ for index $\lambda_1 < \lambda_2$. Note that phonon branches $\lambda$ and $j'_1$ belong to the subsystem I (with index from 1 to $3r/2$) and subsystem II (with index from 1 to $1 + 3r/2$ to $3r$), respectively, we choose $j' = \lambda + \frac{3r}{2}$ to make sure that $E_{qj'} = E_{q\lambda}$ since phonon energy of two uncoupled system is identical. Eq. (5.45) is the Fermi's golden rule for inter-phonon coupling.